\documentclass[12pt]{article}
\usepackage{xr}
\usepackage{authblk}
\usepackage{amsmath,amsthm,amssymb}
\usepackage{graphicx}
\usepackage{enumerate}
\usepackage{natbib}
\usepackage{tabularx}
\usepackage[table]{xcolor}
\usepackage{url} 

\usepackage{tcolorbox}
\newtcolorbox[auto counter,number within=section]{pabox}[2][]{%
    colback=black!5!white,colframe=black!75!black,fonttitle=\bfseries,
title=Bound ~\thetcbcounter: #2,#1}

\usepackage[colorlinks=true,linkcolor=black,anchorcolor=black,citecolor=black,filecolor=black,menucolor=black,runcolor=black,urlcolor=black]{hyperref}

\usepackage{microtype}
\usepackage{graphicx}
\graphicspath{{./../}} 
\usepackage{booktabs} 
\usepackage{commath}
\usepackage{bbm}
\usepackage{xspace}

\usepackage{comment} 
\usepackage{float} 
\usepackage{subcaption}

\usepackage [autostyle, english = american]{csquotes}
\MakeOuterQuote{"}

\newcommand{\R}{\mathbb{R}}

\newcommand{\pr}{\mathbb{P}}
\newcommand{\N}{\mathbb{N}}

\newcommand{\E}{\mathbb{E}}
\newcommand{\Var}{\mathrm{Var}}
\newcommand{\indep}{\raisebox{0.05em}{\rotatebox[origin=c]{90}{$\models$}}}

\newcommand{\inv}{^{-1}}
\newcommand{\msqrt}{^{\frac{1}{2}}}
\newcommand{\nsqrt}{^{-\frac{1}{2}}}
\newcommand{\diag}{\mathrm{diag}}
\newcommand{\trace}{\mathrm{tr}}
\newcommand{\ones}[1]{\mathbf{1}_{#1}}
\newcommand{\opnorm}[1]{\|#1\|_{\mathrm{OP}}}
\usepackage[capitalize]{cleveref}
\crefname{supp}{Supplement}{Supplements}

\def\[#1\]{\begin{align}#1\end{align}}        
\def\(#1\){\begin{align*}#1\end{align*}}     

\newcommand{\Revision}[1]{#1}
\newcommand{\BLT}[1]{\textcolor{red}{ {\small \bf (!BLT)} #1}}

\DeclareMathOperator*{\argmax}{arg\,max}
\DeclareMathOperator*{\argmin}{arg\,min}

\newtheorem{theorem}{Theorem}[section]

\newtheorem{prop}[theorem]{Proposition}

\newtheorem{bound}{Bound}[section]
\newtheorem{approxbound}{Approximate Bound}[section]
\newtheorem{lemma}{Lemma}[section]
\theoremstyle{remark}
\newtheorem{remark}[theorem]{Remark}

\newcommand{\blind}{1}

\begin{document}

\def\spacingset#1{\renewcommand{\baselinestretch}%
{#1}\small\normalsize} \spacingset{1}


\if1\blind
{
    \title{\bf Confidently Comparing Estimat\Revision{es} with the c-value}
  \author[1]{Brian L.~Trippe}
  \author[2]{Sameer K.~Deshpande}
  \author[3]{Tamara Broderick}
  \affil[1]{Department of Statistics, Columbia University}
  \affil[2]{Department of Statistics, University of Wisconsin--Madison}
  \affil[3]{Laboratory for Information and Decision Systems, Massachusetts Institute of Technology}
  \maketitle
} \fi

\if0\blind
{
  \bigskip
  \bigskip
  \bigskip
  \begin{center}
      {\LARGE\bf Confidently Comparing Estimat\Revision{es} with the c-value}
\end{center}
  \medskip
} \fi

\bigskip
\begin{abstract}
Modern statistics provides an ever-expanding toolkit for estimating unknown parameters.
Consequently, applied statisticians frequently face a difficult decision:
retain a parameter estimate from a familiar method or replace it with an estimate from a newer or more complex one.
While it is traditional to compare estimates using risk, such comparisons are rarely conclusive in realistic settings.

In response, we propose the ``c-value'' as a measure of confidence that a new estimate achieves smaller loss than an old estimate on a given dataset.
We show that it is unlikely that a \Revision{large c-value coincides with a larger loss for the new estimate.}
Therefore, just as a small \emph{p}-value supports rejecting a null hypothesis, a large c-value supports using a new estimate in place of the old.
For a wide class of problems and estimates, we show how to compute a c-value by first constructing a data-dependent high-probability lower bound on the difference in loss.
The c-value is frequentist in nature, but we show that it can provide validation of \Revision{shrinkage estimates derived from Bayesian models} in real data applications involving hierarchical models and Gaussian processes.

\end{abstract}

\noindent%
{\it Keywords:}  Decision Theory, Normal Means, Model Selection, Shrinkage, Empirical Bayes 
\vfill

\newpage
\spacingset{1.5}

\section{Introduction}\label{sec:intro}
Modern statistics provides an expansive toolkit of sophisticated methodology for estimating unknown parameters.
However, the abundance of different estimators often presents practitioners with a difficult challenge:
choosing between the output of a familiar method (e.g.\ a maximum likelihood estimate (MLE)) and that of a more complicated method (e.g.\ the posterior mean of a hierarchical Bayesian model).
From a practical perspective, abandoning a familiar approach in favor of a newer alternative is unreasonable without some assurance that the latter provides a more accurate estimate.
Our goal is to determine whether it is safe to abandon a default estimate in favor of an alternative, and to provide an assessment of the degree of confidence we should have in this decision.

Traditionally decisions between estimators are based on risk, the loss averaged over all possible realizations of the data \Revision{with respect to a pre-specified likelihood model} \citep[Chapters 4-5]{lehmann2006theory}. 
We note two limitations of using risk.
First, it is rare that one estimator within a given pair will have smaller risk across all possible parameter values.
Instead, it is more often the case that one estimator will have smaller risk for some unknown parameter values but larger risk for other parameter values.
Second, one estimator may have lower risk than another but incur higher loss on a majority of datasets; see \Cref{sec:lindley_and_smith_risk} for an example
\Revision{in which an estimator with smaller risk has larger loss on nearly 70\% of simulated datasets.}


In this work we propose a framework for choosing between estimators based on their performance \textit{on the observed dataset} rather than their risk.
Specifically, we introduce the ``c-value'' (``c'' for confidence in the new estimate), which we construct using a data-dependent high-probability lower bound on the difference in loss.
We show that it is unlikely that simultaneously the c-value is large and the alternative estimate has larger loss than the default.
\Revision{
For the c-value to be useful, it must meet two desiderata:
\begin{enumerate}
\item{The c-value must not frequently guide practitioners to incorrectly report the alternative estimate when the default estimate has smaller loss.}
\item{
The c-value must, in some cases, allow one to correctly identify that the alternative estimate has smaller loss.}
\end{enumerate}
We demonstrate that the c-value meets the first desideratum with theory showing how to use the c-value to select between two estimates in a principled, data-driven way.
Critically, the c-value requires no assumptions on the unknown parameter; our guarantees hold uniformly across the parameter space.
We demonstrate that the c-value can meet the second desideratum with case studies;
we provide an overview of these next as motivating examples,
and then proceed to present our general methodology.
}
\paragraph{Shrinkage estimates on educational testing data.}
\Revision{We revisit  \citet{hoff2019smaller}'s estimates of average student reading ability at several schools in the 2002 Educational Longitudinal Study.
These estimates are obtained from a hierarchical Bayesian model that ``shares strength'' by partially pooling data across similar schools.
\citet{hoff2019smaller}'s analysis relied on a simplifying and subjectively chosen prior.
A practitioner might wonder whether the resulting estimates are more accurate than the MLE in terms of squared error loss.
As we will see, a large c-value provides confidence that Hoff's estimate is indeed more accurate.
We additionally consider a clearly inappropriate prior and verify that our methodology does not always favor more complex alternative estimators.
Although these estimates have a Bayesian provenance, the use of the c-value to justify these estimates does not require subjective belief in the prior.
}

\paragraph{Estimating violent crime density at the neighborhood level.}
Considerable empirical evidence links a community's exposure to violent crime and adverse behavioral, mental, and physical health outcomes among its residents \citep{Buka2001, Kondo2018}.
Although overall violent crimes rates in the U.S. have decreased over the last two decades, there is considerable variation in time trends at the neighborhood level \citep{BalocchiJensen2019, Balocchi2019}.
A critical first step in understanding what drives neighborhood-level variation is accurate estimation of the actual amount of violent crime that occurs in each neighborhood.

Typically, researchers rely on the reported counts of violent crime aggregated at small spatial resolutions (e.g.\ at the census tract level). 
However, in light of sampling variability due to the relative infrequency of certain crime types in small areas, it is natural to wonder if auxiliary data could be used to improve estimates of violent crime incidence.

As a second application of our framework, we analyze the number of violent crimes reported per square mile in several neighborhoods in the city of Philadelphia.
Our analysis suggests that one can obtain improved estimates of the violent crime density by using \Revision{a shrinkage estimate} that incorporates information about non-violent crime incidence.
Further c-value analysis reveals that leveraging spatial information on top of non-violent incidence does not provide additional improvement.

\paragraph{Gaussian process kernel choice: modeling ocean currents.}
Accurate estimation of ocean current dynamics is critical for forecasting the dispersion of oceanic contaminations \citep{poje2014submesoscale}. 
While it is commonplace to model ocean flow dynamics at or above the \textit{mesoscale} (roughly 10 km), \citet{lodise2020investigating} have recently advocated modeling dynamics at both the mesoscale and the \textit{submesoscale} (roughly 0.1--10 km).
They specifically proposed a Gaussian process model that accounts for variation across multiple resolutions to estimate ocean currents from positional data taken from hundreds of free-floating buoys.

In a third application of our framework, we find that the multi-resolution procedure produces a large c-value, indicating that accounting for variation across multiple scales enables more accurate estimates than are obtained when accounting only for mesoscale variation.
    \subsection{Organization of the article and contributions}
We formally present our general framework and define the c-value in \Cref{sec:c_values}.
In \Cref{sec:background_and_related_work} we highlight similarities and differences between our framework and existing work on preliminary testing and post-selection inference.
Our approach to computing c-values depends on the availability of high-confidence lower bounds on the difference in the losses of the two estimates that holds uniformly across the parameter space.
\Cref{sec:simpler_cases,sec:affine,sec:more_cases} provide these bounds for several models and classes of estimators for squared error loss.
In \Cref{sec:simpler_cases}, we illustrate our general strategy in the canonical normal means problem.
Then, in \Cref{sec:affine}, we generalize this strategy to compare affine estimates of normal means with correlated observations.
\Revision{\Cref{sec:more_cases} shows how to extend the framework to cover two nonlinear cases:
a nonlinear shrinkage estimator and regularized logistic regression.
We provide simulations validating our approach in these settings.}
We apply our framework to the aforementioned motivating examples in \Cref{sec:applications}.
\Revision{In our discussion in \Cref{sec:discussion}, we outline ways to extend our framework beyond the estimates considered here.}
Software that implements the c-value computation, and code that reproduces our analyses is available at: \url{https://github.com/blt2114/c_values}.
\section{Introducing the c-value}\label{sec:c_values}
We now describe our approach for quantifying confidence in the statement that one estimate of an unknown parameter is superior to another.
We begin by introducing some notation and building up to a definition of the c-value,
before stating our main results.
This development is very general, and we defer practical considerations to the subsequent sections.
We include proofs of the results of this section in \Cref{sec:appendix}.

Suppose that we observe data $y$ drawn from some distribution that depends on an unknown parameter $\theta.$
We consider deciding between two estimates, $\hat{\theta}(y)$ and $\theta^{*}(y),$ of $\theta$ on the basis of a loss function $L(\theta,\cdot).$
Our focus is on asymmetric situations in which $\hat{\theta}(\cdot)$ is a standard or more familiar estimator while $\theta^{*}(\cdot)$ is a less familiar estimator. 
For simplicity, we will refer to $\hat{\theta}(\cdot)$ as the default estimator and $\theta^{*}(\cdot)$ as the alternative estimator.

We next define the ``win'' obtained by using $\theta^{*}(y)$ rather than $\hat{\theta}(y)$ as the difference in loss, $W(\theta, y) := L(\theta, \hat{\theta}(y)) - L(\theta, \theta^{*}(y)).$
While a typical comparison based on risk would proceed by taking the expectation of $W(\theta, y)$ over all possible datasets drawn for fixed $\theta,$ we maintain focus on the single observed dataset.
Notably, the win is positive whenever the alternative estimate achieves a smaller loss than the default estimate. 
As such, if we knew that $W(\theta, y) > 0$ for the given dataset $y$ and unknown parameter $\theta,$ then we would prefer to use the alternative $\theta^{*}(y)$ instead of the default $\hat{\theta}(y).$

Since $\theta$ is unknown, determining whether $W(\theta, y) > 0$ is impossible.
Nevertheless, for a broad class of estimators, we can determine whether the win is positive with high probability. 
To start, we construct a lower bound, $b(y,\alpha),$ depending only on the data and a pre-specified level $\alpha \in [0,1],$ that satisfies for all $\theta$
\begin{equation}
\label{eqn:bya_bound}
\pr_{\theta}\left[W(\theta, y) \ge b(y,\alpha) \right] \ge \alpha.
\end{equation}
For values of $\alpha$ close to $1,\, b(y,\alpha)$ is a high-probability lower bound on the win that holds uniformly across all possible values of the unknown parameter $\theta.$
Loosely speaking, if $b(y,\alpha) > 0$ for some $\alpha$ close to $1$, then we can be confident that the alternative estimate has smaller loss than the default estimate.

To make this intuition more precise, we define a measure of confidence that $\theta^*(y)$ is superior to $\hat{\theta}(y).$
We call our measure the c-value $c(y)$:
\begin{equation}
\label{eq:c_value}
c(y) := \inf_{\alpha \in [0, 1]}\left\{\alpha \mid b(y,\alpha) \le 0 \right\}.
\end{equation}
The c-value marks a meaningful boundary in the space of confidence levels;
it is the largest value such that for every $\alpha < c(y),$ we have confidence $\alpha$ that the win is positive.
\begin{remark}\label{remark:c_value_inf_vs_sup}
An alternative definition for the c-value is
$c^+(y) = \sup_{\alpha\in [0, 1] }\{ \alpha | b(y, \alpha) \ge 0\}.$
Although $c^+(y)=c(y)$ when $b(y, \cdot)$ is continuous and strictly decreasing in $\alpha,$ 
$c^+(\cdot)$ may be overconfident otherwise.
We detail a particularly pathological example in \Cref{sec:c_value_inf_vs_sup}.
\end{remark}

Our first main result formalizes the interpretation of $c(y)$ as a measure of confidence.
\begin{theorem}
\label{thm:c_values}
Let $b(\cdot, \cdot)$ be any function satisfying the condition in \Cref{eqn:bya_bound}.
Then for any $\theta$ and $\alpha \in[0,1]$ and $c(y)$ as defined in \Cref{eq:c_value},
\begin{equation}
\label{eqn:c_value_guarantee}
\pr_\theta\left[ W(\theta, y) \le 0 \textrm{ and } c(y) > \alpha \right] \le 1- \alpha.
\end{equation}
\end{theorem}
The result follows directly from the definition of $c(\cdot)$ and the condition on $b(\cdot, \cdot).$
Informally, \Cref{thm:c_values} assures us that it is unlikely that simultaneously (A) the $c$-value is large and (B) $\theta^{*}(y)$ does not provide smaller loss than $\hat{\theta}(y).$
Just as a small \emph{p}-value supports rejecting a null hypothesis,
a large c-value supports abandoning the default estimate in favor of the alternative.

The strategy described above necessarily uses the data twice,
once to compute the two estimates and once more to compute the c-value to choose between them.
Accordingly, one might justly ask how such double use of the data affects the quality of the resulting procedure.
To address this question, we formalize this two-step procedure with a single estimator,
\begin{equation}
\label{eqn:theta_tilde}
\theta^\dagger(y,\alpha) := \mathbbm{1}[c(y) \le \alpha]\hat{\theta}(y) + \mathbbm{1}[c(y) > \alpha]\theta^{*}(y).
\end{equation}
$\theta^\dagger(y,\alpha)$ picks between the two estimates $\hat{\theta}(y)$ and $\theta^{*}(y)$ based on the value $c(y)$ and a pre-specified level $\alpha \in [0,1].$
\Revision{We can characterize the possible outcomes when using $\theta^\dagger(\cdot, \alpha)$ with a contingency table (\Cref{table:contingency_outcomes}), where rows correspond to the estimate with smaller loss, and the columns correspond to the reported estimate.}
{\Revision{ \begin{table}[h!]
\caption{
    \Revision{Contingency table with possible outcomes when using the two-stage estimator $\theta^\dagger(\cdot, \alpha)$.
    $\theta^\dagger(\cdot, \alpha)$ controls the probability of the shaded event (\Cref{thm:theta_tilde_guarantee}).}}
\label{table:contingency_outcomes}
\centering
\begin{tabular}{ c}
\begin{tabularx}{0.90\textwidth}{| X | X | X|}
\hline 
    & Default reported & Alternative reported \\
\hline 
    Default has lower loss &  Correct & \cellcolor{blue!25} Incorrect\\
\hline
    Alternative has lower loss &  Incorrect & Correct\\
\hline
\end{tabularx}
\end{tabular}
\end{table}
}}

\Revision{
Recall that we are interested in an asymmetric situation where the alternative estimator is less familiar than the default estimator.
This asymmetry makes desirable the reassurance that $\theta^\dagger(\cdot, \alpha)$ does not incur greater loss than $\hat \theta (\cdot).$
As such, we focus on the upper right hand entry of the table.
Our second main result formalizes that when we use $\theta^\dagger(\cdot, \alpha)$ with $\alpha$ close to 1, the probability of the event represented by this table entry is small.
}
\begin{theorem}
\label{thm:theta_tilde_guarantee}
Let $b(\cdot, \cdot)$ be any function that satisfies the condition in \Cref{eqn:bya_bound}.
Then for any $\theta$ and $\alpha\in[0,1]$,
\begin{equation}
\label{eqn:guarantee}
\pr_\theta\left[L\left(\theta, \theta^\dagger(y, \alpha)\right) > L\left(\theta, \hat \theta(y)\right) \right]  \le 1-\alpha.
\end{equation}
\end{theorem}

\paragraph{Overview of the remainder of the paper.}
The c-value is useful insofar as the lower bound $b(y,\alpha)$ is sufficiently tight and readily computable.
It remains to show that such practical bounds exist.
A primary contribution of this work is the explicit construction of these bounds in settings of practical interest.
In what follows, we
(A) illustrate one approach for constructing and computing $b(y,\alpha)$,
(B) explore our proposed bounds' empirical properties on simulated data, and
(C) demonstrate their practical utility on real-world data.
\subsection{Related work}\label{sec:background_and_related_work}

\paragraph{Hypothesis testing, \emph{p}-values, and pre-test estimation.}
Our proposed c-value bears a resemblance to the \emph{p}-value in hypothesis testing, but with a few key differences.
Indeed, just as a small \emph{p}-value can support rejecting a simple null hypothesis in favor of a possibly more complex alternative,
\Revision{a large c-value can support rejecting a familiar default estimate in favor of a less familiar alternative.}
Furthermore both tools provide a frequentist notion of confidence based on the idea of repeated sampling.
From this perspective, the two-step estimator $\theta^\dagger(\cdot, \alpha)$ resembles a preliminary testing estimator.
Preliminary testing links the choice between estimators to the outcome of a hypothesis test for the null hypothesis that $\theta$ lies in some pre-specified subspace \citep{wallace1977pretest}.

The similarities to hypothesis testing go only so far.
\Revision{Notably, we consider decisions made about a \emph{random} quantity, $W(\theta, y)$.}
Hypothesis tests, in contrast, concern only fixed statements about parameters, with nulls and alternatives corresponding to disjoint subsets of an underlying parameter space \citep[Definition 8.1.3]{casella2002statistical}.
Our approach does not admit an interpretation as testing a fixed hypothesis.

Nevertheless, the connection to \emph{p}-values can help us understand some limitations of the c-value.
First, just as hypothesis tests may incur Type II errors (i.e.\ failures to reject a false null),
for certain models and estimators there may be no bound $b(\cdot, \cdot)$ that consistently detects improvements by the alternative estimate.
\Revision{
Accordingly, the two stage estimator $\theta^\dagger(\cdot, \alpha)$ does not control the probability that we report the default estimate when the alternative in fact has smaller loss.
In such situations, our approach may consistently report the default estimate even though it has larger loss.}
Second, even if good choices of $b(\cdot, \cdot)$ exist,
it could be challenging to derive them analytically.
This analytical challenge is reminiscent of difficulties for hypothesis testing in many models,
wherein conservative \Revision{\emph{p}-values that are stochastically larger than uniform under the null are used when analytic quantile functions are unavailable.}
Third, we note that it may be tempting to interpret a c-value as the conditional probability that an alternative estimate is superior to a default;
however, just as it is incorrect to interpret a \emph{p}-value as a probability that the null hypothesis is true, such an interpretation for a c-value is also incorrect.

\paragraph{Post-selection inference.}
In recent years, there has been considerable progress on understanding the behavior of inferential procedures that, like $\theta^{\dagger}(\cdot, \alpha)$, use the data twice, first to select amongst different models and then again to fit the selected model.
Important recent work has focused on computing \emph{p}-values and confidence intervals for linear regression parameters that are valid after selection with the lasso \citep{Lockhart2014, Lee2016, taylor2018post} and arbitrary selection procedures \citep{berk2013valid}.
Somewhat more closely related to our focus on estimation are \citet{Tibshirani2019} and \citet{Tian2020}, which both bound prediction error after model selection.
Unlike these papers, which study the effects of selection on downstream inference, we effectively perform inference on the selection itself.

\section{Special case: c-values for estimating normal means}
\label{sec:simpler_cases}
In this section, we derive a bound $b(y,\alpha)$ and compute the c-value in a simple case:
we compare a certain class of shrinkage estimators to maximum likelihood estimates (MLE) of the mean of a multivariate normal from a single vector observation (i.e.\ the normal means problem).
Our goal is to illustrate a simple strategy for lower bounding the win that we will later generalize to more complex estimators and models.
In \Cref{sec:ls_setup}, we define the model and the estimators that we consider.
In \Cref{sec:ls_bound_construction}, we introduce our lower bound $b(\cdot,\cdot)$ and present a theorem that guarantees this bound satisfies \Cref{eqn:bya_bound}.
Then, in \Cref{sec:normal_means_simulation}, we examine the resulting c-value empirically and study the performance of the estimator $\theta^\dagger(\cdot, \alpha)$ that chooses between the default and alternative estimators based on the c-value (\Cref{eqn:theta_tilde}).
Several details, including the proof of \Cref{thm:lindley}, are left to \Cref{sec:simpler_cases_supp}.

\subsection{Normal means: notation and estimates}\label{sec:ls_setup}
Let $\theta \in \R^{N}$ be an unknown vector and consider estimating $\theta$ from a noisy vector observation $y = \theta + \epsilon$ where $\epsilon \sim \mathcal{N}(0,I_N)$ under squared error loss $L(\theta,\hat{\theta}) := \lVert \hat{\theta} - \theta \rVert^{2}.$
For simplicity, we focus on the case of isotropic noise with variance one; we remove this restriction in \Cref{sec:affine}.
For our demonstration, we take the MLE $\hat{\theta}(y) = y$ to be the default estimate.
As the alternative estimator, we consider a shrinkage estimator that was first studied extensively by \citet{lindley1972bayes},
$$
\theta^{*}(y) = \frac{y + \tau^{-2}\bar y \ones{N}}{1+\tau^{-2}},
$$ 
where $\ones{N}$ is the vector of all ones, $\tau > 0$ is a fixed positive constant, and $\bar y:=N\inv \ones{N}^{\top}y$ is the mean of the observed $y_{n}$'s.
Operationally, $\theta^{*}(y)$ shrinks each coordinate of the MLE towards the grand mean $\bar y.$

\subsection{Construction of the lower bound}\label{sec:ls_bound_construction}
{\sloppy To lower bound the win, we first rewrite $\theta^*(y) = \hat\theta(y) - Gy$ where $G:=(1+\tau^2)\inv P_1^\perp$ and $P_1^\perp := I_N - N\inv \ones{N}\ones{N}^\top $ is the projection onto the subspace orthogonal to $\ones{N}$.}
The win in squared error loss may then be written as
\begin{equation}
\label{eqn:win_form}
W(\theta, y):= \lVert \hat{\theta}(y) - \theta \rVert^{2} - \lVert \theta^{*}(y) - \theta \rVert^{2} =  2 \epsilon^\top Gy - \|Gy \|^2.
\end{equation}

Observe that we can compute $\|Gy\|$ directly from our data.
As a result, in order to lower bound the win $W(\theta, y),$ it suffices to lower bound $2\epsilon^{\top}Gy.$
As we detail in \Cref{sec:nc_chi_dist_in_simple_case}, $2\epsilon^\top Gy$ follows a scaled and shifted non-central chi-squared distribution,
$$
2 \epsilon^\top Gy \sim \frac{2}{1+\tau^2}\left[\chi^2_{N-1}(\frac{1}{4} \|P_1^\perp \theta \|^2)
-\frac{1}{4}\|P_1^\perp\theta\|^2 \right],
$$
where $\chi^2_{N-1}(\lambda)$ denotes the non-central chi-squared distribution with $N-1$ degrees of freedom and 
non-centrality parameter $\lambda$.
Thus for any $\alpha\in(0,1)$ and any fixed value of $\|P_1^\perp \theta\|^2$,
\[\label{eqn:nc_chi_bound}
W(\theta, y) \ge 
\frac{2}{1+\tau^2} F^{-1}_{N-1} (1-\alpha;\frac{1}{4}\| P_1^\perp \theta \|^2)
- \frac{\|P_1^\perp \theta\|^2}{2(1+\tau^2)} - \|Gy\|^2
\]
with probability $\alpha$, where $F^{-1}_{N-1}( 1-\alpha; \lambda)$ 
denotes the inverse cumulative distribution function of $\chi^2_{N-1}(\lambda)$ evaluated at $1-\alpha.$
Were $\|P_1^\perp \theta\|^2$ known, the right hand side of \Cref{eqn:nc_chi_bound} would immediately provide a valid bound.
However since $\|P_1^\perp \theta\|^2$ is not typically known, we use the data to address our uncertainty in this quantity.
We obtain our bound by forming a one-sided confidence interval for $\|P_1^\perp \theta\|^2$ 
that holds simultaneously with \Cref{eqn:nc_chi_bound}.

\begin{bound}[Normal means: Lindley and Smith estimate vs.\ MLE]\label{bound:lindley_and_smith}
Observe $y = \theta+ \epsilon$ with $\epsilon \sim \mathcal{N}(0, I_N)$ and consider 
$\hat \theta(y) =y$ vs.\ $\theta^*(y) = (y + \tau^{-2}\bar y \ones{N})/(1 + \tau^{-2}).$
We propose
\[\label{eqn:LS_bya}
    b(y, \alpha) := \inf_{\lambda \in [0,U(y, \frac{1-\alpha}{2})]}\left\{ \frac{2}{1+\tau^2} F^{-1}_{N-1} \left(\frac{1-\alpha}{2};\frac{\lambda}{4}\right)
- \frac{\lambda }{2(1+\tau^2)} - \frac{\|P_1^\perp y \|^2 }{(1+\tau^2)^2} \right\}
\]
as an $\alpha$-confidence lower bound on the win,
where
\begin{equation}\label{eqn:lindley_upper_conf}
    U\left(y, \frac{1-\alpha}{2}\right) := \inf_{\delta>0} \left\{ \delta \Big|
\| P_1^\perp y\|^2 \le 
F^{-1}_{N-1}\left(\frac{1-\alpha}{2}; \delta\right)
\right\}
\end{equation}
is a high-confidence upper bound on $\|P_1^\perp \theta\|^2$.
\end{bound}

\Cref{bound:lindley_and_smith} relies on a high-confidence upper bound on $\|P_1^\perp \theta\|^2,$ but a two-sided interval could in principle provide a valid bound as well.
In \Cref{sec:lindley_why_upper} we provide an intuitive justification for the choice of an upper bound.
\Cref{thm:lindley} justifies the use of \Cref{bound:lindley_and_smith} for computing c-values.
\begin{theorem}\label{thm:lindley}
    Define $c(y) := \inf_{\alpha\in[0,1]} \{ \alpha | b(y, \alpha) \le 0\}$ for $b(\cdot, \cdot)$ in \Cref{bound:lindley_and_smith}.
Then $c(y)$ is a valid c-value, satisfying the guarantees of \Cref{thm:c_values,thm:theta_tilde_guarantee}.
\end{theorem}

\begin{remark}[Computability of the bound]\label{remark:bound_computability}
\Cref{eqn:LS_bya} in \Cref{bound:lindley_and_smith} can be readily computed.
Notably, many standard statistical software packages provide numerical approximation to non-central $\chi^{2}$ quantiles.
Further, the one-dimensional optimization problems in \Cref{eqn:LS_bya,eqn:lindley_upper_conf} can be solved numerically.
\end{remark}
\begin{remark}[Unknown variance]
For cases when the noise variance $\sigma^2$ is unknown but a confidence interval is available, one can adapt the procedure above by replacing $b(y, \alpha)$ with its infimum with respect to $\sigma^2$ over the confidence interval and reducing the confidence level $\alpha$ accordingly.
\end{remark}

\begin{remark}
The alternative estimator $\theta^{*}(y)$ considered in this section is the posterior mean of $\theta$ corresponding to the hierarchical prior $\theta \vert \mu \sim \mathcal{N}(\mu\ones{N}, \tau^{2}I_{N})$ with further improper hyper-prior on $\mu.$
This prior encodes a belief that $\theta$ lies close to the one-dimensional subspace spanned by $\ones{N}.$
Using a similar approach to the one above, we can derive lower bounds on the win for a more general class of estimators that shrink the MLE towards a pre-specified $D$-dimensional subspace.
See \Cref{sec:morris_ext} for details and an application to a real dataset on which a large computed c-value indicates an improved estimate.

\end{remark}

\subsection{Empirical verification}
\label{sec:normal_means_simulation}
To explore the empirical properties of \Cref{bound:lindley_and_smith}, we simulated 500 datasets with $N = 50$ as $y\sim \mathcal{N}(\theta, I_N)$ for each of several values of $\theta.$
For each simulated dataset $y,$ we computed the win $W(\theta, y),$ the proposed lower bound $b(y,\alpha),$ and the c-value $c(y).$
Conveniently, for this likelihood, the distributions of $W(\theta,y)$ and $b(y,\alpha)$ depend on $\theta$ only through $N\nsqrt \|P_1^\perp\theta\|.$
Consequently, we can exhaustively assess how our procedure behaves for different $\theta$ by varying this norm.
Throughout our simulation study, we fixed $\tau = 1.$
\Revision{With larger $\tau,$ the alternative $\theta^*$ behaves more similarly to the default $\hat \theta$, but the qualitative properties of the c-value and estimators remain similar.}

\begin{figure}[h!]
    \centering
     \begin{subfigure}[b]{0.35\textwidth}
        \centering
        \includegraphics[width=\textwidth]{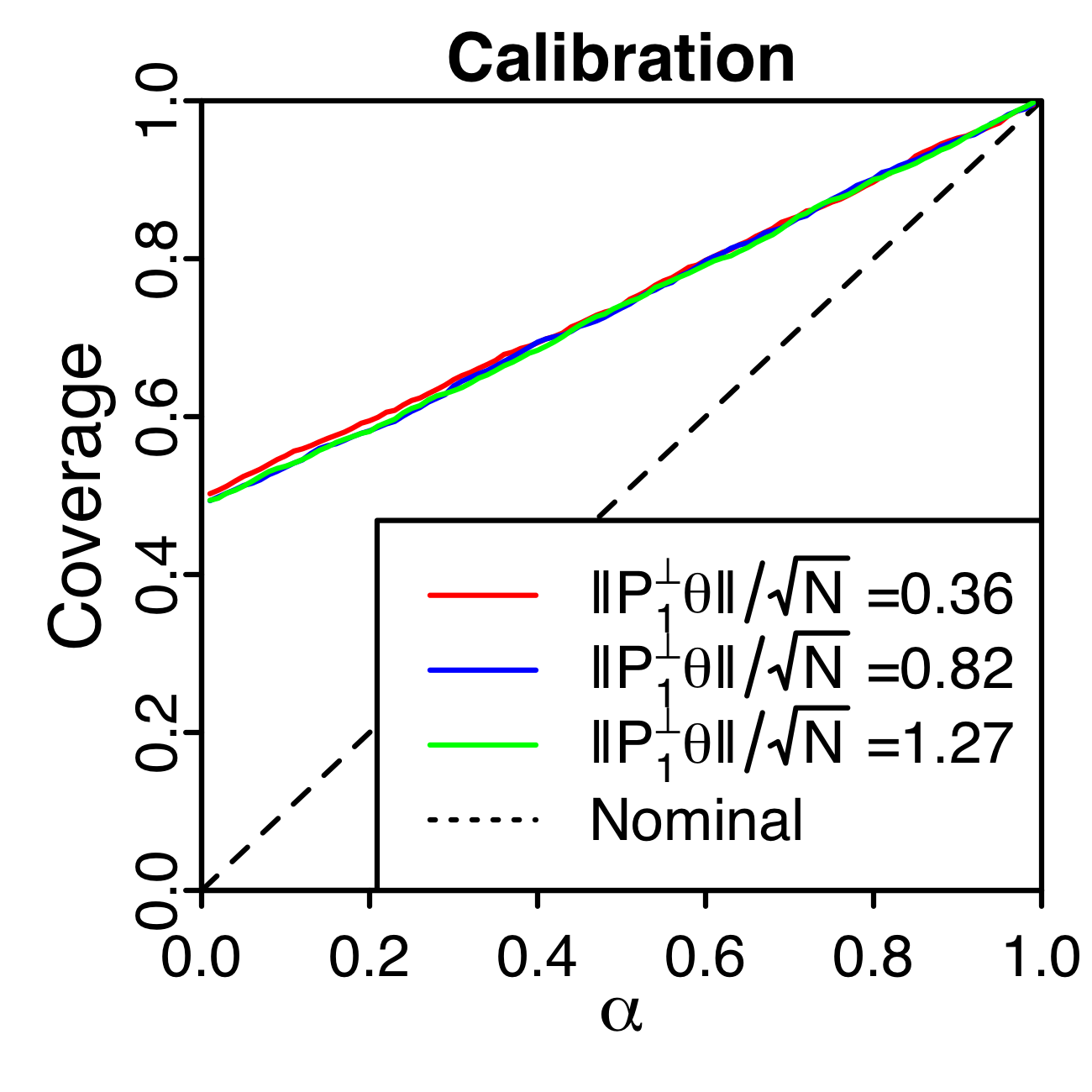}
        \caption{}\label{fig:LS_bound_calibration}
     \end{subfigure}
     \begin{subfigure}[b]{0.35\textwidth}
        \centering
        \includegraphics[width=\textwidth]{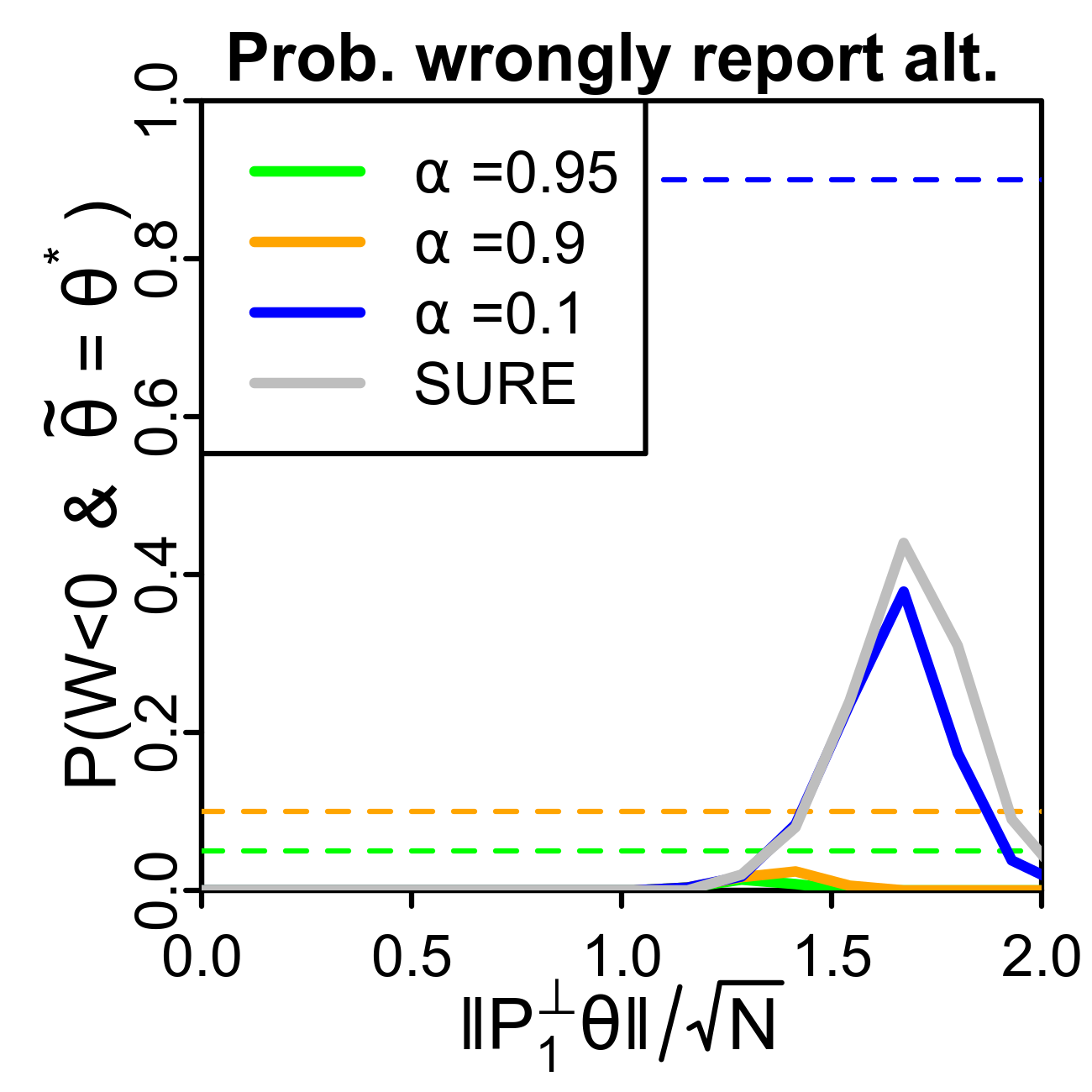}
        \caption{}\label{fig:LS_prob_mistake}
    \end{subfigure}
     \begin{subfigure}[b]{0.35\textwidth}
        \centering
        \includegraphics[width=\textwidth]{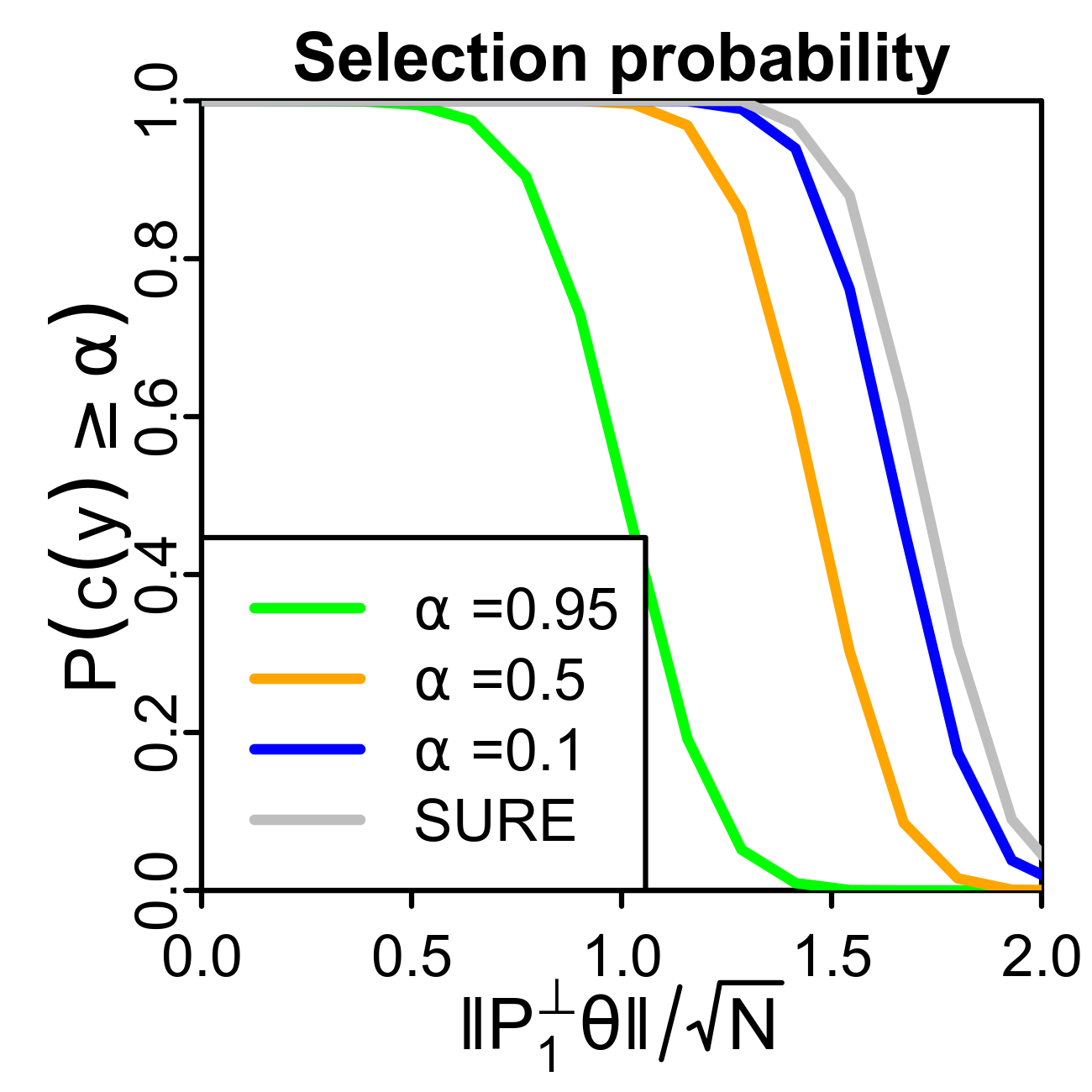}
        \caption{}\label{fig:LS_selection_probability}
    \end{subfigure}
     \begin{subfigure}[b]{0.35\textwidth}
        \centering
        \includegraphics[width=\textwidth]{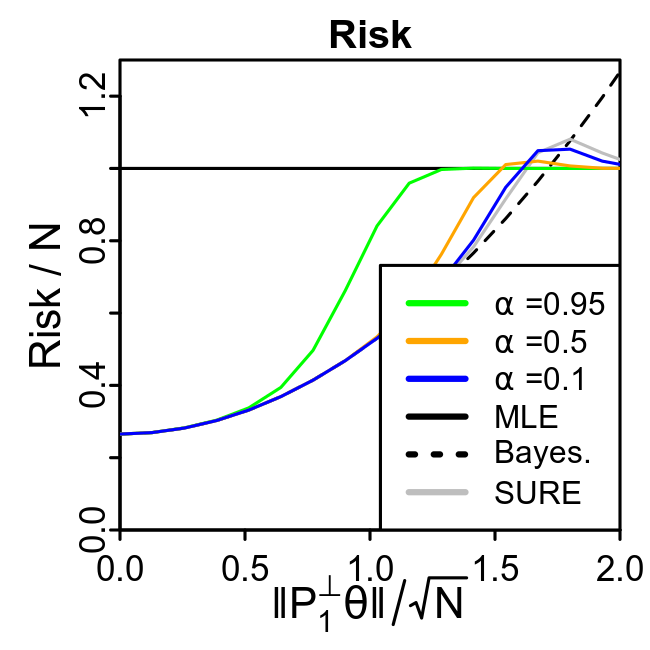}
        \caption{}\label{fig:LS_risk}
    \end{subfigure}
\caption{
Bound calibration and the two-stage estimator for a hierarchical normal model in simulation.
(a) Empirical coverage of the lower bound $b(\cdot, \alpha)$ across different levels $\alpha.$
Coverage is nearly identical across the parameter space.
\Revision{(b) Probability that the default has smaller loss but the alternative estimate is selected across the parameter space,
with dashed lines reflecting nominal coverage.
}
(c) Probability of selecting the alternative estimate.
Selection probability is higher for lower thresholds $\alpha.$
(d) Risk profiles of the two-stage estimators for different choices of $\alpha,$
as well as the MLE $\hat \theta(\cdot)$ and the shrinkage estimator $\theta^*(\cdot).$
Each data point is computed from {500} replicates with $N=50$.
}
\end{figure}

We first checked that the empirical probability that the win $W(\theta, y)$ exceeded the bound $b(y, \alpha)$ in \Cref{bound:lindley_and_smith} was at least as large as the nominal probability $\alpha$ (\Cref{fig:LS_bound_calibration}).
Across various choices of $N\nsqrt \|P_1^\perp \theta\|,$ we see that $b(\cdot,\alpha)$ is conservative, typically providing higher than nominal coverage.
Surprisingly, the gap between the actual and nominal coverages does not seem to depend heavily on $\theta$, suggesting we could potentially obtain a tighter bound by calibrating $b(y, \alpha)$ to its actual coverage.

\Revision{
We next examined the probability that the alternative estimate is selected on the basis of a large c-value but obtains higher loss than the default estimate.
\Cref{thm:theta_tilde_guarantee} upper bounds this probability, and in \Cref{fig:LS_prob_mistake} we confirm this bound holds in practice across different thresholds $\alpha$.
\Cref{fig:LS_prob_mistake} additionally compares our proposed approach to using Stein's unbiased estimate of the risk \citep{stein1981estimation} of $\theta^*(\cdot)$ to select between the estimates.
This approach, which we label ``SURE'', returns $\theta^*(\cdot)$ if the risk estimate exceeds $N$ and returns $\hat \theta(\cdot)$ otherwise, and is akin to the focused information criterion \citep{claeskens2003focused}.
However, in contrast to the two-stage estimator $\theta^\dagger(\cdot, \alpha),$ SURE does not provide tunable control over the probability that the alternative estimator $\theta^*(\cdot)$ is mistakenly returned.}

{\Revision{ 
%
%

\begin{table}[h]
\centering
\caption{
    \Revision{Contingency tables of simulation outcomes with $\|P_1^\perp \theta\|/\sqrt{N} = 1.7$ when using 
Stein's unbiased risk estimate (SURE), $\theta^\dagger(\cdot, \alpha=0.95)$, or $\theta^\dagger(\cdot, \alpha=0.5)$ to choose between the default and alternative estimates.
DLL: \textbf{d}efault has \textbf{l}ower \textbf{l}oss,
ALL: \textbf{a}lternative has \textbf{l}ower \textbf{l}oss,
DR: \textbf{d}efault \textbf{r}eported,
AR: \textbf{a}lternative \textbf{r}eported.
    }
   }\label{table:contingency_main}
\begin{tabular}{c}
    \begin{tabularx}{0.90\textwidth}{| X | X | X | X | X | X | X |}
\hline
    ~ & \multicolumn{2}{c |}{SURE} & \multicolumn{2}{c|}{$\theta^\dagger(\cdot, \alpha=0.95)$} & \multicolumn{2}{c|}{$\theta^\dagger(\cdot, \alpha = 0.5)$} \\ 
~ & DR & AR & DR & AR & DR & AR \\ \hline
DLL & 2\% & 44\% & 46\% & 0\% &  37\% & 9\% \\ \hline
ALL & 36\% & 18\% & 54\% & 0\% & 54\% & 0.1\% \\ \hline
\end{tabularx}
\end{tabular}
\end{table}
}}

\Revision{In the case that $\|P_1^\perp \theta\| / \sqrt{N}=1.7,$ choosing based on SURE gives the wrong estimate 80\% of the time.
Moreover, in the majority of these cases it is the alternative that is incorrectly returned (\Cref{table:contingency_main}, \Cref{fig:LS_prob_mistake}).
By contrast, the estimator that chooses based on the c-value (with a threshold $\alpha=0.95$) conservatively returns the default estimate in every replicate for this $\|P_1^\perp \theta\| / \sqrt{N}$ (\Cref{fig:LS_selection_probability}).
While this approach provides the estimate with greater loss in 54\% of cases, it incorrectly reports the alternative in 0\% of cases (\Cref{table:contingency_main}).
This behavior is expected as \Cref{thm:theta_tilde_guarantee} provides an upper bound of $100*(1-\alpha)\%=5\%$.
An estimator using the unbiased risk estimate satisfies no such guarantee.
}

We next checked that our computed c-values successfully detected improvements by the alternative estimate.
Recall that the alternative estimate $\theta^{*}(y)$ shrinks all components of $y$ towards the global mean $\overline{y}.$
Further, recall that by construction $\theta^{\dagger}(y,\alpha) = \theta^{*}(y)$ if and only if $c(y) > \alpha.$
Intuitively, then, we would expect the alternative estimator to improve over the MLE and for the two-stage $\theta^{\dagger}(\cdot,\alpha)$ to select $\theta^{*}(\cdot)$ when $\theta$ is close to the subspace spanned by $\ones{N}$ and $N\nsqrt \|P_1^\perp \theta\|$ is small.
\Cref{fig:LS_selection_probability}, which plots the probability that $\theta^{\dagger}(\cdot, \alpha)$ selects $\theta^{*}(\cdot)$ across different values of $\theta$ and $\alpha,$ confirms this intuition; when $N\nsqrt \|P_1^\perp \theta\|$ is small, we very often obtain large c-values and select the alternative estimator.

For completeness, we also considered the risk profile of the two-stage estimator $\theta^\dagger(\cdot, \alpha)$ (\Cref{fig:LS_risk}).
Specifically, for different choices of $\theta$ we computed a Monte Carlo estimate of the expected squared error loss.
For the most part, the risk of $\theta^{\dagger}(\cdot, \alpha)$ lies between the risks of $\hat{\theta}(\cdot)$ and $\theta^{*}(\cdot).$
However, the risk of the two-stage estimator appears to exceed the risks of the default and alternative estimators for a narrow range of values of $\|P_1^\perp \theta\|.$
While it is tempting to characterize this excess risk as the price we must pay for ``double-dipping'' into our data, we note that the bump in risk appears to be non-trivial only for very small values of $\alpha.$
Recall again that we recommend choosing $\theta^{*}(y)$ in place of $\hat \theta(y)$ only when $c(y)$ is close to $1$.
As such, we do not expect this type of risk increase to be much of a concern in practice.

\Revision{Interpreted together, \Cref{fig:LS_selection_probability,fig:LS_risk} illustrate the conservatism of the two stage approach with $\alpha=0.95$.
For $\|P_1^\perp \theta\|$ between 1 and 1.5, $\theta^\dagger(\cdot, \alpha)$ only rarely evaluates to $\theta^*(\cdot)$ even though this estimator has lower risk and typically has smaller loss.}

Unlike conventional \emph{p}-values under a null hypothesis, we should not expect the distribution of informative c-values to be uniform;
indeed for parameters such that the win is consistently positive or negative, c-values can concentrate near $1$ or $0,$ respectively.
\section{Comparing affine estimates with correlated noise}\label{sec:affine}
We now generalize the situation described in the previous section in two ways.
First, we consider correlated Gaussian noise with covariance $\Sigma$,
where $\Sigma$ is any $N\times N$ positive definite covariance matrix rather than restricting to $\Sigma=I_N$.
Second, we let our default and alternative estimates, $\hat \theta(y)$ and $\theta^*(y)$, be arbitrary affine transformations of the data $y$.
Though these two estimates take similar functional forms in this section, we remain concerned with asymmetric comparisons wherein 
$\theta^*(y)$ is less familiar than $\hat \theta(y).$

Although such generalization introduces considerable analytical challenges beyond those encountered in \Cref{sec:simpler_cases},
we nevertheless can construct an \emph{approximate} lower bound on the win that works well in practice.
Specifically, for \Cref{bound:lindley_and_smith}, we used the tractable quantile function of the non-central $\chi^2$
to guarantee exact coverage in \Cref{thm:lindley}.
Now we encounter sums of differently scaled non-central $\chi^2$ random variables,
which do not admit analytically tractable quantiles.
However, by approximating these sums with Gaussians with matched means and variances, 
we can proceed in essentially the same manner as in \Cref{sec:simpler_cases} to derive an approximate lower bound on the win.
After introducing the bound, 
we comment on the key steps in its derivation to highlight the approximations involved, but leave details of intermediate steps to \Cref{sec:affine_supp}.
We conclude with a non-asymptotic bound on the error introduced by these approximations
on the coverage of the proposed bound on the win.

\begin{approxbound}[Correlated Gaussian likelihood: arbitrary affine estimates]\label{approxbound:affine}
Observe $y = \theta + \epsilon$ with $\epsilon \sim \mathcal{N}(0, \Sigma)$ and consider
$\hat \theta(y) = Ay + k$ vs.\ $\theta^*(y) = Cy + \ell,$
where $A, C \in \R^{N\times N}$ are matrices and $k, \ell \in \R^N$ are $N$-vectors.
We propose
\begin{align}\label{eqn:affine_bya}
\begin{split}
    b(y,\alpha) &=  \|\hat \theta - y\|^2 - \| \theta^* - y\|^2 + 2\trace[ (A-C) \Sigma ] + { }\\
    &2 z_{\frac{1-\alpha}{2}}\sqrt{
        U(\|G(y)\|_\Sigma^2, \frac{1-\alpha}{2})  
        + \frac{1}{2}\|\Sigma\msqrt(A+A^\top - C - C^\top)\Sigma\msqrt\|_F^2
}
\end{split}
\end{align}
as an approximate high-probability lower bound for the win.
In this expression, 
$\trace[\cdot]$ denotes the trace of a matrix,
$G(y) := (A-C)y + (k-\ell)$,
$\| \cdot\|_\Sigma$ denotes the $\Sigma$ quadratic norm of a vector ($\|v\|_\Sigma := \sqrt{v^\top \Sigma v}$),
$\|\cdot\|_F$ denotes the Frobenius norm of a matrix, and
$z_\alpha$ denotes the $\alpha$-quantile of the standard normal.
\begin{align}\label{eqn:affine_norm_upper_bound}
\begin{split}
    U(\|G(y)\|_\Sigma^2, &1-\alpha) := \inf_{\delta>0} \bigg\{ \delta \, \bigg|\, 
    \|G(y)\|_\Sigma^2 \le 
    (\delta + \|\Sigma\msqrt (A-C)\Sigma\msqrt\|_F^2) +  { }\\
    &z_{1-\alpha}\sqrt{
        2\|\Sigma\msqrt (A-C) \Sigma (A-C)^\top \Sigma\msqrt\|_F^2
        + 4 \opnorm{\Sigma^{\frac{1}{2}} (A-C) \Sigma^{\frac{1}{2}}}^2\delta 
    } 
\bigg\}
\end{split}
\end{align}
is an approximate high-confidence upper bound on $\|G(\theta)\|_\Sigma^2$ where 
$\opnorm{\cdot}$ denotes the L2 operator norm of a matrix.
\end{approxbound}

To derive \Cref{approxbound:affine} we again start by rewriting the alternative estimate as $\theta^*(y) = \hat \theta(y) - G(y)$,
where now $G(\cdot)$ is an affine transformation of $y$,
$
    G(y) := (A-C)y + (k-\ell).
$
We next write the squared error win of using $\theta^*(y)$ in place of $\hat \theta(y)$ as 
\begin{align}\label{eqn:affine_win_decomp}
\begin{split}
    W(\theta, y) 
                 &= 2\epsilon^\top G(y) + \left(\|\hat \theta(y) - y \|^2 - \|\theta^*(y) - y \|^2\right) \\
\end{split}
\end{align}
and observe that it suffices to obtain a high-probability lower bound for this first term.
For tractability, we approximate the distribution of $\epsilon^\top G(y)$ by a normal with matched mean and variance.
As we will soon see, this approximation is accurate when $N$ is large and $A-C$ is well conditioned;
in this case $\epsilon^\top G(y)$ may be written as the sum of many of uncorrelated terms of similar size.
The mean and variance may be expressed as
\begin{align}\label{eqn:affine_mean_and_var}
\begin{split}
    \E[ \epsilon^\top G(y) ]  = \trace[ (A-C) \Sigma ]  
    \ , \ 
    \Var[ \epsilon^\top G(y) ] = \|G(\theta)\|_\Sigma^2 + \frac{\|\Sigma\msqrt(A+A^\top - C - C^\top)\Sigma\msqrt\|_F^2}{2}.
\end{split}
\end{align}
With these moments in hand, we form a probability $\alpha$ lower bound approximately as 
\begin{align}\label{eqn:win_bound_affine}
\begin{split}
    W(&\theta, y) 
    \ge \|\hat \theta(y) -y\|^2 - \|\theta^*(y) - y\|^2 + 2\trace[ (A-C) \Sigma ] + { }\\
     &2 z_{1-\alpha}\sqrt{
        \|G(\theta)\|_\Sigma^2 + \frac{1}{2}\|\Sigma\msqrt(A+A^\top - C - C^\top)\Sigma\msqrt\|_F^2
    }.
\end{split}
\end{align}

However, as before, in order to use this approximate bound we require a simultaneous upper bound on a norm of a transformation of the unknown parameter, 
in this case $\|G(\theta)\|_\Sigma^2.$
We compute one by considering the test statistic $\|G(y)\|_\Sigma^2 $ and again appealing to approximate normality.
In particular we characterize the dependence of the distribution of this statistic on $\|G(\theta)\|_\Sigma^2 $ through its mean and variance.
We find its mean as
\begin{align}\label{eqn:affine_mean_Gy}
\begin{split}
    \E[\|G(y)\|_\Sigma^2] = \|G(\theta) \|_\Sigma^2 + \|\Sigma\msqrt (A-C) \Sigma\msqrt \|_F^2
\end{split}
\end{align}
and upper bound its variance by
\begin{align}\label{eqn:affine_var_ub}
\begin{split}
    \Var[\|G(y)\|_\Sigma^2] &\le 2\|\Sigma\msqrt (A-C) \Sigma (A-C)^\top \Sigma\msqrt\|_F^2
    + 4 \opnorm{\Sigma^{\frac{1}{2}} (A-C) \Sigma^{\frac{1}{2}}}^2 \|G(\theta)\|_\Sigma^2.
\end{split}
\end{align}
Using the two quantities above and an appeal to approximate normality, we propose the approximate high-confidence upper bound, $U(\|G(y)\|_\Sigma^2, 1-\alpha),$ in \Cref{eqn:affine_norm_upper_bound}.
As before, by splitting our $\alpha$ across these two bounds we obtain the desired expression, \Cref{eqn:affine_bya} in \Cref{approxbound:affine}.

\paragraph{Approximation Quality.}
Due to the two Gaussian approximations, \Cref{approxbound:affine} does not provide nominal coverage by construction.
Our next result shows that little error is introduced when $N$ is large enough and the problem is well conditioned.

\begin{theorem}[Berry--Esseen bound]\label{thm:berry_esseen}
Let $\alpha \in(0,1)$ and consider $b(\cdot,\alpha)$ in \Cref{approxbound:affine}.
If both $A$ and $C$ are symmetric, then
\[\label{eqn:berry_esseen_affine_bound}
\pr_\theta\left[
    W(\theta, y) \ge b(y, \alpha)\right] 
    \ge \alpha  - \frac{10\sqrt{2}}{\sqrt{N}}C_1\cdot\kappa(\Sigma\msqrt(A-C)\Sigma\msqrt)^2
\]
where $\kappa(\cdot)$ denotes the condition number of its matrix argument (i.e. the ratio of its largest to smallest singular values) and
$C_1\le 1.88$ is a universal constant \citep[Theorem 1]{berry1941accuracy}.
\end{theorem}
\begin{remark}
\Cref{thm:berry_esseen} is a special case of a more general result that we provide in \Cref{sec:berry_esseen_supp}, which does not require $A$ and $C$ to be symmetric.
We highlight this special case here because the bound takes a simpler form from which the dependence on the conditioning of $A-C$ is clearer,
and because this condition is satisfied for many important estimates.
Notably $A$ and $C$ are symmetric in all applications discussed in this paper.
\end{remark}
Though \Cref{thm:berry_esseen} provides an expected $O(N\nsqrt)$ drop in approximation error, 
the bound itself may be too loose to be useful in practice.
In \Cref{sec:small_area_education} we show in simulation that \Cref{approxbound:affine} provides sufficient coverage even without this correction.
This conservatism likely owes to slack from (A) the operator norm bound in \Cref{eqn:affine_var_ub}
and (B) the union bound ensuring that the confidence interval for $\|G(\theta)\|_\Sigma^2$ and the quantile in \Cref{eqn:win_bound_affine} hold simultaneously.

\begin{remark}[Fast computation of $b(y, \alpha)$]
A naive approach to computing $b(y,\alpha)$ in \Cref{eqn:affine_bya} involves finding
$U(\|G(y)\|_\Sigma^2, \frac{1-\alpha}{2})$ with a binary search.
For more rapid computation,
we can recognize $U(\|G(y)\|_\Sigma^2, \frac{1-\alpha}{2})$ as the root of a quadratic.
Specifically, define
$\gamma:=\|G(y)\|_\Sigma^2 -\|\Sigma\msqrt (A-C)\Sigma\msqrt\|_F^2$,
$\eta:=z_{\frac{\alpha}{2}}$, 
$\rho:=2\|\Sigma\msqrt (A-C) \Sigma (A-C)^\top \Sigma\msqrt\|_F^2$, and
$\nu:=4 \opnorm{\Sigma^{\frac{1}{2}} (A-C) \Sigma^{\frac{1}{2}}}^2$; then
from \Cref{eqn:affine_norm_upper_bound} we have that the $\delta$ that achieves the supremum satisfies
$\gamma = \delta + \eta\sqrt{\rho + \nu \delta}.$
Rearranging, we find that $U(\|G(y)\|_\Sigma^2, \frac{1-\alpha}{2})$ is the larger root of 
$x^2 -(2\gamma + \eta^2 \nu)x + (\gamma^2 - \eta^2 \rho)=0.$
\end{remark}
\section{Extending the reach of the c-value}
\label{sec:more_cases}

Up to this point, we focused on estimating normal means with fixed affine estimators.
Now we extend our c-value framework in two important directions\Revision{, which we support with both theoretical and empirical results}.
In \Cref{sec:empirical_bayes}, we derive c-values for a nonlinear shrinkage estimator of normal means.
We then move beyond Gaussian likelihoods in \Cref{sec:logistic_regression} and derive c-values for regularized logistic regression.
In contrast to the earlier cases, these settings introduce nonlinear estimates and non-Gaussian models.
To gain analytical tractability, we approximate the estimates by linear transformations of a statistic that is asymptotically Gaussian.
This approximation allows us to derive bounds $b(y,\alpha)$ that we show have the correct coverage in an asymptotic regime.
Our approach provides a template that can be followed for other nonlinear estimates and models for which the MLE is asymptotically Gaussian.
We defer all proofs and details of synthetic data experiments to \Cref{sec:empirical_bayes_supp,sec:logistic_regression_supp}.


\subsection{Empirical Bayes shrinkage estimates}\label{sec:empirical_bayes}
Many Bayesian estimates are affine in the data for fixed settings of prior parameters.
But when prior parameters are chosen using the data, the resulting \emph{empirical Bayesian} estimates are not affine in general.
We next explore computation of approximate high-confidence lower bounds on the win of empirical Bayesian estimators.
In particular, we consider an approach that essentially amounts to ignoring the randomness in estimated prior parameters and computing the bound as if the prior were fixed.
For simplicity, we focus on a particularly simple empirical Bayesian estimator for the normal means problem that coincides with the James--Stein estimator \citep{efron1973stein}.
We find that, in the high-dimensional limit, bounds obtained with this naive approach achieve at least the desired nominal coverage.
Finally, we show in simulation that the approximate bound has favorable finite sample coverage properties.

\paragraph{Empirical Bayes for estimation of normal means.}
Consider a sequence of real-valued parameters $\theta_1, \theta_2, \dots,$
and corresponding observations $y_n \overset{indep}{\sim} \mathcal{N}(\theta_n, 1)$.
For each $N\in\N$, let $\Theta_N := [\theta_1, \theta_2, \dots, \theta_N]^\top$ and $Y_N := [y_1, y_2, \dots, y_N]^\top$ denote the first $N$ parameters and observations, respectively.

We consider the MLE for $\Theta_N$ (i.e.\ $Y_N$) as our default, which we denote by $\hat \Theta_N(Y_N) = Y_N,$ 
and we take the James--Stein estimate as our alternative; we compare on the basis of squared error loss.
We write the James--Stein estimate on the first $N$ data points as
$\Theta^*_N(Y_N) := \left( 1 - (1+\hat \tau_N^2(Y_N))\inv\right)Y_N,$
where $\hat \tau_N^2(Y_N) := \|Y_N\|^2/(N-2) -1$.
$\Theta^{*}_N(Y_N)$ corresponds to the Bayes estimate under the prior $\theta_n\overset{i.i.d.}{\sim}\mathcal{N}(0, \hat \tau_N^2)$ \citep{efron1973stein}.
For this comparison, the win is 
$W_N(Y_N, \Theta_N) := \|\hat \Theta_N(Y_N) - \Theta_N \|^2 - \|\Theta^*_N(Y_N) - \Theta_N\|^2$,
and \Cref{sec:empirical_bayes_supp} details the associated bound $b_N(Y_N,\alpha)$ obtained with \Cref{bound:morris}.
In the following theorem, we lower bound the win by applying our earlier machinery for Bayes rules with fixed priors.
We find that the desired coverage is obtained in the high-dimensional limit.
\begin{theorem}\label{thm:JS_asymptotic_coverage}
For each $N\in \N$, let $\tau_N^2 := N\inv\sum_{n=1}^N\theta_n^2$.
If the sequence $\tau_1, \tau_2, \dots$ is bounded, then for any $\alpha\in[0,1],\ \ 
\lim_{N\rightarrow \infty} \pr\left[ W_N(Y_N, \Theta_N) \ge b_N(Y_N, \alpha) \right] \ge \alpha.$
\end{theorem}

The key step in the proof of \Cref{thm:JS_asymptotic_coverage} is establishing an $O_p(N\nsqrt)$ rate of convergence of $\hat \tau_N^2 - \tau_N^2$ to zero;
under this condition the empirical Bayes estimate and bound converge to the analogous estimates and bounds computed with the prior variance fixed to 
$\tau_N^2.$
Accordingly, we expect similar results to hold for other models and empirical Bayes estimates when the standard deviations of the empirical Bayes estimates of the prior parameters drop as $O_p(N\nsqrt)$.

\begin{remark}
\Cref{thm:JS_asymptotic_coverage} easily extends to cover the case in which we consider a sequence of random (rather than fixed) parameters drawn i.i.d.\ from a Bayesian prior,
which is a more classical setup for guarantees of empirical Bayesian methods; see e.g.\ \citet{robbins1964empirical}.
Specifically, our proof goes through in this Bayesian setting so long as the sequence $\tau^2_1, \tau^2_2, \dots$ is bounded in probability.
This condition is satisfied, for example, 
when the $\theta_n$ are i.i.d.\ from any prior with a finite second moment.
\end{remark}

To check finite sample coverage, we performed a simulation and evaluated calibration of the associated c-values (\Cref{fig:JS_calibration} in \Cref{sec:empirical_bayes_supp}).
Despite the empirical Bayes step, the c-values appear to be similarly conservative to those computed with the exact bound in \Cref{fig:LS_bound_calibration}.
Furthermore, this calibration profile does not appear to be sensitive to the magnitude of the unknown parameter.
\subsection{Logistic regression}\label{sec:logistic_regression}
In this subsection we illustrate how to compute an approximate high-confidence lower bound on the win in squared error loss with a logistic regression likelihood.
Our key insight is that by appealing to limiting behavior,
we can tackle the non-Gaussianity using the machinery developed in \Cref{sec:affine}.

\paragraph{Notation and estimates.}
Consider a collection of $M$ data points with random covariates $X_M:=[x_1, x_2, \dots, x_M]^\top\in\R^{M\times N}$ and responses
$Y_M :=[y_1, y_2, \dots, y_M]^\top\in \{1, -1\}^M$.
For the $m$th data point, assume
\[\label{eqn:logistic_likelihood}
    y_m \overset{indep}{\sim} p( \cdot \mid x_m; \theta) := (1+\exp\{-x_m^\top \theta \})\inv\delta_{1} + (1+\exp\{x_m^\top \theta \})\inv\delta_{-1},
\]
where $\theta \in \R^N$ is an unknown parameter of covariate effects and $\delta_1$ and $\delta_{-1}$ denote Dirac masses on $1$ and $-1$, respectively.

In this section, we choose the MLE as our default, $\hat \theta(X_M, Y_M):= \argmax_\theta \log p(Y_M \mid X_M; \theta)$. And we choose our alternative  to be a Bayesian maximum a posteriori (MAP) estimate under a standard normal prior ($\theta \sim \mathcal{N}(0, I_N)$):
$$\theta^*(X_M, Y_M) := \argmax_\theta \left\{ \log p(Y_M\mid X_M; \theta) - \frac{1}{2}\|\theta\|^2\right\}.$$
While a first choice for a Bayesian estimate might be the posterior mean, 
the MAP is an effective and widely used alternative to the MLE in practice.
\Revision{Furthermore, $\theta^*(X_M, Y_M)$ is also of interest as an L2 regularized logistic regression estimate.}

\paragraph{Approximating $\theta^*$ by an affine transformation.}
In moving away from a Gaussian likelihood we forfeit prior-to-likelihood conjugacy.
In previous sections, conjugacy provided analytically convenient expressions for Bayes estimates.
In order to regain analytical tractability, we appeal to a Gaussian approximation of the likelihood,
defined with a second order Taylor approximation of the log likelihood around the MLE.
Under this approximation,
$\hat \theta(X_M, Y_M) \sim \mathcal{N}(\theta,\tilde \Sigma_M),$ where  
$\tilde \Sigma_M := -\nabla_\theta^2 \log p(Y_M \mid X_M;\theta)\big|_{\theta = \hat \theta(X_M, Y_M)}.$
As such, we regain conjugacy, and we obtain an approximate Bayes estimate as an affine transformation of the MLE,
\[\label{eqn:approx_MAP}
    \tilde \theta^*(X_M, Y_M) = \left[ I_N + \tilde \Sigma_M\right]\inv \hat \theta(X_M, Y_M).
\]
As we show in \Cref{sec:logistic_regression_supp},
$\tilde \theta^*(X_M, Y_M)$ is a very close approximation of $\theta^*(X_M, Y_M),$ with distance decreasing at an $O_p(M^{-2})$ rate.

\paragraph{An approximate bound and an asymptotic guarantee.}
We leverage the form in \Cref{eqn:approx_MAP} to compute \Cref{approxbound:affine} as a lower bound on the win in squared error of using the MAP estimate in place of the MLE.
In particular, we take $y:=\hat \theta(X_M, Y_M)$ as the data in \Cref{approxbound:affine} (this corresponds to $A=I_N$ and $k=0$) and approximate the distribution of $\epsilon := \hat \theta(X_M, Y_M) - \theta$ as $\mathcal{N}(0, \tilde \Sigma_M).$
Further, to compute the bound, we approximate $\theta^*(X_M, Y_M)$ by $\tilde \theta^*(X_M, Y_M)$ as in \Cref{eqn:approx_MAP},
corresponding to $C=\left[I_N + \tilde \Sigma_M \right]\inv$ and $\ell=0$.

While the precise coverage of this bound is difficult to analyze, our next result reveals favorable properties in the large sample limit.
\begin{theorem}\label{thm:logistic_asymptotic_coverage}
Consider a sequence of random covariates $x_1, x_2, \dots$ and responses $y_1, y_2,\dots$ distributed as in \Cref{eqn:logistic_likelihood}.
For each $M\in \N,$ let $W_M := \|\hat \theta(X_M, Y_M) - \theta\|^2 - \|\theta^*(X_M, Y_M) - \theta\|^2$
be the win of using the MAP estimate in place of the MLE.
Finally, let $b_M(\alpha)$ be the level-$\alpha$ approximate bound on $W_M$ described above.
If $x_1, x_2, \dots$ are i.i.d.\ with finite third moment and with positive definite covariance, then for any $\alpha\in (0,1)$,
$\lim_{M\rightarrow \infty} \pr_\theta\left[ W_M \ge b_M(\alpha) \right] \ge \alpha.$
\end{theorem}
\Cref{thm:logistic_asymptotic_coverage} guarantees that in the large sample limit, $b_M(\cdot)$ has at least nominal coverage.
We provide a proof of the theorem and demonstrate its favorable empirical properties in simulation in \Cref{sec:logistic_regression_supp}.
\section{Applications}\label{sec:applications}
We now demonstrate our approach on the three applications introduced in \Cref{sec:intro}.
Our goal in this section is to demonstrate how one can compute and interpret c-values in realistic workflows.
In analogy to hypothesis testing, where a \emph{p}-value cutoff of 0.05 is standard for rejecting a null, we require a c-value of at least 0.95 to accept the alternative estimate;
with this threshold, we expect to incorrectly reject the default estimate in at most 5\% of our decisions.
\Revision{This choice, instead of 0.5 for example, reflects the presumed asymmetry of the comparisons;
we demand strong support to adopt the alternative over the default.}
For all applications, we provide substantial additional details in \Cref{sec:applications_supp}.
    \subsection{Estimation from educational testing data and empirical Bayes}\label{sec:small_area_education}
In this section we apply our methodology to a model and dataset considered by \citet[Section 3.2]{hoff2019smaller},
in which the goal is to estimate the average student reading ability at different schools in the 2002 Educational Longitudinal Study.
At each of $N=676$ schools, between $5$ and $50$ tenth grade students were given a standardized test of reading ability.
We let $y=[y_1, y_2,\dots, y_N]^\top$ denote the average scores,
and for each school, indexed by $n$, model $y_n\overset{indep}{\sim} \mathcal{N}(\theta_n, \sigma_n^2),$
where $\theta=[\theta_1, \theta_2, \dots, \theta_N]^\top$ denotes the school-level means
and each $\sigma_n$ is the school-level standard error;
specifically $\sigma_n:=\sigma/\sqrt{N_n}$ where $\sigma$ denotes a student-level standard deviation and $N_n$ is the number of students tested at school $N_n$.
For convenience, we let $\Sigma := \diag([\sigma_1^2, \sigma_2^2, \dots, \sigma_N^2])$ so that we may write $y\sim \mathcal{N}(\theta, \Sigma).$
The goal is to estimate the school-level performances $\theta.$

Following \citet{hoff2019smaller}, we perform small area inference with the Fay-Herriot model \citep{fay1979estimates}
to estimate $\theta$ under the assumption that similar schools may have similar student performances.
Specifically, we consider a vector of $D=8$ attributes of each school $X=[x_1, x_2, \dots, x_N]^\top$; these include
participation levels in a free lunch program, enrollment, and other characteristics such as region and school type.
We model the school-level mean as \emph{a priori} distributed as 
$\theta \sim \mathcal{N}(X\beta, \tau^2 I_N)$
where $\beta$ is an unknown $D$-vector of fixed effects and $\tau^2$ is an unknown scalar that describes variation in $\theta$ not captured by the covariates.
Following \citet{hoff2019smaller}, we take an empirical Bayesian approach and estimate $\beta, \tau$, and $\sigma$ with \texttt{lme4} \citep{bates2014fitting}.
We then compare the posterior mean --- which is affine in $y$ for fixed $\beta,\tau$, and $\sigma$ --- as an alternative to the MLE as a default; we use 
\Cref{approxbound:affine}.
Specifically, we take $\theta^*(y) := \E[\theta \mid y; \beta, \tau, \sigma] = [I_N + \tau^{-2}\Sigma]\inv y + [I_N + \tau^2 \Sigma\inv]\inv X\beta$ 
and $\hat \theta(y) = y.$
We compute a large c-value ($c=0.9926$); its closeness to one strongly suggests that $\theta^*(y)$ is more accurate than $\hat \theta(y).$

We should not always expect to obtain a large c-value for any alternative estimate, however.
We next describe a case where we expect the alternative estimate to be less accurate than the default, and we check that we obtain a small c-value.
In particular, we now let our alternative estimate be the posterior mean under the same model as above but with the covariates, $X,$ randomly permuted across schools.
In this situation, the responses $y$ have no relation to the covariates, and we should not expect an improvement.
Indeed, on this dataset we compute a c-value of exactly zero.
However, we recall that just as a large p-value in hypothesis testing does not provide support that a null hypothesis is true,
a small c-value does not provide direct support that the alternative estimate is less accurate than the default.

We provide additional details for all parts of this application in \Cref{sec:small_area_estimation_supp}.
There, we demonstrate in a simulation study that our bounds remain substantially conservative for these estimators and model even with an empirical Bayes step.
    \subsection{Estimating violent crime density in Philadelphia}
\label{sec:philadelphia}

As a second application, we consider estimating the areal density of violent crimes (i.e. counts per square mile) reported in each of Philadelphia's $N = 384$ census tracts.
Following \citet{Balocchi2019}, we work with the inverse hyperbolic sine transformed density.
Letting $y_{n}$ be the observed transformed density of reported violent crimes in census tract $n,$ we model $y_{n} \overset{indep}{\sim} \mathcal{N}(\theta_{n}, \sigma_{y}^{2})$ where $\theta_{n}$ represents the underlying transformed density and $\sigma^{2}_{y}$ is the noise variance.
While one might interpret $\theta_n$ as the true density of violent crime in census tract $n$, we note that the implicit assumption of zero-mean error in each tract may not be realistic.
Namely, systematic biases may impact the rates at which police receive and respond to calls and file incident reports in different parts of the city.
Unfortunately, we are unable to probe this possibility with the available data.
Nevertheless, our goal is to estimate the vector of unknown rates,
$\theta = [\theta_1, \theta_2, \dots, \theta_N]^\top$ from $y=[y_1, y_2, \dots, y_N]^\top.$
The observations $y$ are a simple proxy of transformed violent crime density, but they are noisy.
So it is natural to wonder if we might obtain a more accurate estimate of $\theta.$

\begin{figure}[h]
\centering
\begin{subfigure}[b]{0.42\textwidth}
\includegraphics[width = \textwidth]{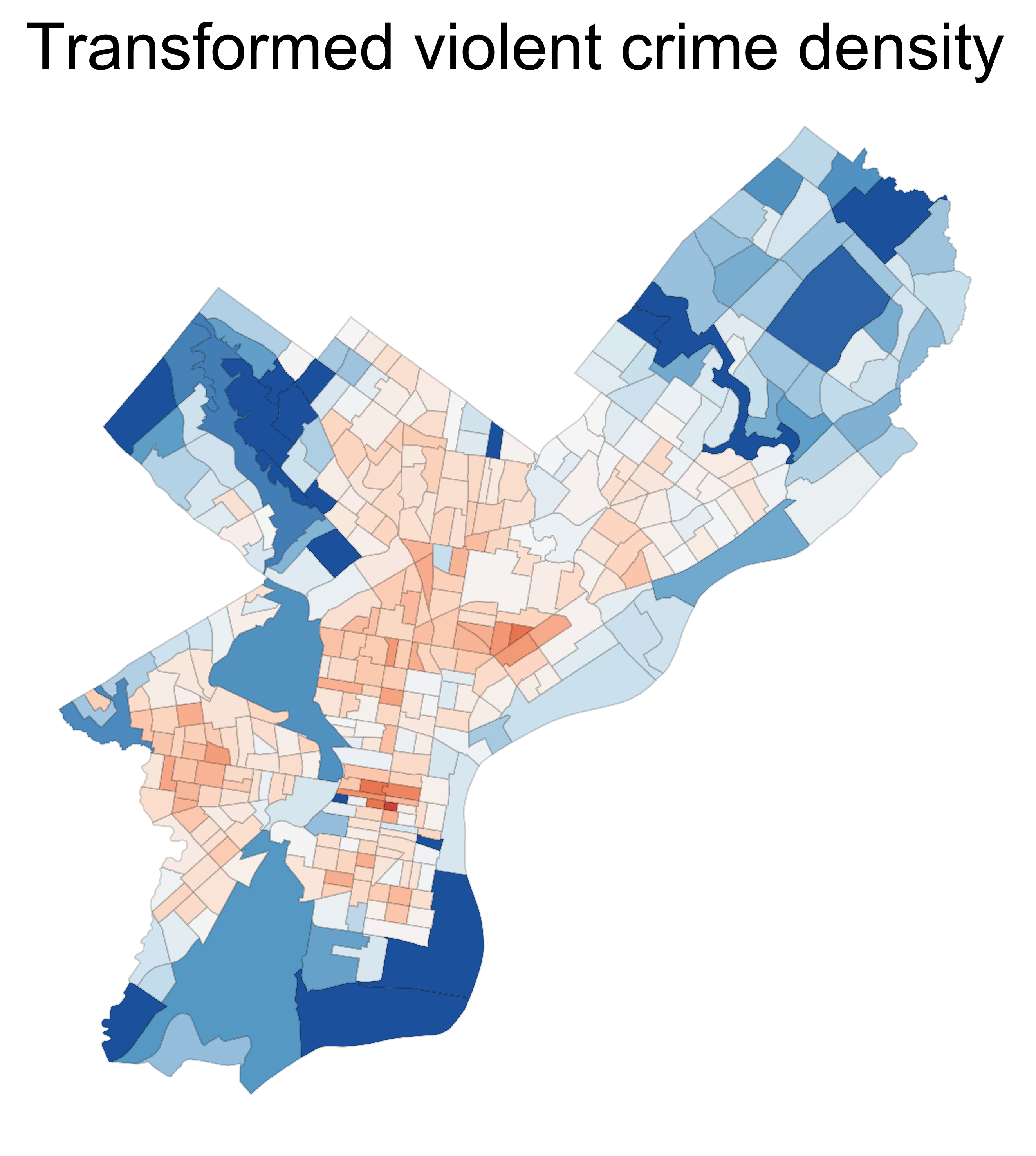}
\caption{}\label{fig:philly_viol}
\end{subfigure}
\begin{subfigure}[b]{0.43\textwidth}
\includegraphics[width=\textwidth]{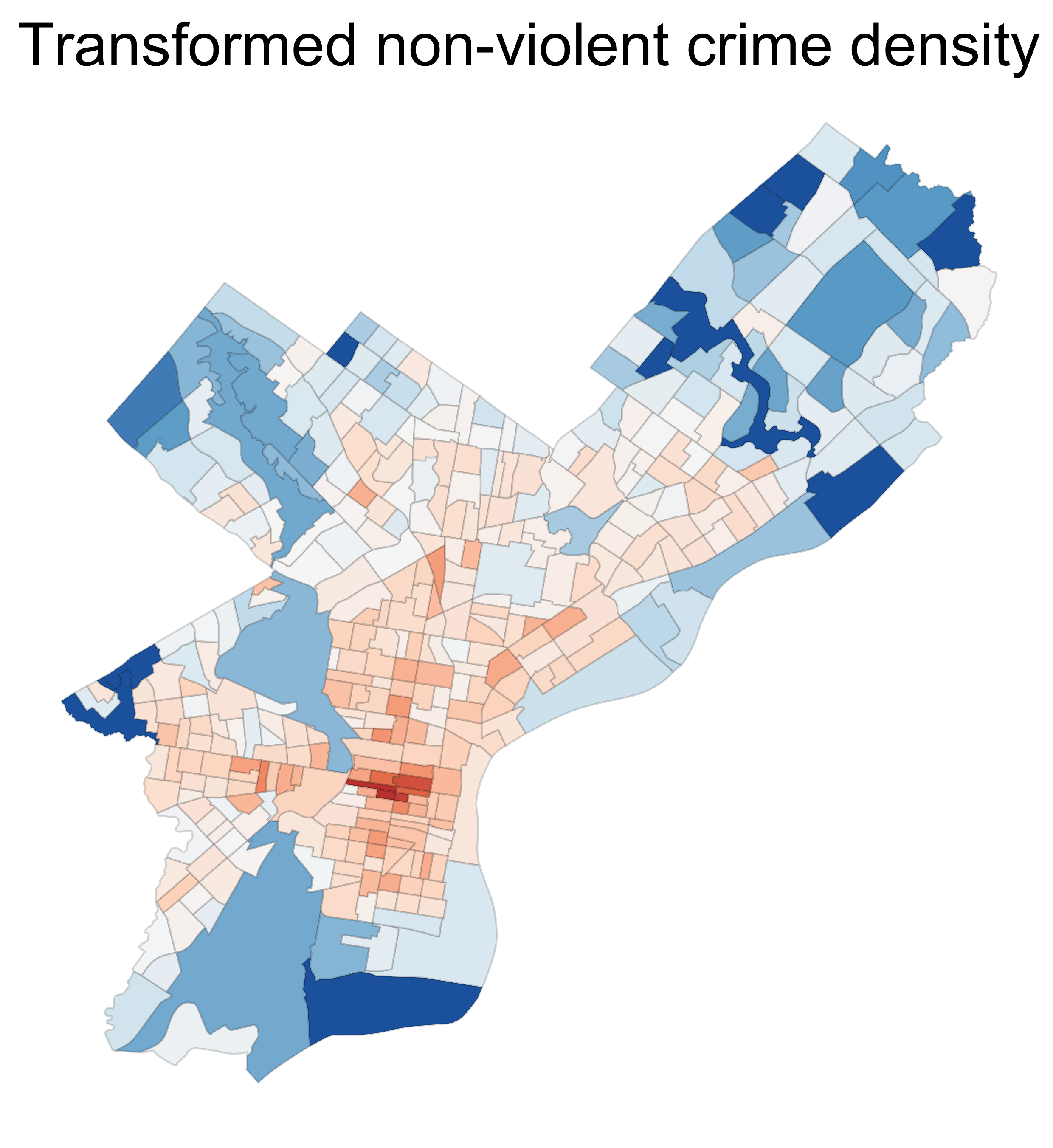}
\caption{}\label{fig:philly_nonviol}
\end{subfigure}
\begin{subfigure}[b]{0.08\textwidth}
\includegraphics[width=\textwidth]{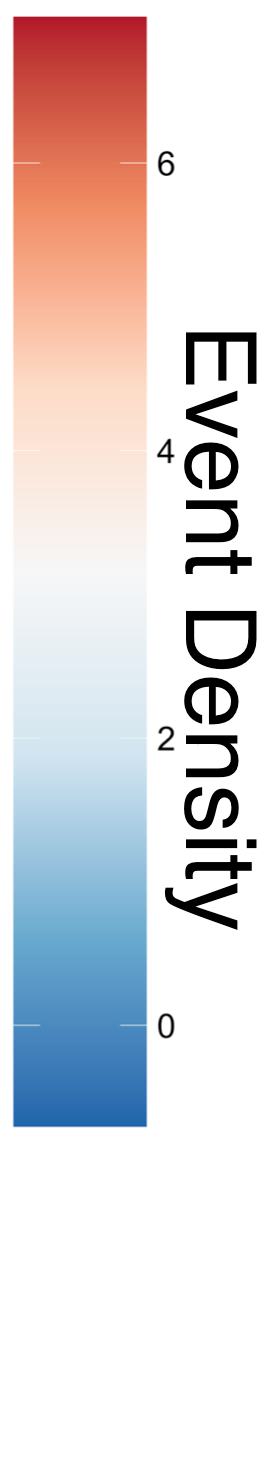}
\end{subfigure}
\caption{Transformed densities of reported (a) violent and (b) non-violent crimes in each census tract in Philadelphia in October 2018.}
\label{fig:philly}
\end{figure}

\Cref{fig:philly} plots the transformed densities of both violent and non-violent crimes reported in October 2018 in each census tract.
Immediately, we see that, for any particular census tract, the observed densities of the two types of crime are similar.
Further, we observe considerable spatial correlation in each plot.
It is tempting to use a Bayesian hierarchical model that exploits this structure in order to produce more accurate estimates of $\theta.$
In this application, we consider iteratively refining an estimate of $\theta$ by (A) incorporating the observed non-violent crime data and then by (B) carefully accounting for the observed spatial correlation. 
At each step of our refinement, we use a c-value to decide whether to continue.
Before proceeding, we make a remark about our sequential approach.

\begin{remark}\label{remark:comparing_3_estimates}
Consider using $c$-values and a chosen level $\alpha$ to choose one of three estimates (say $\hat \theta(y), \theta^*(y)$, and $\theta^\circ(y)$) in two stages. 
Suppose we first choose $\theta^*(y)$ over $\hat \theta(y)$ only if the associated c-value is greater than $\alpha$.
Second, only if we chose $\theta^*(y)$, we next choose $\theta^\circ(y)$ over $\theta^*(y)$ only if the new c-value associated with those estimates exceeds $\alpha$. Then a union bound guarantees that $\theta^\circ(y)$ will be incorrectly chosen with probability at most $2(1-\alpha)$.
\end{remark}

We begin by seeing if we can improve upon the MLE, $\hat \theta(y)=y,$ by leveraging the auxiliary dataset of transformed non-violent crimes in each tract, $z_1, z_2, \dots, z_N.$
To this end, we model these auxiliary data analogously to $y$;
in each tract $n,$  we let $\eta_n$ be the unknown transformed density and independently model $z_n \overset{indep}{\sim} \mathcal{N}(\eta_n, \sigma_z^2).$ 
We next introduce a hierarchical prior that captures the apparent similarity between $\theta$ and $\eta$ within each tract.
Specifically, for each tract $n$ we decompose
$\theta_n = \mu_n + \delta^y_n$ and 
$\eta_n = \mu_n + \delta^z_n,$
where $\mu_n$ is a shared mean for the transformed densities of violent and non-violent reports and $\delta_n^y$ and $\delta_n^z$ represent deviations from the shared mean specific to each crime type.
Rather than encode explicit prior beliefs about $\mu_n,$ we express ignorance in these quantities with an improper uniform prior.
Additionally, we model $\delta^y_n, \delta^z_n\overset{i.i.d}{\sim}\mathcal{N}(0, \sigma_\delta^2)$.
We fix the values of $\sigma_{y}, \sigma_{z},$ and $\sigma_{\delta}$ using historical data.
We then compute the posterior mean of $\theta$ as an alternative estimate, $\theta^*(y).$
Thanks to the Gaussian conjugacy of this model, $\theta^{*}(y)$ is affine in the data $y$, and a closed form expression is available. 
See \Cref{sec:philadelphia_supp} for additional details.
\Revision{The resulting c-value exceeded 0.999, suggesting} that we should be highly confident that $\theta^{*}(y)$ is a more accurate estimate of $\theta$ than $\hat{\theta}(y).$ 

We next consider additionally sharing strength amongst spatially adjacent census tracts.
To this end, consider a second model with spatially correlated variance components:
$
\theta_n = \mu_n + \delta_n^y + \kappa_n^y \text{  and  }
\eta_n= \mu_n + \delta_n^z + \kappa_n^z.
$
The additional terms $\kappa^y=[\kappa^y_1, \kappa^y_2, \dots, \kappa^y_N]^\top$ and $\kappa^z=[\kappa^z_1, \kappa^z_2, \dots, \kappa^z_N]^\top$ capture a priori spatial correlations;
we model $\kappa^y, \kappa^z \overset{i.i.d.}{\sim}\mathcal{N}(0, K),$
where $K$ is an $N\times N$ covariance matrix determined by a squared exponential covariance function \citep[Chapter 4]{rasmussen2006gaussian} that depends on the distance between the centroids of the census tracts.
Once again, we exploit conjugacy in this second hierarchical model to derive the posterior mean $\theta^{\circ}(y)$ in closed form.
As $\theta^{\circ}(y)$ is also an affine transformation of $y,$ we can use \Cref{approxbound:affine} to compute the c-value for comparing $\theta^{\circ}(y)$ to $\theta^{*}(y).$ 
The c-value for this comparison is only 0.843, providing much weaker support for using $\theta^{\circ}(y)$ over $\theta^{*}(y).$ 
Because this c-value is less than 0.95, we conclude our analysis content with $\theta^{*}(y)$ as our final estimate.
    \subsection{Gaussian process kernel choice: modeling ocean currents}\label{sec:submesoscale}
Accurate understanding of ocean current dynamics is important for forecasting the dispersion of oceanic contaminations, such as after the Deepwater Horizon oil spill \citep{poje2014submesoscale}.
\citet{lodise2020investigating} have recently advocated for a statistical approach to inferring ocean currents from observations of free-floating, GPS-trackable buoys.
Their approach seeks to provide improved estimates by incorporating 
variation at the \emph{submesoscale} (roughly 0.1--10 km) in addition to
more commonly considered \emph{mesoscale} variation (roughly 10 km and above).
In this section we apply our methodology to assess if this approach provides improved estimates relative to a baseline including only mesoscale variation.

In our analysis, we consider a segment of the Carthe Grand Lagrangian Drifter (GLAD) deployment dataset \citep{ozgokmen2013glad}.
Specifically, we model a set of $50$ buoys with velocities estimated at $3$ hour intervals over one day ($N=400$ observations total).
Each observation $n$ consists of latitudinal and longitudinal ocean current velocity measurements $y_n =[y^{(1)}_n, y^{(2)}_n]^\top\in \R^2$
and associated spatio-temporal coordinates $[\text{lat}_n, \text{lon}_n, t_n].$
Following \citet{lodise2020investigating}, we model each measurement as a noisy observation of an underlying time varying vector-field distributed independently as 
$y_n\overset{indep}{\sim}\mathcal{N}\left(F(\text{lat}_n, \text{lon}_n, t_n), \sigma_\epsilon^2 I_2\right),$
where $F: \R^3 \rightarrow \R^2$ denotes the time evolving vector-field of ocean currents and $\sigma_\epsilon^2$ is the error variance.
Our goal is to estimate $F$ at the observation points
$\theta := [\theta_1, \theta_2, \dots, \theta_N]^\top,$
where for each $n,  \theta_n=[\theta_n^{(1)}, \theta_n^{(2)}]^\top =F(\text{lat}_n, \text{lon}_n, t_n).$

Following \citet{lodise2020investigating}, we place a Gaussian process prior on $F$ to encode expected spatio-temporal structure while allowing for variation at multiple scales.
Specifically, we model 
$F \sim \mathcal{GP}\left(0, k(\cdot, \cdot)\right),$ where
\[\label{eqn:two_variance_component_kernel}
    k(\theta_n^{(i)}, \theta_{n^\prime}^{(i)}) =  k_1(\theta_n^{(i)}, \theta_{n^\prime}^{(i)}) + k_2(\theta_n^{(i)}, \theta_{n^\prime}^{(i)}),
\quad i \in \{1, 2\}.
\]
Here $k_1$ and $k_2$ are squared exponential kernels with spatial and temporal length-scales that reflect mesoscale and submesoscale variations, respectively; see \Cref{sec:submesoscale_supp} for details.
For simplicity, we model the latitudinal and longitudinal components of $F$ independently.
We take the posterior mean of $\theta$ under this model as the alternative estimate, $\theta^*(y).$

As a baseline, we consider an analogous estimate with covariance function
$k(\theta_n^{(i)}, \theta_{n^\prime}^{(i)}) =  k_1(\theta_n^{(i)}, \theta_{n^\prime}^{(i)}) + k_2(\theta_n^{(i)}, \theta_{n^\prime}^{(i)})\mathbbm{1}[n=n^\prime],$
which maintains the same marginal variance but excludes submesoscale covariances.
We take the posterior mean under this model as the default estimate $\hat \theta(y)$.
Both $\theta^*(y)$ and $\hat \theta(y)$ may be written as affine transformations of $y.$

Using \Cref{approxbound:affine}, we compute a c-value of $0.99981.$ This large c-value allows us to confidently conclude that modeling both mesoscale and submesocale variation can yield more accurate estimates of ocean currents than mesocale modeling alone. 
\section{Discussion}\label{sec:discussion}
We have provided a simple method for quantifying confidence in improvements provided by \Revision{a wide class of shrinkage estimates} without relying on subjective assumptions about the parameter of interest.
Our approach has compelling theoretical properties, and we have demonstrated its utility on several data analyses of recent interest.
However, the scope of the current work has several limitations.
The present paper has explored the use of the c-value only for problems of moderate dimensionality ($N$ between $20$ and $700$).
Loosely speaking, we suspect c-values may be underpowered to robustly identify substantial improvements provided by estimates in lower dimensional problems.
Further investigation into such dimension dependence is an important direction for future work.
In addition, our approach depends crucially on a high-probability lower bound that is inherently specific to the underlying model of the data, a loss function, and the pair of estimators.
In the present work, we have shown how to derive and compute this bound for models with general Gaussian likelihoods,
when accuracy may be measured in terms of squared error loss,
and when both estimates are affine transformations of the data.
We have provided a first step to extending beyond simple Gaussian models with the application to logistic regression;
while we have not yet explored the efficacy of this extension on real data,
we view our work as an important starting point for generalizing to broader model classes and estimation problems.
We believe that further extensions to the classes of models, estimates, and losses for which c-values can be computed provide fertile ground for future work.

\Revision{
One direction we believe is promising is to construct the bound $b(y, \alpha)$ in a model and loss agnostic manner using, for example, the parametric bootstrap.
Constructing an informative c-value is possible because in some cases the distribution of the win depends on the unknown parameter only through some low-dimensional projection (or at least approximately so).
We suspect that this phenomenon may extend to more complex models and estimates.
In such cases, when this low-dimensional characteristic sufficiently captures the distribution of the win and is estimated well enough, a parametric bootstrap may present a powerful solution.
In particular, one would begin by forming an initial estimate of the parameter, and simulate a collection of bootstrap datasets by sampling data from the likelihood parameterized by the initial estimate, compute the win for each simulated dataset, and return for each $b(y, \alpha)$ the $1-\alpha$ quantile of this distribution.
We expect that this method may work in many important settings; indeed, much of modern statistics and nonlinear methods are predicated on the assumption that low-dimensional structure (e.g.\ sparsity) exists and may be inferred.
We leave further development of this more flexible approach, including an investigation of the theoretical properties, to follow-up work.
}

\if1\blind
{
\section*{Acknowledgements}
The authors thank Jonathan H. Huggins for the suggestions to consider Berry--Esseen bounds and the extension to logistic regression,
Lorenzo Masoero and Hannah Diehl for insightful comments on the manuscript,
and Matthew Stephens and Lucas Janson for useful early conversations.
This work was supported in part by an ARPA-E project with program director David Tew, and an NSF CAREER Award.
We are grateful to the Office of Naval Research for partial support under grant N00014-20-1-2023 (MURI ML-SCOPE) to the Massachusetts Institute of Technology.
BLT is supported by NSF GRFP.
SKD is supported by the Wisconsin Alumni Research Foundation.

} \fi

\appendix

\renewcommand{\theequation}{S\arabic{equation}}
\renewcommand{\thesection}{S\arabic{section}}  
\renewcommand{\thefigure}{S\arabic{figure}}  
\renewcommand{\thetable}{S\arabic{table}} 
\section{Appendix}\label{sec:appendix}

\textbf{Proof of \Cref{thm:c_values}}
\begin{proof}
The result follows directly from the definition of $c(y)$ and the conditions on $b(\cdot, \cdot)$.
More explicitly,
\(
    \pr_\theta\left[ W(\theta, y) \le 0  \text{ and }c(y) > \alpha \right] &\le 
    \pr_\theta\left[ W(\theta, y) \le 0 \text{ and } b(y,\alpha) > 0 \right]\\
    &\le \pr_\theta\left[ W(\theta, y) < b(y,\alpha) \right]\\
&\le 1-\alpha,
\)
where the first line follows from the definition of the c-value 
and the final line follows from \Cref{eqn:bya_bound}.
\end{proof}

\noindent
\textbf{Proof of \Cref{thm:theta_tilde_guarantee}}
\begin{proof}
    The condition $L(\theta, \theta^\dagger(y, \alpha)) > L(\theta, \hat \theta(y))$ can occur only when both (A) $0>W(\theta, y)$ and
(B) $\theta^\dagger(\cdot, \alpha)$ evaluates to $\theta^*(\cdot)$ rather than $\hat \theta(\cdot).$
Event (B) implies $c(y)>\alpha$ and therefore $b(y,\alpha)>0$.
By transitivity, $b(y, \alpha)>0 \text{ and } 0 > W(\theta, y) \implies b(y,\alpha) > W(\theta, y)$.
By assumption, the event $b(y,\alpha) > W(\theta, y)$ occurs with probability at most $1-\alpha$.
\end{proof}

\bibliographystyle{agsm}
\bibliography{references}

@article{Kondo2018,
	Author = {Michelle C Kondo and Elena Andreyeva and Eugenia C South and John M MacDonal and Charles C Branas},
	Journal = {Annual Review of Public Health},
	Title = {Neighborhood interventions to reduce violence},
	Volume = {39},
	Year = {2018}}

@article{Buka2001,
	Author = {S. L. Buka and T. L. Stichick and I. Birdthistle and F.J. Earls},
	Journal = {American Journal of Orthopsychiatry},
	Number = {3},
	Title = {Youth exposure to violence: prevalance, risks, and consequences},
	Volume = {71},
	Year = {2001}}

@article{bates2014fitting,
	Author = {Bates, Douglas and M{\"a}chler, Martin and Bolker, Ben and Walker, Steve},
	Journal = {Journal of Statistical Software},
	Number = {1},
	Title = {Fitting Linear Mixed-Effects Models Using lme4},
	Volume = {67},
	Year = {2015}}

@misc{ozgokmen2013glad,
	Author = {{\"O}zg{\"o}kmen, Tamay M},
	Doi = {10.7266/N7VD6WC8},
	Title = {{GLAD experiment CODE-style drifter trajectories (low-pass filtered, 15 minute interval records), Northern Gulf of Mexico near DeSoto Canyon, July-October 2012. Harte Research Institute, Texas A\&M University-Corpus Christi}},
	Url = {https://data.gulfresearchinitiative.org/data/R1.x134.073:0004},
	Year = {2013}}

@article{fay1979estimates,
	Author = {Fay, Robert E and Herriot, Roger A},
	Journal = {Journal of the American Statistical Association},
	Number = {366a},
	Publisher = {Taylor \& Francis},
	Title = {{Estimates of income for small places: an application of James-Stein procedures to census data}},
	Volume = {74},
	Year = {1979}}

@article{weisburd2019proactive,
	Author = {Weisburd, David and Majmundar, Malay K and Aden, Hassan and Braga, Anthony and Bueermann, Jim and Cook, Philip J and Goff, Phillip Atiba and Harmon, Rachel A and Haviland, Amelia and Lum, Cynthia and others},
	Journal = {Asian Journal of Criminology},
	Number = {2},
	Pages = {145--177},
	Publisher = {Springer},
	Title = {Proactive policing: A summary of the report of the National Academies of Sciences, Engineering, and Medicine},
	Volume = {14},
	Year = {2019}}

@article{slocum2020enforcement,
	Author = {Slocum, Lee A and Huebner, Beth M and Greene, Claire and Rosenfeld, Richard},
	Journal = {Journal of Community Psychology},
	Number = {1},
	Pages = {36--67},
	Publisher = {Wiley Online Library},
	Title = {{Enforcement trends in the city of St. Louis from 2007 to 2017: Exploring variability in arrests and criminal summonses over time and across communities}},
	Volume = {48},
	Year = {2020}}

@article{lodise2020investigating,
	Author = {Lodise, John and {\"O}zg{\"o}kmen, Tamay and Gon{\c{c}}alves, Rafael C and Iskandarani, Mohamed and Lund, Bj{\"o}rn and Horstmann, Jochen and Poulain, Pierre-Marie and Klymak, Jody and Ryan, Edward H and Guigand, Cedric},
	Journal = {Fluids},
	Number = {3},
	Publisher = {Multidisciplinary Digital Publishing Institute},
	Title = {Investigating the Formation of Submesoscale Structures Along Mesoscale Fronts and Estimating Kinematic Quantities Using Lagrangian Drifters},
	Volume = {5},
	Year = {2020}}

@article{gonccalves2019reconstruction,
	Author = {Gon{\c{c}}alves, Rafael C and Iskandarani, Mohamed and {\"O}zg{\"o}kmen, Tamay and Thacker, W Carlisle},
	Journal = {Journal of Physical Oceanography},
	Number = {4},
	Pages = {941--958},
	Title = {Reconstruction of submesoscale velocity field from surface drifters},
	Volume = {49},
	Year = {2019}}

@article{poje2014submesoscale,
	Author = {Poje, Andrew C and {\"O}zg{\"o}kmen, Tamay M and Lipphardt, Bruce L and Haus, Brian K and Ryan, Edward H and Haza, Angelique C and Jacobs, Gregg A and Reniers, AJHM and Olascoaga, Maria Josefina and Novelli, Guillaume and Griffa, Annalisa and Beron-Vera, Francisco J. and Chen, Shuyi S. and Coelho, Emanuel and Hogan, Patrick J. and Kirwan, Albert D. Jr. and Huntley, Helga S. and Mariano, Arthur J.},
	Journal = {Proceedings of the National Academy of Sciences},
	Number = {35},
	Publisher = {National Acad Sciences},
	Title = {Submesoscale dispersion in the vicinity of the {Deepwater Horizon} spill},
	Volume = {111},
	Year = {2014}}

@book{casella2002statistical,
	Author = {Casella, George and Berger, Roger L},
	Publisher = {Duxbury Pacific Grove, CA},
	Title = {{Statistical Inference}},
	Year = {2002}}

@article{robbins1964empirical,
	Author = {Robbins, Herbert},
	Journal = {The Annals of Mathematical Statistics},
	Number = {1},
	Publisher = {JSTOR},
	Title = {{The empirical Bayes approach to statistical decision problems}},
	Volume = {35},
	Year = {1964}}

@article{efron1973stein,
	Author = {Efron, Bradley and Morris, Carl},
	Journal = {Journal of the American Statistical Association},
	Number = {341},
	Publisher = {Taylor \& Francis Group},
	Title = {{Stein's estimation rule and its competitors --- an empirical Bayes approach}},
	Volume = {68},
	Year = {1973}}

@book{rasmussen2006gaussian,
	Author = {Rasmussen, Carl Edward and Williams, Christopher KI},
	Publisher = {MIT Press},
	Title = {{Gaussian processes for machine learning}},
	Year = {2006}}

@article{berry1941accuracy,
	Author = {Berry, Andrew C},
	Journal = {Transactions of the American Mathematical Society},
	Number = {1},
	Pages = {122--136},
	Publisher = {JSTOR},
	Title = {{The accuracy of the Gaussian approximation to the sum of independent variates}},
	Volume = {49},
	Year = {1941}}

@article{trippe2019lr,
	Author = {Trippe, Brian L and Huggins, Jonathan H and Agrawal, Raj and Broderick, Tamara},
	Booktitle = {International Conference on Machine Learning},
	Pages = {6315--6324},
	Title = {{LR}-{GLM}: High-Dimensional {B}ayesian Inference Using Low-Rank Data Approximations},
	Volume = {97},
	Year = {2019}
}

@book{van2000asymptotic,
	Author = {Van der Vaart, Aad W},
	Publisher = {Cambridge University Press},
	Title = {{Asymptotic Statistics}},
	Year = {2000}}

@article{hoff2019smaller,
	Author = {Hoff, Peter D},
	Number = {0},
	Volume = {0},
	Journal = {Journal of the American Statistical Association},
	Publisher = {Taylor \& Francis},
	Title = {Smaller $ p $-values via indirect information},
	Year = {2021}}

@book{mathai1992quadratic,
	Author = {Mathai, Arakaparampil M and Provost, Serge B},
	Publisher = {Dekker},
	Title = {{Quadratic Forms in Random Variables: Theory and Applications}},
	Year = {1992}}

@article{Lee2016,
	Author = {Jason D Lee and Dennis L. Sun and Yuekai Sun and Jonathan E. Taylor},
	Journal = {Annals of Statistics},
	Number = {3},
	Title = {Exact post-selection inference, with application to the lasso},
	Volume = {44},
	Year = {2016}}

@article{Tian2020,
	Author = {Xiaoying Tian},
	Journal = {Annals of Statistics},
	Number = {2},
	Title = {Prediction error after model search},
	Volume = {48},
	Year = {2020}}

@article{Lockhart2014,
	Author = {Richard Lockhart and Jonathan Taylor and Ryan J. Tibshirani and Robert Tibshirani},
	Journal = {Annals of Statistics},
	Number = {2},
	Title = {A significance test for the lasso},
	Volume = {42},
	Year = {2014}}

@article{Tibshirani2019,
	Author = {Ryan Tibshirani and Saharon Rosset},
	Journal = {Journal of the American Statistical Association},
	Number = {526},
	Pages = {697 -- 712},
	Title = {Excess optimism: how biased is the apparent error of an estimator tuned by {SURE}?},
	Volume = {114},
	Year = {2019}}

@article{Burbidge1988,
	Author = {Burbidge, John B and Magee, Lonnie and Robb, A Leslie},
	Journal = {Journal of the American Statistical Association},
	Number = {401},
	Pages = {123--127},
	Publisher = {Taylor \& Francis},
	Title = {Alternative transformations to handle extreme values of the dependent variable},
	Volume = {83},
	Year = {1988}}

@article{Balocchi2019,
	Author = {Cecilia Balocchi and Sameer K. Deshpande and Edward I. George and Shane T. Jensen},
	Journal = {Journal of the American Statistical Association},
	Title = {{`Crime in {P}hiladelphia: {B}ayesian clustering with particle optimization'}},
	Year = {2022}}

@article{BalocchiJensen2019,
	Author = {Cecilia Balocchi and Shane T. Jensen},
	Journal = {The Annals of Applied Statistics},
	Number = {4},
	Title = {Spatial modeling of trends in crime over time in {P}hiladelphia},
	Volume = {13},
	Year = {2019}}

@book{lehmann2006theory,
	Author = {Lehmann, Erich L and Casella, George},
	Publisher = {Springer Science \& Business Media},
	Title = {Theory of point estimation},
	Year = {2006}}

@article{lindley1972bayes,
	Author = {Lindley, Dennis V and Smith, Adrian FM},
	Journal = {Journal of the Royal Statistical Society: Series B},
	Number = {1},
	Publisher = {Wiley Online Library},
	Title = {Bayes estimates for the linear model},
	Volume = {34},
	Year = {1972}}

@article{wallace1977pretest,
	Author = {Wallace, T Dudley},
	Journal = {American Journal of Agricultural Economics},
	Number = {3},
	Publisher = {Wiley Online Library},
	Title = {Pretest estimation in regression: a survey},
	Volume = {59},
	Year = {1977}}

@article{taylor2018post,
	Author = {Taylor, Jonathan and Tibshirani, Robert},
	Journal = {Canadian Journal of Statistics},
	Number = {1},
	Publisher = {Wiley Online Library},
	Title = {Post-selection inference for penalized likelihood models},
	Volume = {46},
	Year = {2018}}

@article{berk2013valid,
	Author = {Berk, Richard and Brown, Lawrence and Buja, Andreas and Zhang, Kai and Zhao, Linda},
	Journal = {The Annals of Statistics},
	Number = {2},
	Publisher = {Institute of Mathematical Statistics},
	Title = {Valid post-selection inference},
	Volume = {41},
	Year = {2013}}

@article{morris1983parametric,
	Author = {Morris, Carl N},
	Journal = {Journal of the American Statistical Association},
	Number = {381},
	Pages = {47--55},
	Publisher = {Taylor \& Francis Group},
	Title = {{Parametric empirical Bayes inference: theory and applications}},
	Volume = {78},
	Year = {1983}}

@article{claeskens2003focused,
  title={The focused information criterion},
  author={Claeskens, Gerda and Hjort, Nils Lid},
  journal={Journal of the American Statistical Association},
  volume={98},
  number={464},
  year={2003},
  publisher={Taylor \& Francis}
}

@article{stein1981estimation,
  title={Estimation of the mean of a multivariate normal distribution},
  author={Stein, Charles M},
  journal={The Annals of Statistics},
  pages={1135--1151},
  year={1981},
  publisher={JSTOR}
}
\newpage
\bigskip
\begin{center}
{\large\bf SUPPLEMENTARY MATERIAL}
\end{center}
%
%
%
%


\section{Pitfalls of risk when choosing between estimators}
\label{sec:lindley_and_smith_risk}
Before proceeding, we require some additional notation and definitions.
We denote the risk of an arbitrary estimator $\theta'(\cdot)$ by $R(\theta, \theta') = \E_{\theta}\left[L\left(\theta, \theta'(y)\right)\right].$
Given two estimators $\theta'(\cdot)$ and $\theta^{\dagger}(\cdot)$ we say that $\theta'(\cdot)$ \textit{dominates} $\theta^{\dagger}(\cdot)$ if, for all values of $\theta,$ $R(\theta, \theta') \leq R(\theta, \theta^{\dagger})$ and $R(\theta, \theta') < R(\theta, \theta^{\dagger})$ for at least one value of $\theta.$

If we were able to show that one of $\hat{\theta}(\cdot)$ or $\theta^{*}(\cdot)$ dominates the other, it would be tempting to always select the dominating estimator.
Unfortunately, it is very often the case that neither estimator dominates the other.
In other words, it may be the case that $R(\theta, \theta^{*}) < R(\theta, \hat{\theta})$ for all values of $\theta$ in some non-trivial subset of the space $\Theta_{0}$ but $R(\theta, \theta^{*}) > R(\theta,\hat{\theta})$ for some $\theta \notin \Theta_{0}.$
\citet{lindley1972bayes} provide a simple illustration of this dilemma in the following normal means problem.
Suppose that we observe an $N$-vector normally distributed about its mean and with identity covariance, $I_N$, as  $y \sim \mathcal{N}(\theta, I_N),$ and wish to compare the default estimate $\hat{\theta}(y) = y$ of $\theta$ and the alternative estimate
$$
\theta^{*}(y) = \frac{y + \overline{y}\mathbf{1}_{N}/\tau^{2}}{1 + 1/\tau^{2}}
$$
for a fixed value of $\tau > 0,$ where $\overline{y}:=N\inv \sum_{n=1}^N y_n$
and $\ones{N}$ is the $N$-vector of ones.
\citet{lindley1972bayes} showed that $R(\theta, \theta^{*}) < R(\theta, \hat{\theta})$ if and only if 
\begin{equation}
\label{eq:lindley_smith_condition}
\lVert \theta - \overline{\theta}\ones{N}\rVert_{2} < \sqrt{(N-1)(2 + \tau^{2})},
\end{equation}
where $\overline{\theta} := N\inv \sum_{n=1}^N\theta_n.$
Without strong assumptions about the value of $\theta,$ which we may be unable or unwilling to make, a simple comparison of risk functions can prove inconclusive. 
Interestingly, in the setting considered by \citet{lindley1972bayes}, it is possible to construct $\theta$ so that (A) $R(\theta, \theta^{*}) < R(\theta, \hat{\theta})$ but (B) $\mathbb{P}_{\theta}[L(\theta, \theta^{*}(y)) > L(\theta, \hat{\theta}(y))] > 0.5.$
In particular, for $N=2, \tau=1,$ and $ \|\theta - \overline{\theta} \ones{N}\|^2 = 2.999, \theta^*(\cdot)$ has slightly smaller risk than the MLE, but the MLE has smaller loss in $3397$ out of $5000$ simulated datasets, or about $68\%$ of the time.
In other words, even if we were to assume that $\theta$ satisfied \Cref{eq:lindley_smith_condition}, for the majority of datasets $y$ that we might observe, the alternative estimator incurs higher loss than the default.
The situation above highlights an important, but in our mind under-discussed, limitation of risk: the loss averaged over all possible unrealized datasets may not be close to the loss incurred on an observed dataset.

This disagreement between risk and the probability of having smaller loss can be especially pronounced when the distribution of the loss of one of the estimators is heavy-tailed.
For example, consider a scalar parameter $\theta=0,$
a deterministic default estimate $\hat \theta = 1,$ and 
an alternative estimate distributed as
$\theta^* \sim \frac{1}{\alpha}\delta_{\sqrt{\alpha(1 + \epsilon)}} + (1-\frac{1}{\alpha})\delta_0,$
where $\delta_x$ denotes a Dirac mass on $x$ and $\epsilon>0$.
Then $\theta^*(\cdot)$ has larger risk than $\hat \theta(\cdot)$ ($1+\epsilon$ rather than $1$),
but has smaller loss with probability $1-\frac{1}{\alpha}.$
By taking $\alpha\rightarrow\infty,$ we see that $\theta^*(\cdot)$ may have smaller loss than $\hat \theta(\cdot)$ with arbitrarily high probability.
This example is particularly extreme;
our intent is merely to illustrate that large disagreements could, at least in principal, arise in practical settings.
\section{Defining c-values as a supremum vs.\ infimum}\label{sec:c_value_inf_vs_sup}
In this section we describe a pathological model and construction of a lower bound function for which the two possible definitions of the c-value described in \Cref{remark:c_value_inf_vs_sup} lead to notably different behaviours.

Consider a variant of the normal means problem.
Let $\theta\in\R$ be an unknown mean and observe
$$
y := \begin{bmatrix}
\theta + \epsilon \\
u
\end{bmatrix},
$$
where $\epsilon \sim \mathcal{N}(0, 1)$ and $u\sim \mathcal{U}([0, 1])$ is a uniform random variable on $[0, 1]$.
Note that $u$ is ancillary to $\theta$ (i.e.\ its distribution does not depend on $\theta$).
We will construct a pathological $b(y,\alpha)$ that depends on $y$ only through $u$ and will therefore be ancillary to $\theta$ as well.
We begin by constructing a countably infinite collection of independent uniform random variables from $u,$ indexed by the rationals $\mathbb{Q}$, $S(u) := \{ u_r\}_{r\in \mathbb{Q}}.$
Such a countably infinite collection may be obtained by segmenting the decimal expansion of $u$;
for example, if we let $d_i$ denote the $i^{th}$ digit of $u,$ we could obtain this sequence by defining uniform random variables with decimal expansions
\(
u^1 :&= [d_1, d_2, d_4, d_7, d_{11} \dots],\\
u^2 :&= [d_3, d_5, d_8, d_{12} \dots],\\
u^3 :&= [d_6, d_9, d_{13}\dots],\\
u^4 :&= [d_{10}, d_{14}, \dots],\\
u^5 :&= [d_{15}, \dots],
\)
and so on, and then mapping from $\{u^i \}_{i \in \N}$ to $S(u)$.

Next, define 
\(
  b(y, \alpha) :=
    \begin{cases}
       (-1)^{\mathbbm{1}[u_\alpha < \alpha]} \infty & \text{if } \alpha \in \mathbb{Q} \\
      -\infty  & \text{otherwise}.
    \end{cases}
\)
For any bounded default and alternative estimators, the win will be finite and
the bound $b(y, \alpha)$ holds if an only if it evaluates to $-\infty.$
Because $b(y,\alpha)=-\infty$ with probability at least $\alpha,$
even though $b(y,\alpha)$ is ancillary to $\theta$, it still satisfies the condition in \Cref{eqn:bya_bound} for every $\theta$ and $\alpha \in [0, 1]$.
However, consider two possible definitions of the c-value,
$$
c^+(y) := \sup_{\alpha \in [0,1]} \{\alpha | b(y,\alpha) \ge 0\}
\text{ vs.\ }c^-(y) := \inf_{\alpha \in[0,1]} \{\alpha | b(y, \alpha) \le 0\},
$$
where $c^-(y) = c(y)$ is the definition we have chosen in \Cref{sec:c_values}.
Note that $c^-(y) \le c^+(y),$ and that if $b(y, \alpha)$ is continuous and strictly decreasing in $\alpha$ for every $y,$ then $c^-(y)=c^+(y).$
In this almost surely discontinuous case, however, we have that $c^+(y)\overset{a.s.}{=}1.$ and $c^-(y) \overset{a.s.}{=}0.$
Since estimators exist for which $W(\theta, y)<0$ with positive probability, the guarantees of \Cref{thm:c_values,thm:theta_tilde_guarantee} are not met by $c^+(y).$

In the present paper, $c^-(y) = c^+(y)$ for all bounds considered.
Our preference for defining the c-value as $c^-(y)$ derives from simplicity; we may disregard edge cases like the one above, which would complicate our proofs.
However for the reason described in this section, we emphasize that using $c^-(y)$ rather than $c^+(y)$ may have practical implications when these quantities differ.
\section{Additional details related to \Cref{sec:simpler_cases}}\label{sec:simpler_cases_supp}

\subsection{Distribution of win term}\label{sec:nc_chi_dist_in_simple_case}
We here provide a derivation of the distributional form of $2\epsilon^\top Gy$ given in \Cref{sec:ls_bound_construction}.
In \Cref{sec:ls_bound_construction} we found that 
$$
2 \epsilon^\top Gy \sim \frac{2}{1+\tau^2}\left[\chi^2_{N-1}(\frac{1}{4} \|P_1^\perp \theta \|^2)
-\frac{1}{4}\|P_1^\perp\theta\|^2 \right],
$$
where $\chi^2_{N-1}(\lambda)$ denotes the non-central chi-squared distribution with $N-1$ degrees of freedom and 
non-centrality parameter $\lambda$.

Recall that $Gy=(1 + \tau^2)\inv P_1^\perp(\theta + \epsilon).$
As such we can rewrite 
\(
2\epsilon^\top Gy &= \frac{2}{1+\tau^2} \left[\epsilon^\top P_1^\perp \epsilon + \epsilon^\top P_1^\perp \theta\right] \\
                  &= \frac{2}{1+\tau^2} \left[(P_1^\perp\epsilon)^\top (P_1^\perp \epsilon) +  (P_1^\perp\epsilon)^\top (P_1^\perp\theta)\right] \\
                  &\text{// since } P_1^\perp = P_1^\perp P_1^\perp \\
                  &= \frac{2}{1+\tau^2} \left[\| P_1^\perp\epsilon + \frac{1}{2}P_1^\perp \theta\|^2 - \frac{1}{4} \|P_1^\perp \theta\|^2\right] \\
                  &\text{// by completing the square} \\
&= \frac{2}{1+\tau^2} \left[\chi^2_{N-1}(\frac{1}{4}\|P_1^\perp \theta\|^2) - \frac{1}{4} \|P_1^\perp \theta\|^2\right],
\)
as desired, where in the last line the degrees of freedom parameter is $N-1$ because $P_1^\perp$ projects into an $N-1$ dimensional subspace of $\R^N.$

\subsection{Proof of \Cref{thm:lindley}}
We here provide a proof of \Cref{thm:lindley}.
\begin{proof}
The proof amounts to showing that $b(\cdot, \cdot)$ achieves at least nominal coverage, i.e.\ for any $\theta$ and $\alpha \in [0,1]$, $\pr\left[W(y, \theta) \ge b(y,\alpha)\right]\ge \alpha$.
By construction, $W(\theta, y) \ge b(y, \alpha)$ may be violated only if either (A) $\|P_1^\perp \theta\|^2 \not \in [0, U(y, \frac{1-\alpha}{2})]$ or (B) $W(\theta, y) < \frac{2}{1+\tau^2} F^{-1}_{N-1} (\frac{1-\alpha}{2};\frac{\| P_1^\perp \theta\|^2}{4})
- \frac{\|P_1^\perp \theta\|^2 }{2(1+\tau^2)} - \frac{\|P_1^\perp y \|^2 }{(1+\tau^2)^2}.$
Noticing that $\|P_1^\perp y\|^2 \sim \chi^2_{N-1}(\|P_1^\perp \theta\|^2),$ 
we can recognize $[0, U(\frac{1-\alpha}{2})]$ as valid confidence interval for $\|P_1^\perp \theta\|^2$ and see that (A) occurs with probability at most $\frac{1-\alpha}{2}.$
Next, comparing to \Cref{eqn:nc_chi_bound}, we see that (B) represents $2\epsilon^\top Gy$ falling below its $\frac{1-\alpha}{2}$ quantile and thus occurs with probability at most $\frac{1-\alpha}{2}$.
Therefore the union bound guarantees that $b(y,\alpha)$ obtains at least nominal coverage.
\end{proof}

\subsection{Why an \emph{upper} bound on $\|P_1^\perp \theta\|^2$? }\label{sec:lindley_why_upper}
We here provide justification for the use of a high-confidence upper bound on $\|P_1^\perp \theta\|^2$ in \Cref{bound:lindley_and_smith}.
Recall that \Cref{eqn:nc_chi_bound} provides a lower bound on $W(\theta, y)$ if we can control $\|P_1^\perp \theta \|^2.$
However, it is not immediately obvious what sort of control on $\|P_1^\perp \theta \|^2$ will yield the tightest bound;
should we have derived a two-sided interval or a lower bound instead of an upper bound?
We answer this question by appealing to a normal approximation of the non-central $\chi^2$ for intuition.
This approximation will be close when the degrees of freedom parameter is large.
Specifically, by replacing the non-central $\chi^2$ quantile with that of a normal with matched first and second moments we may approximate the lower bound as
\[\label{eqn:nc_chi_bound_normal_approx}
    W(\theta, y) &\overset{\sim}{\ge}
\frac{2}{1+\tau^2}\left[N-1 -(\| P_1^\perp \theta \|^2 + 2N-2)^{\frac{1}{2}}z_{\alpha}\right] - \frac{\|P_1^\perp y\|^2}{(1+\tau^2)^2},
\]
where $z_\alpha$ is the $\alpha$ quantile of the standard normal.

\Cref{eqn:nc_chi_bound_normal_approx} is monotone decreasing in $\| P_1^\perp \theta\|^2$ for any $\alpha>\frac{1}{2}$.
As such, we can expect this quantile to be smallest for large values of $\| P_1^\perp \theta\|^2,$ 
and for this reason seek to find a high-confidence upper bound on $\|P_1^\perp \theta\|^2$.
Indeed, in agreement with \Cref{eqn:nc_chi_bound_normal_approx} we have found empirically that the infimum in \Cref{eqn:LS_bya} is always achieved at this upper bound,
and conjecture that this is true in general.

\subsection{Shrinking towards an arbitrary subspace}\label{sec:morris_ext}
We now show how the approach developed in \Cref{sec:simpler_cases} immediately extends to a broader class of models in the spirit of those considered by \citet{morris1983parametric}.
In particular, let $\theta$ again be an unknown $N$-vector and $X\in \R^{N\times D}$ be a design matrix where for each $n$,
$X_n$ is a $D$-vector of covariates associated with $\theta_n$.
If we believe that the parameters can be roughly described as scattered around a linear function of these covariates with variance $\tau^2$,
we might consider trying to improve our estimates by estimating the linear dependence and interpolating between the sample estimate 
and the associated linear approximation.
Following \citet{morris1983parametric}, we obtain this type of shrinkage with the estimate
$$
\theta^*(y) := \frac{y + \tau^{-2}X(X^\top X)\inv X^\top y}{1+\tau^{-2}},
$$
which is the posterior mean of the Bayesian model that assumes for each $n$, $\theta_n \sim \mathcal{N}(X_n^\top \beta, \tau^2)$ a priori.
Here $\beta$ is an unknown $D$-vector of coefficients that is given an improper uniform prior.

For this setting, we propose the following bound.
\begin{bound}[Normal Means: Flexible shrinkage estimate vs.\ MLE]\label{bound:morris}
Observe $y = \theta+ \epsilon$ with $\epsilon \sim \mathcal{N}(0, I_N)$ and consider estimates
$$
\hat \theta(y) =y \text{\ \ and \ \ } \theta^*(y) := \frac{y + \tau^{-2}X(X^\top X)\inv X^\top y}{1+\tau^{-2}},
$$
where $\tau$ is a scalar and $X$ is an $N$ by $D$ matrix of covariates.
We propose 
\begin{equation}\label{eqn:morris_b_bound}
b(y, \alpha) = \inf_{\lambda \in [0, U(y, \frac{1-\alpha}{2})]} 
\frac{2}{1+\tau^2} F^{-1}_{N-D} \left(\frac{1-\alpha}{2}, \frac{\lambda}{4}\right)
- \frac{\lambda}{2(1+\tau^2)}- \frac{\|P_X^\perp y\|^2}{(1+\tau^2)^2}
\end{equation}
as a high-probability lower bound on the win.
In this expression, $F^{-1}_{N-D} (1-\alpha, \lambda)$ denotes the inverse cumulative distribution function of the non-central $\chi^2$ with $N-D$ degrees of freedom and non-centrality parameter $\lambda$ evaluated at $1-\alpha.$
$P_X^\perp:=I_N - X(X^\top X)\inv X^\top $ is the projection onto the subspace orthogonal to the column-space of $X.$
\begin{equation}
    U(y, 1-\alpha) := \inf_{\delta>0}\left\{ \delta \Big|
\| P_X^\perp y\|^2 \le F^{-1}_{N-D}(1-\alpha,  \delta)
\right\}
\end{equation}
is a high-confidence upper bound on $\|P_X^\perp \theta\|^2$.
\end{bound}

This bound is identical to \Cref{bound:lindley_and_smith} except that it projects to a different subspace,
and loses $D$ degrees of freedom in the $\chi^2$ random variables, rather than $1$.
Indeed, this is a strict generalization, as we obtain our earlier example when we take $X=\ones{N}$.
\Cref{bound:morris} is also computable (for the same reasons discussed in \Cref{remark:bound_computability}) and valid, as we see in the next proposition.

\begin{prop}\label{prop:morris}
\Cref{eqn:morris_b_bound} in \Cref{bound:morris} satisfies the conditions of \Cref{thm:c_values}.
In particular, for any $\theta$ and $\alpha \in [0,1]$, $\pr_\theta\left[W(y, \theta) \ge b(y,\alpha)\right]\ge \alpha$.
\end{prop}
\begin{proof}
\Cref{prop:morris} follows from an argument very closely analogous to the proof of \Cref{thm:lindley}.
We first rewrite $\theta^*(y)$ as $\theta^*(y) = y - Gy$ for $G:=(1+\tau^2)\inv P_X^\perp$.
\Cref{eqn:win_form} then holds exactly as before (i.e.\ $W(\theta, y)= 2\epsilon^\top Gy - \|Gy\|^2$).
The two terms are treated as in \Cref{thm:lindley}; the only differences are that the norm under consideration is $\|P_X^\perp \theta\|$
rather than $\| P_1^\perp \theta\|$, and the change in degrees of freedom from $N-1$ to $N-D$.
\end{proof}

\begin{figure}[H]
    \centering
    \includegraphics[width=0.7\textwidth]{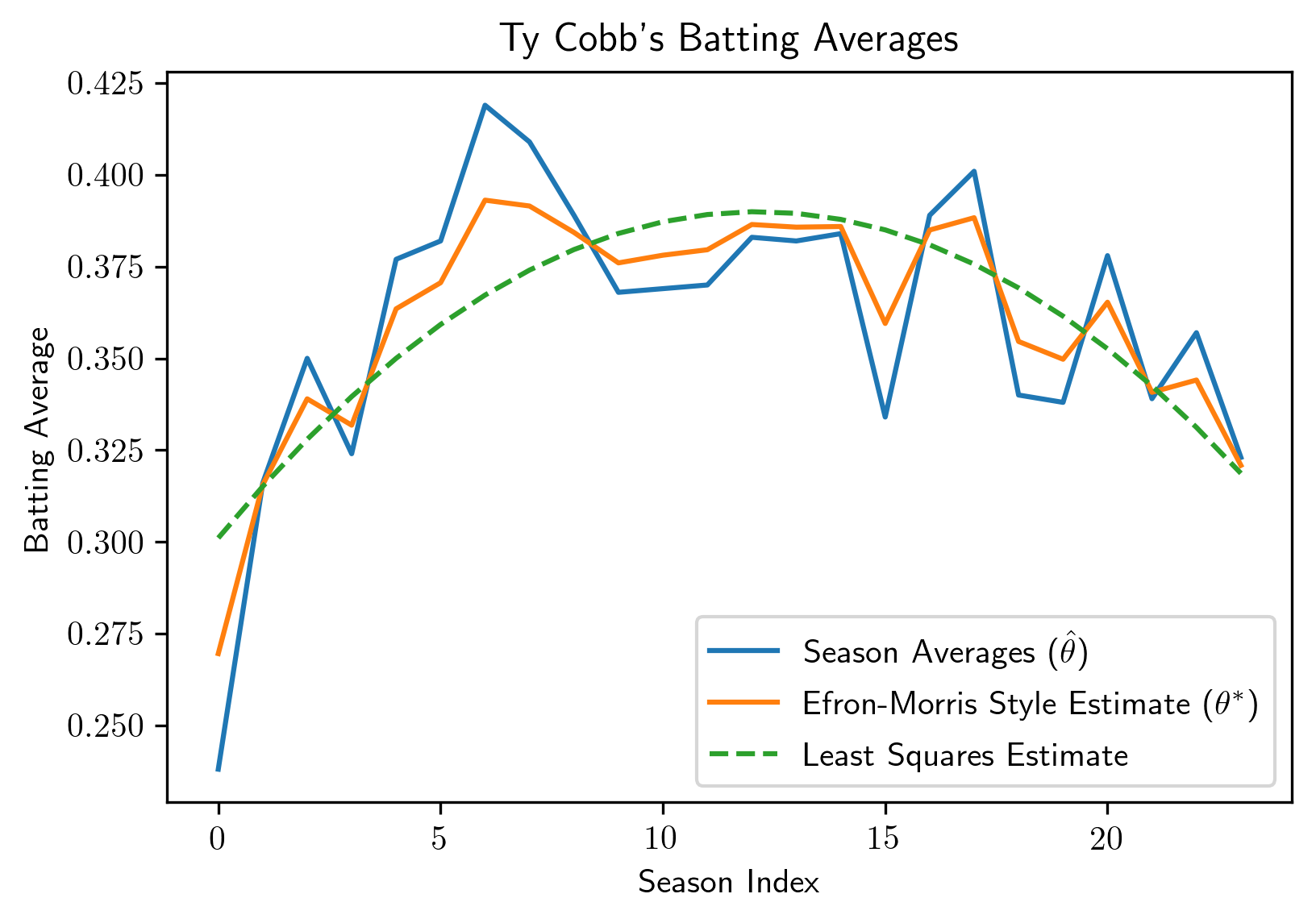}
\caption{
The estimate shrinking towards a quadratic fit provides a significant improvement ($c=0.953$).
The noise and prior standard deviations were set as $\sigma=0.025$ and $\tau=0.025$, respectively.}
    \label{fig:ty_cobb}
\end{figure}

\Cref{fig:ty_cobb} demonstrates an application to Ty Cobb's season batting averages, an example adapted from \citet{morris1983parametric}.
In this analysis, our approach indicates that we should be highly confident ($c=0.953$) that the alternative estimate, which shrinks the observations towards a quadratic fit of the data, outperforms the MLE .
While \citet{morris1983parametric} provides an argument for estimators of this style based on risk,
the present analysis goes a step further by providing a measure of confidence that the estimator improves on this particular dataset.
Even though the risk of the estimator $\theta^*(\cdot)$ may be greater than that of $\hat \theta(\cdot)$ for many possible $\theta$,
this analysis supports the conclusion that for the true unknown $\theta$ and observed $y$, $\theta^*(y)$ is superior.

\section{Affine estimators supplementary information}\label{sec:affine_supp}
\subsection{Step by step derivation of \Cref{eqn:affine_win_decomp} }
The win of using $\theta^*(y)$ in place of $\hat \theta(y)$ may be expressed as
\begin{align}\label{eqn:affine_win_reformulation}
\begin{split}
    W(\theta, y) &= \|\hat \theta(y) -\theta \|^2 - \|\theta^*(y) - \theta\|^2 \\
                 &= \left(\|\hat \theta(y)\|^2 + \|\theta\|^2 -2\theta^\top\hat \theta(y)\right) - \left(\|\theta^*(y)\|^2 + \|\theta\|^2 - 2\theta^\top\theta^*(y)\right) \\
                 &= -2\theta^\top G(y) + \left(\|\hat \theta(y)\|^2 - \|\theta^*(y) \|^2\right)  \\
                 &\text{ // where }G(y):=\hat \theta(y) - \theta^*(y) \\ 
                 &= 2\epsilon^\top G(y) - 2y^\top G(y) + \left(\|\hat \theta(y)\|^2 - \|\theta^*(y) \|^2\right) \\
                 &= 2\epsilon^\top G(y) + \left(\|\hat \theta(y) - y \|^2 - \|\theta^*(y) - y \|^2\right).
\end{split}
\end{align}

\subsection{Derivation of \Cref{eqn:affine_mean_and_var} }
Observe that
\(
    \E[ \epsilon^\top G(y) ] &= \E[ \epsilon^\top G(\theta) + \epsilon^\top (A-C) \epsilon  ]  \\
                             &= \E[ \epsilon]^\top G(\theta) + \E[\trace[(A-C) \epsilon \epsilon^\top  ] ] \\
                             &= \trace[ (A-C) \Sigma ] 
\)
and 
\(
    \Var[ \epsilon^\top G(y) ] &= \Var[ \epsilon^\top G(\theta)] + \Var[ \epsilon^\top (A-C) \epsilon  ]  \\
                               &\text{ // since $\epsilon^\top G(\theta)$ and $\epsilon^\top (A-C) \epsilon$ are uncorrelated}\\
                               &= (G(\theta))^\top \Sigma (G(\theta)) + 2\trace[\frac{A+A^\top - C-C^\top}{2}\Sigma\frac{A+A^\top - C - C^\top}{2}\Sigma]\\
                               &= \|G(\theta)\|_\Sigma^2 + \frac{1}{2}\trace[((A+A^\top - C - C^\top)\Sigma)^2] \\
                               &= \|G(\theta)\|_\Sigma^2 + \frac{1}{2}\|\Sigma\msqrt(A+A^\top - C - C^\top)\Sigma\msqrt\|_F^2,
\)
where $\|\cdot\|_\Sigma$ and $\| \cdot \|_F$ denote the $\Sigma$ quadratic norm and Frobenius norm, respectively.
The third line of the derivation above obtains from recognizing $\Var[\epsilon^\top (A-C)\epsilon]$ as a quadratic form \citep[Chapter 2]{mathai1992quadratic}.

\subsection{Derivations of \Cref{eqn:affine_mean_Gy,eqn:affine_var_ub} }\label{sec:gaussian_approx_of_Gy_norm}
\Cref{eqn:affine_mean_Gy,eqn:affine_var_ub} characterize the dependence of the distribution of $\|G(y)\|_\Sigma^2$ on $\|G(\theta)\|_\Sigma^2 $ through its mean and variance.
Recognizing $\|G(y)\|_\Sigma^2$ as a quadratic form \citep[Chapter 2]{mathai1992quadratic}, with $G(y) \sim \mathcal{N}\left(G(\theta), (A-C)\Sigma (A-C)^\top \right)$, 
we find its mean as
\(
    \E[\|G(y)\|_\Sigma^2] &= G(\theta)^\top \Sigma G(\theta) + \trace[\Sigma ((A-C)\Sigma (A-C)^\top)]  \\
                          &= \|G(\theta) \|_\Sigma^2 + \trace[\Sigma\msqrt (A-C)\Sigma (A-C)^\top \Sigma\msqrt]  \\
                          &= \|G(\theta) \|_\Sigma^2 + \|\Sigma\msqrt (A-C) \Sigma\msqrt \|_F^2.
\)
For the variance, we similarly rely on the known variance of a quadratic form.
Starting from that expression, we upper bound the variance as
\begin{align}\label{eqn:affine_norm_upper_bound_derivation}
\begin{split}
    \Var[\|G(y)\|_\Sigma^2] &= 2\trace\left[\Sigma \left((A-C) \Sigma (A-C)^\top\right) \Sigma \left((A-C)\Sigma (A-C)^\top\right)\right] +  \\
                            &\,\,\,\, 4 G(\theta)^\top \Sigma \left((A-C)\Sigma (A-C)^\top\right) \Sigma G(\theta)\\
    &= 2\|\Sigma\msqrt (A-C) \Sigma (A-C)^\top \Sigma\msqrt\|_F^2
    + 4 \|\left(\Sigma^{\frac{1}{2}} (A-C)^\top \Sigma^{\frac{1}{2}}\right)\Sigma^{\frac{1}{2}} G(\theta)\|_2^2\\
    &\le 2\|\Sigma\msqrt (A-C) \Sigma (A-C)^\top \Sigma\msqrt\|_F^2
    + 4 \opnorm{\Sigma^{\frac{1}{2}} (A-C) \Sigma^{\frac{1}{2}}}^2 \|G(\theta)\|_\Sigma^2,
\end{split}
\end{align}
where $\opnorm{\cdot}$ denotes the $L2$ operator norm.
\subsection{The Berry--Esseen bound: \Cref{thm:berry_esseen}}\label{sec:berry_esseen_supp}

We here prove \Cref{thm:berry_esseen}, a non-asymptotic upper bound on the error introduced by the two Gaussian approximations in \cref{approxbound:affine}.
We begin by restating key notation for convenience.
We then state a more general variant of the bound that removes the restriction that the operators $A$ and $C$ be symmetric,
and we show how it reduces to the simpler quantity stated in \Cref{thm:berry_esseen}.
Finally, we present a proof of the theorem as well as several supporting lemmas.

\paragraph{Notation and statement of the theorem its more general form.}
Recall that we are concerned with the coverage of \Cref{approxbound:affine}
\(
    b(y,\alpha) &=  \|\hat \theta - y\|^2 - \| \theta^* - y\|^2 + 2\trace[ (A-C) \Sigma ] + { }\\
    &2 z_{\frac{1-\alpha}{2}}\sqrt{
        U(\|G(y)\|_\Sigma^2, \frac{1-\alpha}{2})  
        + \frac{1}{2}\|\Sigma\msqrt(A+A^\top - C - C^\top)\Sigma\msqrt\|_F^2
}.
\)
In this equation,
$G(y) := (A-C)y + (k-\ell)$,
$z_\alpha$ denotes the $\alpha$-quantile of the standard normal, and 
\(
U\big( \|G(y)\|_\Sigma^2, &\frac{1-\alpha}{2}\big) = \inf_{\delta>0} \bigg\{ \delta \, \bigg|\, 
    \|G(y)\|_\Sigma^2 \le 
    (\delta + \|\Sigma\msqrt (A-C)\Sigma\msqrt\|_F^2) +  { }\\
    &z_{\frac{1-\alpha}{2}}\sqrt{
        2\|\Sigma\msqrt (A-C) \Sigma (A-C)^\top \Sigma\msqrt\|_F^2
        + 4 \|\Sigma^{\frac{1}{2}} (A-C) \Sigma^{\frac{1}{2}}\|_{\text{OP}}^2\delta 
    } 
\bigg\}
\)
is a high-confidence upper bound on $\|G(\theta)\|_\Sigma^2.$

For convenience, we introduce 
\[\label{eqn:be_approx_inv_cdf}
\tilde F\inv(\|G(\theta)\|_\Sigma^2, \alpha) :=  2\trace[ (A-C) \Sigma ] + 
    2 z_{\alpha}\sqrt{
        \|G(\theta)\|_\Sigma^2
        + \frac{1}{2}\|\Sigma\msqrt(A+A^\top - C - C^\top)\Sigma\msqrt\|_F^2
    },
\]
to denote the inverse CDF of our normal approximation to the distribution of $2\epsilon^\top G(y)$
evaluated at $\alpha.$
As such, we may write 
$$
b(y, \alpha) = \|\hat \theta - y\|^2 - \| \theta^* - y\|^2 +
\tilde F\inv\left(U(\|G(y)\|_\Sigma^2, \frac{1-\alpha}{2}), \frac{1-\alpha}{2}\right).
$$

Finally, recall that to prove the theorem we desire to show
$$
\pr_\theta\left[
    W(\theta, y) \ge b(y, \alpha)
\right] \ge \alpha  - \frac{10\sqrt{2}}{\sqrt{N}}C_1\cdot\kappa(\Sigma\msqrt(A-C)\Sigma\msqrt)^2
$$
for any $\theta$ and $\alpha \in [0, 1],$
where $C_1<1.88$ is a universal constant,
in the case when both $A$ and $C$ are symmetric.
We accomplish this by first proving a more general bound holds even in the non-symmetric case,
\[\label{eqn:more_general_berry_bound}
\pr_\theta\left[
    W(\theta, y) \ge b(y, \alpha)
\right] \ge \alpha  - \frac{5\sqrt{2}}{\sqrt{N}}C_1\left[\kappa(\Sigma\msqrt(A-C)\Sigma\msqrt)^2 + \kappa(\Sigma\msqrt(A+A^\top-C - C^\top)\Sigma\msqrt)\right].
\]
The special case obtains by replacing $A^\top$ and $C^\top$ with $A$ and $C,$ respectively,
and noting that $\kappa(M)^2 \ge \kappa(M)$ for any matrix, $M.$

A key tool in this proof is the classic result of \citet{berry1941accuracy},
which we restate below.
\begin{theorem}[Berry, 1941, Theorem 1]\label{thm:berry_esseen_restatement}
Let $X_1, X_2, \dots, X_N$ be random variables.
For each $n\in \{1, 2, \dots, N\}$, let $\sigma_n^2$ and $\rho_n$ denote the variance and third central moment of $X_n,$ respectively.
Define $\lambda_n := \frac{\rho_n}{\sigma_n^2}$ if $\sigma_n^2>0$ and $\lambda_n=0$ otherwise.
Define $\sigma^2 := \sum_{n=1}^N \sigma_n^2$ and 
$X := N\inv \sum_{n=1}^N X_n$.
Then 
$$
\sup_x  \left|F_X(x) -\Phi\left(\frac{x-\E[X]}{\sigma}\right) \right| < C_1 \frac{\max_n \lambda_n}{\sigma},
$$
where $C_1\le 1.88$ is a universal constant and $F_X(\cdot)$ is the cumulative distribution function of $X$.
\end{theorem}

\paragraph{Proof of \Cref{thm:berry_esseen}}
The desired bound may be stated equivalently as, for any $\alpha \in [0,1],$
\[\label{eqn:be_bound_restatement}
\pr_\theta\left[
    W(\theta, y) < b(y, \alpha)
\right] < (1-\alpha)  + \frac{5\sqrt{2}}{\sqrt{N}}C_1\left[\kappa(\Sigma\msqrt(A-C)\Sigma\msqrt)^2 + \kappa(\Sigma\msqrt(A+A^\top -C - C^\top)\Sigma\msqrt)\right].
\]
We first rewrite the condition $W(\theta, y) < b(y, \alpha)$ as 
$2\epsilon^\top G(y)  < \tilde F\inv\left(U(\|G(y)\|_\Sigma^2, \frac{1-\alpha}{2}), \frac{1-\alpha}{2}\right)$ (recall \Cref{eqn:affine_win_reformulation}).
Since $\tilde F\inv$ is monotonically decreasing in its first argument, this condition may occur only if either 
$2\epsilon^\top G(y)  < \tilde F\inv\left(\|G(\theta)\|_\Sigma^2, \frac{1-\alpha}{2}\right)$
or 
$\|G(\theta)\|_\Sigma^2 > U(\|G(y)\|_\Sigma^2, \frac{1-\alpha}{2}).$

Therefore, by the union bound, we have that 
\begin{align}\label{eqn:be_union_bound}
\begin{split}
\pr_\theta\left[W(\theta, y) < b(y, \alpha)\right] &< \pr_\theta\left[
2\epsilon^\top G(y)  < \tilde F\inv \left(\|G(\theta)\|_\Sigma^2, \frac{1-\alpha}{2}\right)
\right] \\
&+\pr_\theta\left[ 
\|G(\theta)\|_\Sigma^2 > U(\|G(y)\|_\Sigma^2, \frac{1-\alpha}{2})
\right].
\end{split}
\end{align}

\Cref{lemma:win_quantile_lb,lemma:affine_norm_ub} provide that
$\pr_\theta\left[
2\epsilon^\top G(y)  < \tilde F\inv\left(\|G(\theta)\|_\Sigma^2, \frac{1-\alpha}{2}\right)
\right] < \frac{1-\alpha}{2} + \frac{5\sqrt{2}}{\sqrt{N}}C_1\kappa(\Sigma\msqrt(A+A^\top-C -C^\top)\Sigma\msqrt)$
and
$\pr_\theta\left[ 
\|G(\theta)\|_\Sigma^2 > U(\|G(y)\|_\Sigma^2, \frac{1-\alpha}{2})
\right]< \frac{1-\alpha}{2} + \frac{5\sqrt{2}}{\sqrt{N}}C_1\kappa(\Sigma\msqrt(A-C)\Sigma\msqrt)^2,$
respectively.
Substituting these two bounds into \Cref{eqn:be_union_bound}
we obtain \Cref{eqn:be_bound_restatement} as desired.

\begin{lemma}\label{lemma:win_quantile_lb}
Let $y=\theta + \epsilon$ be a random $N$-vector with $\epsilon \sim \mathcal{N}(0, \Sigma).$
Let $\tilde F\inv$ be the normal approximation to the inverse CDF of $2\epsilon^\top G(y)$ in \Cref{eqn:be_approx_inv_cdf}.
Then for any $\alpha \in [0, 1],$
$$\pr_\theta\left[
2\epsilon^\top G(y)  < \tilde F\inv\left(\|G(\theta)\|_\Sigma^2, \alpha \right)
\right] < \alpha + \frac{5\sqrt{2}}{\sqrt{N}}C_1\kappa(\Sigma\msqrt(A + A^\top -C - C^\top)\Sigma\msqrt).$$
\end{lemma}
\begin{proof}
Note first that for any $\alpha$ we may rewrite
\(\pr_\theta\left[
2\epsilon^\top G(y)  < \tilde F\inv\left(\|G(\theta)\|_\Sigma^2,\alpha \right)
\right] &= F\left[  \tilde F\inv\left(\|G(\theta)\|_\Sigma^2, \alpha\right)\right]\\
        &= \alpha + \left\{ F\left[  \tilde F\inv\left(\|G(\theta)\|_\Sigma^2, \alpha\right)\right]- 
\tilde F\left[  \tilde F\inv\left(\|G(\theta)\|_\Sigma^2, \alpha\right)\right]\right\},
\)
where $F$ and $\tilde F$ are the exact and approximate CDFs of $2\epsilon^\top G(y),$ respectively.
Recalling that the normal approximation comes from matching moments to $2\epsilon^\top G(y)$, we have that for any $v,$
$\tilde F(v) = \Phi(\frac{v - \E[2\epsilon^\top G(y)]}{\sqrt{\Var[2\epsilon^\top G(y)]}}).$
Therefore, it will suffice to obtain that for every $v,$
$$ \left|\tilde F(v)- F( v )\right| = \left|F(v) - \Phi\left(\frac{v - \E[2\epsilon^\top G(y)]}{\sqrt{\Var[2\epsilon^\top G(y)]}}\right)\right|
\le \frac{5\sqrt{2}}{\sqrt{N}}C_1\kappa(\Sigma\msqrt(A+ A^\top -C - C^\top)\Sigma\msqrt).$$

We will obtain this result by writing $2\epsilon^\top G(y)$ a sum of independent random variables and using a Berry--Esseen Theorem (\Cref{thm:berry_esseen_restatement}) to bound the error of this normal approximation.

\Cref{lemma:as_indep_nc_chis} allows us to write $2\epsilon^\top G(y)=2\epsilon^\top (A -C) \epsilon + 2\left[(A- C)\theta + (k-\ell)\right]^\top \epsilon$ as a shifted sum of $N$ differently-scaled, independent non-central $\chi^2$ random variables.
We denote these $N$ random variables by $X_1, X_2, \dots, X_N.$
\Cref{lemma:as_indep_nc_chis} additionally tells us that the scaling parameters of these non-central $\chi^2$ random variables will be 
the eigenvalues of $\Sigma\msqrt(A+A^\top - C^\top -C)\Sigma\msqrt,$ which we denote by
$\lambda_1\ge \lambda_2 \ge \dots \ge \lambda_N \ge 0.$

To use \Cref{thm:berry_esseen_restatement}
we require the ratios of the third to second central moments of these random variables, as well as the variance of the sum.
Specifically,
$$\sup_{v\in \R} \left| \Phi(\frac{v - \E[2\epsilon^\top G(y) ]}{\sqrt{\Var[2\epsilon^\top G(y)]}}) - F(v))\right| < C_1\frac{\max_{n} \frac{\rho(X_n)}{\Var[X_n]}}{\sqrt{\Var[2\epsilon^\top G(y)]}},$$
where for each index $n,\, \rho(X_n) := \E[(X_n-\E[X_n])^3]$ is the third central moment of $X_n,$
and $C_1<1.88$ is a universal constant.

Conveniently, as we show in \Cref{lemma:nc_chi_moments}, for each $n, \frac{\rho(X_n)}{\Var[X_n]} \le 10\lambda_n.$
Further, since $\sqrt{\Var[2\epsilon^\top G(y)]}>\sqrt{2\sum_{n=1}^N \lambda_n^2}>\sqrt{2N}\lambda_N$ 
(recall that \Cref{eqn:affine_mean_and_var} provides that
$\Var[2\epsilon^\top G(y)] =  4\|G(\theta)\|_\Sigma^2 + 2\|\Sigma\msqrt(A+A^\top - C - C^\top)\Sigma\msqrt\|_F^2$)
we may additionally see that
\(
\sup_{v\in \R} \left| \Phi\left(\frac{v - \E[2\epsilon^\top G(y) ]}{\sqrt{\Var[2\epsilon^\top G(y)]}}\right) - F(v))\right| &< 
C_1\frac{10}{\sqrt{2N}}\frac{\max_{n} \lambda_n}{\min_n \lambda_n} \\
&=C_1\frac{5\sqrt{2}}{\sqrt{N}}\kappa\left(\Sigma\msqrt (A + A^\top - C - C^\top) \Sigma\msqrt\right)
\)
where $\kappa(\cdot)$ denotes the condition number of its matrix argument,
as desired.
\end{proof}

\begin{lemma}\label{lemma:affine_norm_ub}
Let $y=\theta + \epsilon$ be a random $N$-vector with $\epsilon \sim \mathcal{N}(0, \Sigma).$
Let $U(\|G(y)\|_\Sigma^2, \alpha)$ be the approximate high-confidence upper bound on $\|G(\theta)\|_\Sigma^2$.
Then for any $\alpha \in [\frac{1}{2}, 1],$
$\pr_\theta\left[ 
\|G(\theta)\|_\Sigma^2 > U(\|G(y)\|_\Sigma^2, 1-\alpha)
\right]< 1-\alpha + \frac{5\sqrt{2}}{\sqrt{N}}C_1\kappa(\Sigma\msqrt(A-C)\Sigma\msqrt)^2.$
\end{lemma}
\begin{proof}
Our proof of the lemma follows roughly the same approach taken to prove \Cref{lemma:win_quantile_lb}.
First note that the condition that 
$\|G(\theta)\|_\Sigma^2 > U(\|G(y)\|_\Sigma^2, 1-\alpha)$ implies that 
\(
\|G(y)\|_\Sigma^2 &\le 
(\|G(\theta)\|_\Sigma^2 + \|\Sigma\msqrt (A-C)\Sigma\msqrt\|_F^2) +  { }\\
    &z_{1-\alpha}\sqrt{
        2\|\Sigma\msqrt (A-C) \Sigma (A-C)^\top \Sigma\msqrt\|_F^2
        + 4 \|\Sigma^{\frac{1}{2}} (A-C) \Sigma^{\frac{1}{2}}\|_{\text{OP}}^2\|G(\theta)\|_\Sigma^2
    }\\ 
    &\le \E[G(y)\|_\Sigma^2]+ z_{1-\alpha}\sqrt{\Var[G(y)]}
\)
for any $\alpha \in [\frac{1}{2}, 1],$ where the first line follows from the definition of $U(\|G(y)\|_\Sigma^2, 1-\alpha).$ 
The second line follows from the observations that (A) $z_{1-\alpha}<0$ and (B) the second term in the first line uses an upper bound on the variance of $\|G(y)\|_\Sigma^2$
(\Cref{eqn:affine_var_ub}).

We now proceed to upper bound the probability of the event in the display equation above.
First consider a normal approximation to the distribution of $\|G(y)\|_\Sigma$ with matched moments, and denote its inverse CDF by $F^{\dagger -1}(\theta, \alpha).$
We may then write the probability of the event above as 
\(
\pr\left[ \|G(y)\|_\Sigma^2 \le\E[G(y)\|_\Sigma^2]+ z_\alpha \sqrt{\Var[G(y)]}\right] 
&= F\left[ F^{\dagger -1}( \theta, \alpha)\right] \\
&= \alpha + \left\{ F\left[ \bar F\inv( \theta, \alpha)\right] -
F^\dagger \left[ F^{\dagger -1} (\theta, \alpha) \right]\right\},
\)
where $F(\cdot)$ and $F^\dagger(\cdot)$ denote the exact and approximate CDFs of $\|G(y)\|_\Sigma^2.$
It will suffice to show that for any $v,$
\(
|F(v) - F^\dagger(v)| \le \frac{5\sqrt{2}}{\sqrt{N}} \kappa(\Sigma\msqrt (A-C)\Sigma\msqrt)^2
.\)

As in \Cref{lemma:win_quantile_lb} we obtain this result through the Berry--Esseen theorem.
In this case, the variable of interest is $\|G(y)\|_\Sigma^2 = \epsilon^\top (A-C)^\top \Sigma (A-C) \epsilon + 2\epsilon^\top\left[(A-C)\theta + (k-\ell)\right].$
As in this previous lemma, we use \Cref{lemma:as_indep_nc_chis} to write this variable as a shifted sum of independent, scaled non-central $\chi^2$ random variables, this time with scaling parameters equal to the eigenvalues $\Sigma\msqrt (A-C)^\top \Sigma (A-C) \Sigma\msqrt.$ 
Recognizing that the eigenvalues of the matrix $M^\top M$ are the squares of the singular values of $M$ for any matrix $M,$
we obtain the desired result.
\end{proof}

\begin{lemma}\label{lemma:as_indep_nc_chis}
Let $X$ be a random $N$-vector distributed as
$X\sim 2\epsilon^\top A \epsilon + b^\top \epsilon$ where $A\in \R^{N\times N}, b\in\R^N,$
and $\epsilon \sim \mathcal{N}(0, \Sigma).$
Then $X$ is distributed as a shifted sum of differently scaled, independent non-central $\chi^2$ random variables.
In particular, if we let $U\diag(\lambda)U^\top$
be the eigen-decomposition of $\Sigma\msqrt(A+A^\top)\Sigma\msqrt,$
then we can write
$X\overset{d}{=}\sum_{n=1}^N Y_n - \frac{1}{4}\|\diag(\lambda)\inv U^\top \Sigma\msqrt b\|_2,$
where each 
$Y_n \overset{indep}{\sim} \lambda_n \chi^2_1(\frac{1}{2}\lambda_n\inv e_n^\top U^\top\Sigma\msqrt b),$
where $e_n$ is the $n^{th}$ basis vector.
\end{lemma}
\begin{proof}
The proof of the lemma proceeds through a long algebraic rearrangement.
In particular we rewrite $X$ as 
\(
    X &= 2\epsilon^\top A \epsilon + b^\top \epsilon \\
      &= \delta^\top \Sigma\msqrt (A+A^\top) \Sigma\msqrt \delta + b^\top \Sigma \msqrt \delta \\
    & \text{// defining }\delta := \Sigma^{-\frac{1}{2}}\epsilon \text{ so that }\delta \sim \mathcal{N}(0, I_N).\\
    &= \delta^\top U\diag(\lambda) U^\top \delta + b^\top \Sigma \msqrt U\diag(\lambda)^{-\frac{1}{2}}\diag(\lambda)\msqrt U^\top  \delta \\
    & \text{// Letting }U\diag(\lambda) U^\top := \Sigma\msqrt (A+A^\top) \Sigma\msqrt \text{ be an eigen-decomposition,} \\
    & \text{// with }U^\top U =I_N \text{ and }\lambda \in \R_+^N\\
    &\overset{d}{=} \delta^\top \diag(\lambda) \delta + b^\top \Sigma \msqrt U\diag(\lambda)^{-\frac{1}{2}} \diag(\lambda)\msqrt \delta \\
    &= \sum_{n=1}^N (\lambda_n^{\frac{1}{2}}\delta_n + \frac{1}{2}\lambda_n^{-\frac{1}{2}} e_n^\top U^\top \Sigma\msqrt b)^2 - \frac{1}{4} b^\top \Sigma \msqrt U\diag(\lambda)\inv U^\top \Sigma\msqrt b \\
    &\overset{d}{=} 
     \frac{- b^\top (A + A^\top)\inv b}{4} +
    \sum_{n=1}^N \lambda_n \chi^2_1(\frac{1}{2}\lambda_n\inv e_n^\top U^\top \Sigma\msqrt b),
\)
where each $e_n$ denotes the $n^{th}$ basis vector and each of the scaled non-central $\chi^2$ random variables in the last line are independent.
\end{proof}

\begin{lemma}\label{lemma:nc_chi_moments}
Consider a scaled non-central chi-squared random variable, $X \sim s\chi^2_1(\lambda)$, 
where $s$ and $\lambda$ are scaling and non-centrality parameters, respectively.
Denote the second and third central moments of $X$ by
$\sigma^2 = \Var[X]$ and
$\rho = \E\left[(X - \E[X])^3\right].$
Then
$\frac{\rho}{\sigma^2} \le 10s.$
\end{lemma}
\begin{proof}
Recall that the second and third central moments of the scaled non-central $\chi^2$ have known forms, $\sigma^2 = 2s^2(1+2\lambda)$ and $\rho = 8 s^3(1 + 3\lambda)$.
Therefore we may write
\(
\frac{\rho}{\sigma^2} &= \frac{8 s^3(1 + 3\lambda)}{2s^2(1+2\lambda)} \\
                      &\le 4s\left(\frac{1}{1} + \frac{3\lambda}{2\lambda}\right) \\
                      &= \frac{4\cdot 5}{2}s\\
                      &= 10s,
\)
as desired.
\end{proof}

\section{Empirical Bayes supplementary details}\label{sec:empirical_bayes_supp}

\subsection{Additional figure}
\begin{figure}[H]
\centering
\includegraphics[width=0.45\textwidth]{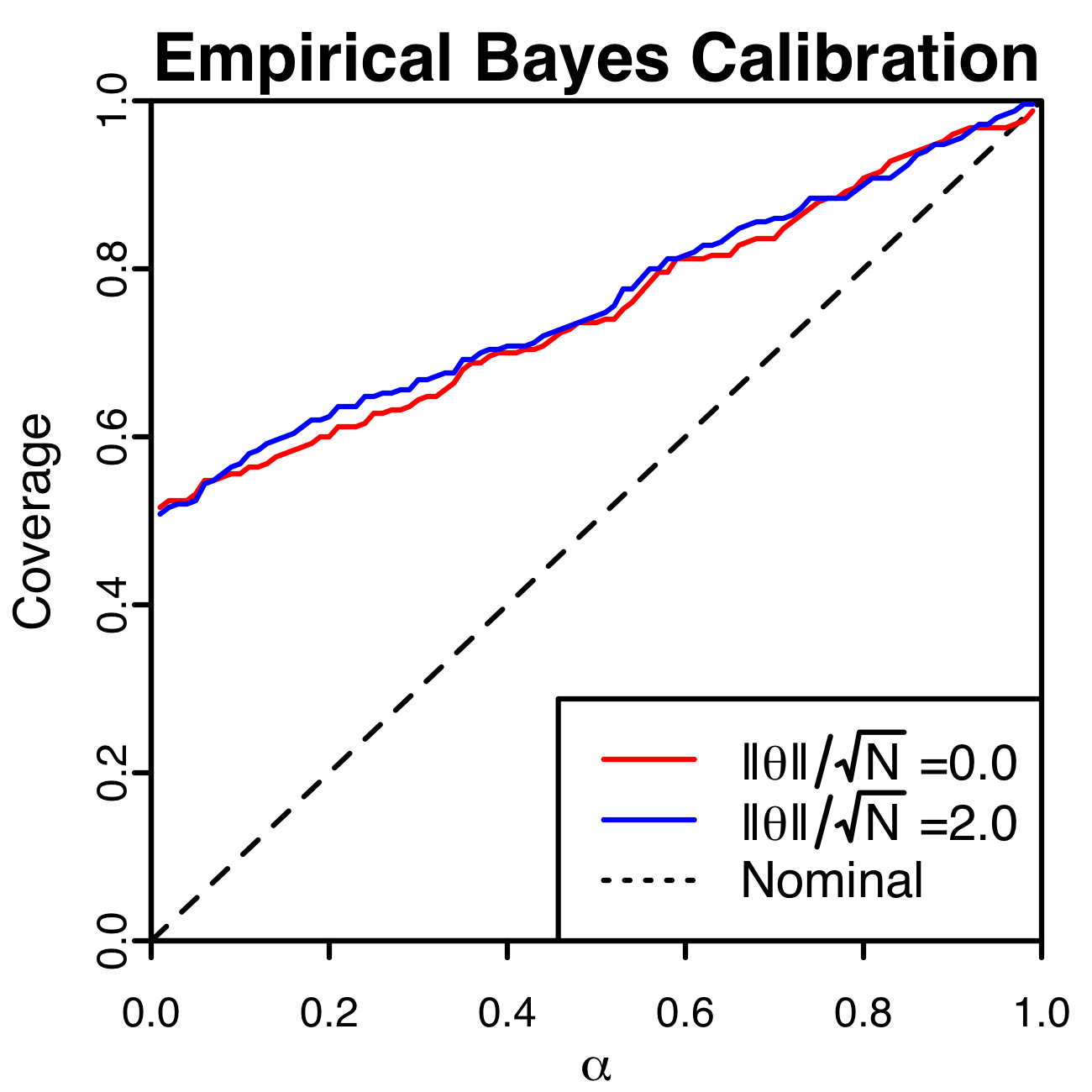}
\caption{
Calibration of approximate high-confidence bounds on the win of an empirical Bayes estimate over the MLE in simulation. 
Each series depicts calibration for a different choice of the parameter $\theta$ ($N=50$).
}
\label{fig:JS_calibration}
\end{figure}

\Cref{fig:JS_calibration} shows the calibration in the simulation experiment described in \Cref{sec:empirical_bayes}.

\subsection{Asymptotic coverage of the empirical Bayes estimate}
\Cref{thm:JS_asymptotic_coverage} shows that we can apply the machinery developed for Bayes rules with fixed priors 
to lower bound the win with at least the desired coverage asymptotically.
We here consider a scaling of win,
$$W_N(\Theta_N, Y_N) := \frac{1}{\sqrt{N}} \left[\|Y_N - \Theta_N \|^2 - \|\Theta^{*}_N(Y_N) - \Theta_N\|^2 \right].$$
We use a special case of \Cref{bound:morris} in \Cref{sec:morris_ext} with no covariates (i.e.\ $D=0$), and we treat the estimate $\hat \tau_N^2(Y_N)$ as if it were fixed rather than estimated from the data.
For each $N$, this bound is
$$b_N(Y_N, \alpha) := \frac{1}{\sqrt{N}} \inf_{\lambda \in [0, U(Y_N, \frac{1-\alpha}{2})]} 
\frac{2}{1+\hat \tau_N^2} F^{-1} \left[ \chi^2_{N}(\frac{\lambda}{4}), \frac{1-\alpha}{2} \right]
- \frac{\lambda}{2(1+\hat \tau_N^2)}- \frac{\|Y_N\|^2}{(1+\hat \tau_N^2)^2}$$
where
$F^{-1} \left[ \chi^2_{N}(\lambda), 1- \alpha \right]$ denotes the inverse cumulative distribution function of the non-central $\chi^2$ with $N$ degrees of freedom and non-centrality parameter $\lambda,$ evaluated at $1-\alpha$ and 
$U(Y_N, 1-\alpha) := \inf_{\delta\ge 0}\left\{ \delta \Big|
\| Y_N\|^2 \le 
F^{-1} \left[ \chi^2_{N}(\delta), 1-\alpha \right]
\right\}$
is a high-confidence upper bound on $\|\theta\|^2$.

For our theorem and its proof, a key quantity is, for each $N$, the sample second moment for the first $N$ parameters, which we denote by
$\tau_N^2 := N\inv \sum_{n=1}^N\theta_n^2.$
We emphasize, however, that while it may be convenient to describe $\tau_N^2$ as a sample moment, $\theta$ is fixed in \Cref{thm:JS_asymptotic_coverage} and throughout this analysis.

\paragraph{Proof of \Cref{thm:JS_asymptotic_coverage}.}
We prove the theorem by showing that for any $\alpha$, the gap between the win $W_N(\Theta_N, Y_N)$ and the bound $b_N(Y_N, \alpha)$ computed for the empirical Bayes estimate 
converges in distribution to the gap between the analogous win and bound computed for the same estimates but with prior variance fixed as $\tau^2 = \tau^2_N$.
We denote these latter quantities by $W^*_N(\Theta_N, Y_N)$ and $b^*_N(Y_N, \alpha),$
and note that since $\tau_N^2$ is fixed $\pr[W^*_N(\Theta_N, Y_N) \ge b^*(Y_N, \alpha)] \ge \alpha$ by construction (\Cref{prop:morris}).
For convenience, we denote $W_N(\Theta_N, Y_N)$ by $W_N,$ $b_N(Y_N, \alpha)$ by $b_N,$
$W_N^*(\Theta_N, Y_N)$ by $W_N^*,$ and $b_N^*(Y_N, \alpha)$ by $b^*_N.$

Observe that we can write
\(
W_N - b_N = \frac{W_N- b_N}{W^*_N -b^*_N }(W^*_N - b^*_N).
\)
By \Cref{lemma:normal_convergence}, $W^*_N - b^*_N$ is asymptotically Gaussian,
and by \Cref{lemma:convergence_of_w_b_ratio} $\frac{W_N- b_N}{W^*_N - b^*_N }\overset{p}{\rightarrow}1$.
As a result, the distribution of $W_N - b_N$ approaches the distribution of $W_N^*-b_N^*$ in supremum norm.
Since $b_N^*$ obtains the desired coverage by construction, the result follows.

\paragraph{Supporting lemmas.}
\begin{lemma}\label{lemma:variance_est_convergence}
If the sequence $\tau_N^2$ is bounded, then 
$\tau_N^2 - \hat \tau_N^2$ is $O_p(N\nsqrt),$ 
where $O_p(\cdot)$ denotes stochastic convergence in probability.
\end{lemma}
\begin{proof}
Note that for each $N$, $\|Y_N\|^2 \sim \chi^2_{N}(N \tau^2_N)$.
Therefore we have that $\E[\|Y_N\|^2] =N + N\tau^2_N$ and 
$\Var[\|Y_N\|^2] =  2(N + 2N\tau^2_N)$.
So, recalling that $\hat \tau^2_N :=\frac{\|Y_N\|^2}{N-2} -1 =\frac{\|Y_N\|^2 - (N-2)}{N-2}$ we may write
\(
\hat \tau_N^2 &= \frac{\|Y_N\|^2 - \E[\|Y_N\|^2]}{N-2}+\frac{(N +N \tau^2_N) - (N-2)}{N-2}\\
                 &= \frac{\|Y_N\|^2 - \E[\|Y_N\|^2]}{N}+\tau^2_N  + O(\frac{1}{N}).
\)
And so
\(
|\hat \tau_N^2 - \tau^2_N| &\le \Big|\frac{\|Y_N\|^2 - \E[\|Y_N\|^2]}{N}\Big| + O(\frac{1}{N})\\
                                 &=\left(\frac{\sqrt{2 + 4\tau_N^2}}{\sqrt{N}}\right) \Big|\frac{\|Y_N\|^2 - \E[\|Y_N\|^2]}{\sqrt{\Var[\|Y_N\|^2]}}\Big| + O(\frac{1}{N}).
\)
By Chebyshev's inequality, $\frac{\|Y_N\|^2 - \E[\|Y_N\|^2]}{\sqrt{\Var[\|Y_N\|^2]}}$ is bounded in probability and we can see that 
$|\hat \tau_N^2 - \tau^2_N| $ is $O_p(N\nsqrt)$.
\end{proof}

\begin{lemma}\label{lemma:convergence_of_w_b_ratio}
    Let $W^*_N$ and $b^*_N$ denote the win and its bound evaluated for $\tau^2=\tau_N^2,$ rather than the empirical Bayes estimate.
Then 
\(
\frac{W_N -b_N }{W^*_N -b^*_N } =  1 + \frac{\tau_N^2 - \hat \tau_N^2}{1 + \hat \tau_N^2}  = 1 + O_p(\frac{1}{\sqrt{N}}).
\)
\end{lemma}
\begin{proof}
Recall that we may decompose $W_N$ as 
$$
W_N(\Theta_N, Y_N) =\frac{1}{\sqrt{N}}\left[ \frac{2}{1+\hat\tau_N^2}\epsilon_N^\top Y_N - \frac{1}{(1+\hat \tau_N^2)^2}\|Y_N\|^2\right]
$$
and that our bound is 
$$
b_N(Y_N, \alpha)  = \frac{1}{\sqrt{N}} \left\{\inf_{\lambda \in [0, U(Y_N, \frac{1-\alpha}{2})]} 
\frac{2}{1+\hat \tau_N^2} F^{-1} \left[ \chi^2_{N}(\frac{\lambda}{4}), \frac{1-\alpha}{2} \right]
- \frac{\lambda}{2(1+\hat \tau_N^2)}- \frac{\|Y_N\|^2}{(1+\hat \tau_N^2)^2}\right\},
$$
where $U(Y_N,\alpha)$ does not depend on $\hat \tau^2_N$.

As such,
$$
W_N - b_N = \frac{2}{\sqrt{N}(1+\hat \tau_N^2)} \left\{\epsilon_N^\top Y_N  - \inf_{\lambda \in [0, U(Y_N, \frac{1-\alpha}{2})]} 
F^{-1} \left[ \chi^2_{N}(\frac{\lambda}{4}), \frac{1-\alpha}{2} \right]
+ \frac{\lambda}{4}\right\},
$$
and we can see that 
\(
\frac{W_N -b_N }{W^*_N -b^*_N } &=  \frac{1 + \tau_N^2}{1 + \hat \tau_N^2} \\
&=  1 + \frac{\tau_N^2 - \hat \tau_N^2}{1 + \hat \tau_N^2}.
\)
By \Cref{lemma:variance_est_convergence} the second term is $O_p(N\nsqrt)$, as desired.
\end{proof}

\begin{lemma}\label{lemma:normal_quantile_approximation}
    Let  $\lambda_1, \lambda_2, \dots$ be a sequence of reals satisfying, for each $N$, $N\inv \lambda_N< \kappa$ for some constant $\kappa$.
Let $F^{-1}_{\chi^2_N}(\lambda_N, \alpha)$ denote the inverse CDF of a non-central $\chi^2$ with $N$ degrees of freedom and non-centrality parameter $\lambda_N$. 
Then for any $\alpha\in(0,1)$,
$$
\frac{1}{\sqrt{N}}\left[ F^{-1}_{\chi^2_N}(\lambda_N, \alpha) - (N + \lambda_N)\right] = \sqrt{2 + 4\frac{\lambda_N}{N}} z_\alpha + O(\frac{1}{\sqrt{N}}),
$$
where $z_\alpha$ is the $\alpha$-quantile of the standard normal.
\end{lemma}
\begin{proof}
Note that a $\chi^2_N(\lambda_N)$ random variable is equal in distribution to a sum of $N$ i.i.d. $\chi^2_1(N\inv \lambda_N)$ random variables.
Let $\sigma^2_N := \Var[\chi^2_1(N\inv \lambda_N)] = 2+4N\inv \lambda_N$ and note that each $\sigma_N^2 \ge 2.$
Let $\rho_N:=8+24N\inv \lambda_N$ be third central moment of these variates and note that each $\rho_N \le 8 +24\kappa$.

Let $F_{\chi^2_N(\lambda_N)}(x)$ denote the CDF of a non-central $\chi^2$ random variable with $N$ degrees of freedom and non-centrality parameter $\lambda_N$ evaluated at $x$.
By the Berry--Esseen theorem \citep[Theorem 1]{berry1941accuracy}, for all $x$
\(
\left| F_{\chi^2_N(\lambda_N)}(x)-\Phi\left[ \frac{x - (N + \lambda_N)}{\sqrt{2N+ 4\lambda_N}}\right] \right|
    &\le \frac{C_1 \rho}{\sigma^3\sqrt{N}} \\
    &\le \frac{C_1 (8+ 24\kappa)}{2^\frac{3}{2} \sqrt{N}} \\
    &=O(\frac{1}{\sqrt{N}}),
\)
where $C_1 \le 1.88$ is a universal constant.
Since $\Phi(\cdot)$ is continuously differentiable and invertible, we obtain the same convergence rate for the inverse CDFs.
That is, for any $\alpha \in (0, 1),$
$$
\frac{ F^{-1}_{\chi^2_N}(\lambda_N, \alpha) - (N + \lambda_N)}{\sqrt{2N + 4\lambda_N}} - z_\alpha =  O(\frac{1}{\sqrt{N}}).
$$
Rescaling these terms by $N\nsqrt \sqrt{2N + 4\lambda_N}$ and rearranging, we find 
$$
\frac{1}{\sqrt{N}}\left[ F^{-1}_{\chi^2_N}(\lambda_N, \alpha) - (N + \lambda_N)\right]= \sqrt{2 + 4\frac{\lambda_N}{N}}z_\alpha +  O(\frac{1}{\sqrt{N}})
$$
as desired.
\end{proof}

\begin{lemma}\label{lemma:normal_convergence}
Let $b^*_N$ and $W^*_N$ again denote the win and bounds evaluated for the variance $\tau^2=\tau_N^2$ rather than the empirical Bayes estimate.
If the sequence $\tau_N^2$ is bounded, then 
$$
\frac{(W^*_N - b^*_N)- c_N}{d_N} \rightarrow \mathcal{N}(0, 1)
$$
for some sequences of constants $c_1, c_2, \dots$ and $d_1, d_2, \dots$.
\end{lemma}
\begin{proof}
Let $\kappa$ be such that for all $N$, $\tau_N^2 < \kappa$. 

Recall that we may write
\[\label{eqn:correct_win_bound_diff_distribution}
W^*_N - b^*_N = \frac{2}{\sqrt{N}(1+\tau_N^2)} \left\{\epsilon_N^\top Y_N  - \inf_{\lambda \in [0, U(Y_N, \frac{1-\alpha}{2})]} 
F^{-1} \left[ \chi^2_{N}(\frac{\lambda}{4}), \frac{1-\alpha}{2} \right]
+ \frac{\lambda}{4}\right\}.
\]

To prove the lemma, we build off of the normal approximation described in \Cref{sec:nc_chi_dist_in_simple_case}.
Note first that an application of Chebyshev's inequality provides that $N\inv U(Y_N, \frac{1-\alpha}{2}) - \tau_N^2$ is $O_p(N\nsqrt)$, so that $N\inv U(Y_N, \frac{1-\alpha}{2})<\kappa$ with probability approaching 1.
Next, by \Cref{lemma:normal_quantile_approximation},
\(
\frac{1}{\sqrt{N}} \left\{ F^{-1} \left[ \chi^2_{N}(\frac{\lambda_N}{4}), \frac{1-\alpha}{2} \right] - 
    \left[\frac{\lambda_N}{4}  + N\right] 
\right\} = \sqrt{2 + \frac{\lambda_N}{N}}z_{\frac{1-\alpha}{2}} + O(\frac{1}{\sqrt{N}}),
\)
for any sequence $\lambda_1, \lambda_2, \dots$ that satisfies, for each $N$, $N\inv \lambda_N < \kappa.$

Notably, since any sequence of $\lambda_N$'s achieving the infima in \Cref{eqn:correct_win_bound_diff_distribution} will satisfy this condition, 
we may substitute this expression in and rewrite $W^*_N -b^*_N$ as
\(
W^*_N - b^*_N &= \frac{2}{1+\tau_N} \left[
\frac{\epsilon_N^\top Y_N}{\sqrt{N}}  -  
\sqrt{N}\left\{\inf_{\lambda_N \in [0, U(Y_N, \frac{1-\alpha}{2})]}
F^{-1} \left[ \chi^2_{N}(\frac{\lambda_N}{4}), \frac{1-\alpha}{2} \right]
- \left[ \frac{\lambda_N}{4}+N\right]\right\} - \sqrt{N} \right] \\
&= \frac{2}{1+\tau_N} \left[
\frac{\epsilon_N^\top Y_N - N}{\sqrt{N}}  - \inf_{\lambda_N \in [0, U(Y_N, \frac{1-\alpha}{2})]}z_{\frac{1-\alpha}{2}} \sqrt{2 + \frac{\lambda_N}{N}} + O_p(\frac{1}{\sqrt{N}})
\right] \\
&= \frac{2}{1+\tau_N} \left[
\frac{\epsilon_N^\top Y_N - N}{\sqrt{N}}  -  z_{\frac{1-\alpha}{2}} \sqrt{2 + \frac{U(Y_N,\frac{1-\alpha}{2})}{N}} + O_p(\frac{1}{\sqrt{N}})
\right] \\
&= \frac{2}{1+\tau_N} \left[
\frac{\epsilon_N^\top Y_N - N}{\sqrt{N}}  -  z_{\frac{1-\alpha}{2}} \sqrt{2 + \tau_N^2} + O_p(\frac{1}{\sqrt{N}})
\right]\\
&\text{// Since } \tau_N^2 - \frac{U(Y_N, \frac{1-\alpha}{2})}{N}  \text{ is } O_p(\frac{1}{\sqrt{N}}).
\)

Finally, note that $\epsilon^\top Y_N$ is approximately normal with mean $N$ and variance $N(2 + \tau_N^2)$.
Furthermore, the distribution of this quantity approaches that of a normal at the same $O(N\nsqrt)$ rate in the supremum norm (one may make this precise with a Berry--Esseen bound).
This allows us to write
\(
W^*_N - b^*_N &\sim \frac{2}{1+\tau_N^2} \left[\sqrt{2 + \tau_N^2}x   -  \sqrt{2 + \tau_N^2} z_{\frac{1-\alpha}{2}}\right] + O_p(\frac{1}{\sqrt{N}}) \\
              &\sim \frac{2\sqrt{2 + \tau_N^2}}{1+\tau_N^2} (x   -  z_{\frac{1-\alpha}{2}})+ O_p(\frac{1}{\sqrt{N}}) 
\)
for $x \sim \mathcal{N}(0, 1)$.
The result obtains by taking  $d_N := (2\sqrt{2 + \tau_N^2})/(1+\tau_N^2)$ and $c_N := -d_N z_{\frac{1-\alpha}{2}},$
and noting that the lower order term does not influence the limiting distribution of $d_N\inv\left[ (W_N^*- b_N^*)-c_N\right].$
\end{proof}
\section{Logistic regression supplementary material}\label{sec:logistic_regression_supp}
This section provides supplementary information related to \Cref{sec:logistic_regression}.
We begin by reviewing notation for convenience in \Cref{sec:logreg_preliminaries}.
In \Cref{sec:logreg_approximation_rate_proof} we then provide a proposition demonstrating the asymptotic rate of convergence of the approximation of the MAP estimate to the exact MAP estimate,
as well a proof and supporting lemmas.
\Cref{sec:logreg_asymptotic_coverage_proof} then provides a proof of \Cref{thm:logistic_asymptotic_coverage}.
\Cref{sec:logreg_simulation_details} gives additional details on the simulation experiments.

\subsection{Preliminaries and notation}\label{sec:logreg_preliminaries}
Consider logistic regression with random $N$-vector covariates $x_1, x_2, \dots$ and responses $y_1, y_2, \dots$, where for each data point $m,$
$y_m \mid x_m, \theta \sim (1+\text{exp}\{-x_m^\top \theta\})\inv\delta_{1} + 
(1+\text{exp}\{x_m^\top \theta\})\inv\delta_{-1}
$
for some unknown parameter $\theta\in\R^N$.
We use $X_M = [x_1, x_2, \dots, x_M]^\top$ and 
$Y_M = [y_1, y_2, \dots, y_M]^\top$ to denote the first $M$ data points.

One choice of an estimate for $\theta$ after observing $M$ observations is the MLE,
$$
\hat \theta_M := \argmax_\theta \log p(Y_M \mid X_M, \theta).
$$

Another possibility is the MAP estimate under a standard normal prior
$$
\theta^*_M := \argmax_\theta \log p(Y_M \mid X_M, \theta)  - \frac{1}{2}\|\theta\|^2 .
$$

The approach in \Cref{sec:logistic_regression} involves an approximation to this estimate involving a Gaussian approximation to the likelihood, defined by a 2nd order Taylor approximation of the log posterior formed at $\hat \theta_M.$
In particular, by Bayes' rule, the log posterior is, up to an additive constant,
$$
\log p_M(\theta) := \log p(Y_M\mid X_M, \theta) - \frac{1}{2} \|\theta\|^2
$$
and we use the approximation
\[\label{eqn:approx_posterior}
\log \tilde p_M(\theta) := \log p(Y_M \mid  X_M, \hat \theta_M) - \frac{1}{2}\|\theta\|^2 
- \frac{1}{2} (\theta - \hat \theta_M)^\top H_M(\hat \theta_M)  (\theta - \hat \theta_M),
\]
where $H_M(\hat \theta_M) = \nabla_\theta^2 -\log p(Y_M \mid X_M, \theta)\big|_{\theta=\hat \theta_M}$ is the Hessian of the negative log likelihood,
computed at the MLE.

The approximation we use for computing our proposed bound is then the maximizer of this approximation
$$
\tilde \theta^*_M := \argmax_\theta \log \tilde p_M(\theta).
$$
In \Cref{sec:logistic_regression} we found that we could express $\tilde \theta^*_M$ as 
$$
\tilde \theta^*_M = \left[ I_N +  \tilde \Sigma_M \right]\inv \hat \theta_M,
$$
where $\tilde \Sigma_M := H_M(\hat \theta_M)\inv$ is an approximation to the covariance of $\hat \theta_M.$
This solution may be seen by considering the first order optimality condition (i.e. setting the gradient of $\log \tilde p_M(\theta)$ to zero).

\subsection{Asymptotic approximation quality}\label{sec:logreg_approximation_rate_proof}
We here show that, in the large sample limit, $\tilde \theta^*$ provides a very close approximation of the MAP estimate, $\theta^*.$

\begin{prop}[Asymptotic approximation quality]\label{prop:logistic_asymptotic_approximation_quality}
Consider Bayesian logistic regression with a Gaussian prior $\theta \sim \mathcal{N}(0,I_N)$.
Let $x_1, x_2, \dots$ be a sequence of random  i.i.d.\ covariates satisfying $\E[x_m x_m^\top] \succ 0$ 
and with bounded third moment,
and let $y_1, y_2, \dots$ be responses distributed as in \Cref{eqn:logistic_likelihood}.
Denote by $X_M:=[x_1, x_2, \dots, x_M]^\top$ and 
$Y_M:=[y_1, y_2, \dots, y_M]^\top$ the covariates and labels of the first $M$ data points.
Consider the MAP estimate of $\theta$ after observing $M$ data points,
\[
\theta^*_M := \argmax_\theta p(\theta | Y_M, X_M)
\text{ and the approximation }
\tilde \theta_M^* := \left[ I_N + \tilde\Sigma_M\right]\inv \hat\theta_M,
\]
where $\hat \theta_M := \argmax_\theta p(Y_M| X_M ; \theta)$ and 
$\tilde \Sigma_M :=   \left[ -\nabla_\theta^2 \log p(Y_M| X_M;\theta)\big|_{\theta=\hat \theta_M}\right]\inv.$
Then $\|\tilde \theta_M^* - \theta_M^*\| \in O_p(M^{-2}),$
where $O_p$ denotes stochastic convergence in probability.
\end{prop}

The $O_p(M^{-2})$ convergence rate established in \Cref{prop:logistic_asymptotic_approximation_quality} is very fast in comparison
to the $O_p(M\nsqrt)$ convergence rate of the MLE,
as well as to the $O_p(M\inv)$ rate of convergence of the MAP to the posterior mean.
Notably, this asymptotic rate is consistent with rates observed in simulation (\Cref{fig:log_reg_convergence}).

\begin{proof}
We here show that $\| \theta^*_M - \tilde \theta^*_M \|$ is $O_p(M^{-2})$.
Our route to proving this relies on \Cref{lemm:strong-convexity} \citep[Lemma E.1]{trippe2019lr},
which will provide a sequence of bounds on $\| \theta^*_M - \tilde \theta^*_M \|$ that depend on the norms of the gradients of $\log p_M(\cdot)$ at $\tilde \theta^*_M,$
$c_M :=\| \nabla_\theta \log p_M(\tilde \theta_M^*)\|,$
and a sequence of strong log-concavity constants $\alpha_M$ for $\log p_M(\cdot)$ which hold on the interval
$\{ t \theta^*_M + (1-t) \tilde \theta^*_M | t \in[0, 1]\}.$
In particular, \Cref{lemm:strong-convexity} provides that $\| \theta^*_M - \tilde \theta^*_M \| \le \frac{c_M}{\alpha_M}$
and we obtain the result by showing that $\alpha_M$ grows as $\Omega_p(M)$ and 
$c_M$ drops as $O_p(M\inv)$.

We first use \Cref{lemma:strong_log_concavity_rate} to show that the strong log-concavity constants of $\log p_M$ in a neighborhood of radius $\epsilon$ of $\theta, B_\epsilon(\theta)$ grow as $\Omega_p(M).$
This allows us to establish that $\|\tilde \theta^*_M - \hat \theta_M\|$ is $O_p(M\inv)$ (\Cref{lemma:rate_of_tilde_to_hat}).
Since both $\hat \theta_M$ and $\theta^*_M$ converge strongly to $\theta$ under these conditions (see e.g. \citet[Theorem 10.10]{van2000asymptotic}),
the interval
$\{ t \theta^*_M + (1-t) \tilde \theta^*_M | t \in[0, 1]\}$
is then contained within $B_\epsilon(\theta)$ with probability approaching $1.$
Consequently, the constants of strong log concavity of $\log p_M$ on this interval,
which we take as $\alpha_1, \alpha_2, \dots,$ must grow as $\Omega_p(M)$ as well.

Now all that remains is to show that $c_M$ drops as $O_p(M\inv).$
Recall from above that $\|\tilde \theta_M^* - \hat \theta_M\|$ is $O(M\inv).$
This fact and the boundedness of the higher derivatives of $ \nabla \log p_M$ 
will allow us to use Taylor's theorem to obtain the desired rate.

However, before proceeding to a more detailed derivation of this rate, we introduce some additional notation.
Let $\phi(y, a)$ denote the GLM mapping function, such that
\(
\phi(y, a=x^\top \theta) &= \log p(y | x, \theta) \\
&= -\log(1 + \text{exp}\{ -y x^\top \theta \})
\)
and note that all higher derivatives with respect to $a$ are bounded.
In particular, third derivative satisfies
$$
\phi^{\prime\prime\prime}(a) := \frac{d^3}{d a^3} \phi(y, a) \le \frac{1}{6 \sqrt{3}},
$$
where we have dropped $y$ as an argument, because these higher derivatives do not depend on $y$.

We now proceed to derive a stochastic rate of convergence of $\|\nabla_\theta \log p_M(\tilde \theta_M^*)\|$.
We obtain this through a long derivation involving a series of upper bounds.
\(
\|\nabla_\theta \log p_M(\tilde \theta_M^*)\| 
&= \|\nabla_\theta (\log p_M - \log \tilde p_M)(\tilde \theta^*_M)\|  \\
&= \|\nabla_\theta (\log p_M - \log \tilde p_M)(\hat \theta_M) 
+ (\tilde \theta_M^* - \hat \theta_M)^\top \nabla_\theta^2 (\log p_M - \log \tilde p_M)(\theta_M^\prime) \|  \\
&\text{ // By Taylor's theorem, for some } \theta_M^\prime \in \{t\hat \theta_M + (1-t) \tilde \theta^*_M | t \in [0, 1]\} \\
&= \| (\tilde \theta_M^* - \hat \theta_M)^\top \nabla_\theta^2 (\log p_M(\theta_M^\prime) - \log \tilde p_M(\theta_M^\prime) ) \|  \\
&\text{ // Since } \nabla_\theta \log \tilde p_M(\hat \theta) = \nabla_\theta \log p_M(\hat \theta)\\
&= \| (\tilde \theta_M^* - \hat \theta_M)^\top\left[ 
\nabla_\theta^2 \log p(Y_M | X_M, \theta_M^\prime) - \nabla_\theta^2 \log p(Y_M | X_M, \hat \theta_M)\right] \|  \\
&\text{ // Since } \log \tilde p_M \text{ is a second degree approximation defined at }\hat \theta_M \\
&\le \| \tilde \theta_M^* - \hat \theta_M\|\left[ \sum_{m=1}^M
\opnorm{\nabla_\theta^2 \log p(y_m | x_m, \theta^\prime_M) - \nabla_\theta^2 \log p(y_m | x_m, \hat \theta_M)} \right]   \\
&= \| \tilde \theta_M^* - \hat \theta_M\|\left[ \sum_{m=1}^M \|\theta_M^\prime - \hat \theta_M\|\cdot
\opnorm{\int_{t=0}^1 \frac{\partial}{\partial t} \nabla_\theta^2 \log p(y_m | x_m, \theta)\big|_{\theta = t \hat \theta_M + (1-t)\theta_M^\prime}} \right]   \\
&\text{ // By the fundamental theorem of calculus} \\
&\le \| \tilde \theta_M^* - \hat \theta_M\|^2\left[ \sum_{m=1}^M
\opnorm{\int_{t=0}^1 \frac{\partial}{\partial t} \nabla_\theta^2 \log p(y_m | x_m, \theta)\big|_{\theta = t \hat \theta_M + (1-t)\theta_M^\prime}} \right]   \\
&\le \| \tilde \theta_M^* - \hat \theta_M\|\left[ \sum_{m=1}^M
\|x_m\|^3 (\text{max}_{a} \phi^{\prime\prime\prime}(a))\right] \\
&= \frac{1}{6\sqrt{3}}\| \tilde \theta^*_M - \hat \theta_M\|^2\left[ \sum_{m=1}^M \|x_m\|^3 \right] \\
&\le O_p(\frac{1}{M^2})O_p(M) = O_p(\frac{1}{M}),
\)
where the final line requires that the covariates have bounded third moment.
\end{proof}

\paragraph{Supporting Lemmas}
\begin{lemma}[Trippe et al., 2019, Lemma E.1]\label{lemm:strong-convexity}
    Let $f, g$ be twice differentiable functions mapping $\R^N \rightarrow \R$ and attaining minima at $\theta_f = \argmin_\theta f(\theta)$ and $\theta_g= \argmin_\theta g(\theta)$, respectively.
    Additionally, assume that $f$ is $\alpha$--strongly convex for some $\alpha >0$ on the set $\{t \theta_f + (1-t)\theta_g | t \in [0, 1] \}$ and that $\| \nabla_\theta f(\theta_g) - \nabla_\theta g(\theta_g)\|_2= \|\nabla_\theta f(\theta_g)\|_2 \le c$. Then
\[
\| \theta_f - \theta_g \|_2 \le \frac{c}{\alpha}.
\]
\end{lemma}

\begin{lemma}[uniform law of large numbers]\label{lemm:uniform_law}
Let $H_M(\theta)$ be as defined in \Cref{eqn:approx_posterior} and define $H(\theta) := \E[\nabla_\theta^2 \log p(y_1 | x_1 ; \theta)]$,
where the expectation is taken under the true $\theta$.
If $\E[x_1 x_1^\top]$ exists and is positive definite then
$$
\sup_{\theta^\prime \in B_\epsilon(\theta)} \| \frac{1}{M}H_M(\theta^\prime) - H(\theta^\prime)\|_2 \overset{a.s.}{\rightarrow} 0.
$$
according to $p$,
where $B_\epsilon(\theta)$ is a closed neighborhood of $\theta$ of radius $\epsilon,$ for any $\epsilon>0.$
\end{lemma}
\begin{proof}
    Since the each of the $M$ data points $\{(x_m, y_m)\}_{m=1}^\infty$ are i.i.d.\ by assumption, $M\inv H_M$ converges point-wise by the law of large numbers.
However, we are additionally interested in uniform convergence; a number of different uniform laws of large numbers suffice for this.
Because $H$ is continuously differentiable in $\theta$ (recall that for any $x_m$, $\frac{d^3}{d\theta^3} \log p(y_m | x_m, \theta)$ is bounded) it is therefore Lipschitz continuous on the bounded set $B_\epsilon(\theta)$.
As such one can construct a bounded envelope for $H$ on this set, which amounts to a sufficient condition for uniform convergence on $B_\epsilon$, see \citet[Theorem 19.4 - Glivenko-Cantelli]{van2000asymptotic}.
We refer the reader to \citet[Chapter 19]{van2000asymptotic} for technical background,
and in particular to \citet[Example 19.8]{van2000asymptotic} which walks through an example closely related to the present case.
\end{proof}

\begin{lemma}\label{lemma:strong_log_concavity_rate}
{\sloppy Consider logistic regression with random covariates, $x_1, x_2, \dots.$ 
Let $B_\epsilon(\theta)$ be a closed neighborhood of radius $\epsilon>0$ around $\theta$ and for each $M$ define 
$$\alpha_M := \inf_{\theta^\prime \in B_\epsilon(\theta)} \lambda_{min}\left[\nabla_\theta^2 \log p_M(\theta^\prime)\right]$$
to be the constant of strong log-concavity constant of $\log p_M(\cdot)$ on $B_\epsilon(\theta),$
where $\lambda_{min}(\cdot)$ denotes the smallest eigenvalue of its matrix argument.
If the covariates are i.i.d.\ and satisfy $\E[x_1 x_1^\top ] \succ 0,$
then $\alpha_M$ is $\Omega_p(M).$}
\end{lemma}
\begin{proof}
Consider the scaled Hessians of $\log p_M(\cdot)$, $M\inv H_M(\cdot).$
By \Cref{lemm:uniform_law}, $M\inv H_M(\cdot)$ converges uniformly to its expectation, $H(\theta) := \E[\nabla_\theta^2 \log p(y_1| x_1, \theta)]$ on $B_\epsilon(\theta).$
Since $H(\theta)\succ 0$ on $B_\epsilon(\theta),$ we have that 
$$\inf_{\theta^\prime \in B_\epsilon(\theta)} \lambda_{min}(\frac{1}{M}H_M(\theta)) \overset{a.s.}{\rightarrow} 
\inf_{\theta^\prime \in B_\epsilon(\theta)} \lambda_{min}(H_M(\theta)) >0.$$
Therefore 
$\alpha_M := \inf_{\theta^\prime \in B_\epsilon(\theta)} \lambda_{min}\left[\nabla_\theta^2 \log p_M(\theta^\prime)\right]$
is $\Omega_p(M).$
\end{proof}

\begin{lemma}\label{lemma:rate_of_tilde_to_hat}
Let $\hat \theta$ and $\tilde \theta^*$ be the MLE and the approximation to the MAP defined in \Cref{eqn:approx_MAP}, respectively.
If the covariates, $x_1, x_2, \dots$ are i.i.d.\ and satisfy $\E[x_1 x_1^\top ] \succ 0,$
then $\|\hat \theta_M - \tilde \theta^*_M\|$ is $O_p(M\inv).$
\end{lemma}
\begin{proof}
Recall that
$$
\tilde \theta^*_M = \left[ I_N +  \tilde \Sigma_M \right]\inv \hat \theta_M,
$$
where $\tilde \Sigma_M := H_M(\hat \theta_M)\inv.$
\Cref{lemma:strong_log_concavity_rate} provides that the constants of strong log-concavity for $\log p_M$ grow as $\Omega_p(M)$ in a neighborhood of $\theta.$
Therefore, since $\hat \theta_M$ converges strongly to $\theta,$ we can see that $\lambda_{min}(H_M(\hat \theta_M))$ is $\Omega_p(M).$
Next, we rewrite 
\(
\|\tilde \theta^*_M - \hat \theta_M\| &= 
\|\left[I_N + \tilde \Sigma_M \right]\inv \hat \theta_M- \hat \theta_M\| \\
&=\|\left[I_N + H_M(\hat \theta_M) \right]\inv \hat\theta_M\| \\
&\le\opnorm{\left[I_N + H_M(\hat \theta_M) \right]\inv} \|\hat \theta_M\| \\
&\le\frac{\|\hat \theta_M\|}{\lambda_{min}\left(H_M(\hat \theta_M)\right)}.
\)
which one can see is $O_p(M\inv)$ since $\|\hat \theta_M\|$ is bounded in probability.
\end{proof}

\subsection{Proof of \Cref{thm:logistic_asymptotic_coverage}}\label{sec:logreg_asymptotic_coverage_proof}
Before proving the theorem we begin by explicitly writing out the win and our proposed bound defined in \Cref{sec:logistic_regression}.
For clarity, we introduce a subscript $M$ to index the size of the dataset on which these quantities are computed.
Specifically, recalling that in this case we have $A=I_N$ and $C=(I_N + \tilde \Sigma_M)\inv,$ and noting that
therefore
$A-C = (I_N + \tilde \Sigma_M\inv)\inv,$
we have
\(
b_M(\alpha) &= 2\trace[ (I_N + \tilde \Sigma_M\inv)\inv \tilde \Sigma_M ] + { }\\
    &2 z_{\frac{1-\alpha}{2}}\sqrt{
U_M(\| G_M(\hat \theta_M) \|_{\tilde \Sigma_M}^2, \frac{1-\alpha}{2})  
+ 2\|\tilde \Sigma_M\msqrt (I_N + \tilde \Sigma_M\inv)\inv\tilde \Sigma_M \msqrt\|_F^2
}- \| \tilde \theta_M^* - \hat \theta_M\|^2 
\)
where  $G_M(\hat \theta_M) := (I_N + \tilde \Sigma_M\inv)\inv \hat \theta_M$ and
\[\label{eqn:logistic_U}
    U_M&(\|G_M(\hat \theta_M)\|_{\tilde\Sigma_M}^2, 1-\alpha)   
:= \inf_{\delta>0} \bigg\{ \delta \, \bigg|\, 
    \|G_M(\hat \theta_M)\|_{\tilde \Sigma_M}^2 \le 
    (\delta + \|\tilde \Sigma_M\msqrt (I_N + \tilde \Sigma_M\inv)\inv
    \tilde \Sigma_M\msqrt\|_F^2) +  { }\\
    &z_{1-\alpha}\sqrt{
        2\|\tilde \Sigma_M\msqrt  (I_N + \tilde \Sigma_M\inv)\inv
        \tilde \Sigma_M  (I_N + \tilde \Sigma_M\inv)\inv
        \tilde \Sigma_M \msqrt\|_F^2
        + 4 \opnorm{\tilde \Sigma_M\msqrt (I_N + \tilde \Sigma_M\inv)\inv
        \tilde\Sigma_M\msqrt
    }^2\delta 
    } 
\bigg\}
\]
is an approximate high-confidence upper bound on $\|G_M(\hat \theta_M)\|_{\tilde \Sigma_M}^2.$
For convenience, we abbreviate $U_M(\|G_M(\hat \theta_M)\|_{\tilde\Sigma_M}^2, 1-\alpha)$ by $U_M.$

Next, we recall that we may decompose the win in squared error loss for using $\theta^*_M$ in place of $\hat \theta_M$ as 
$$
W_M(\theta) = 2\epsilon_M^\top (I_N + \tilde\Sigma_M\inv)\inv \hat \theta - \| \theta^*_M - \hat \theta_M\|^2,
$$
where $\epsilon_M := \hat \theta_M - \theta.$

\begin{proof}
Proving the theorem amounts to showing that for any $\theta$ and $\alpha \in (0,1)$,   
$$\lim_{M\rightarrow \infty} \pr_\theta\left[W_M(\theta) \ge b_M(\alpha)\right] \ge \alpha.$$
\Cref{lemma:logistic_W_b_dist} provides that $M^{1.5}(W_M(\theta) -b_M(\alpha))$ converges in distribution to 
$2\sqrt{\theta^\top H(\theta)^{-3}\theta}(\delta- z_{\frac{1-\alpha}{2}}),$ for $\delta \sim \mathcal{N}(0,1).$
Thus for any $\theta,$ 
$\pr_\theta\left[W_M(\theta)-b_M(\alpha) > 0\right] \rightarrow (1-\Phi(z_{\frac{1-\alpha}{1}})) = 1 - \frac{1-\alpha}{2}>\alpha.$
This establishes that $b_M(\cdot)$ has above nominal coverage asymptotically, as desired.
\end{proof}

\begin{lemma}\label{lemma:U_convergence_rate}
    $|U_M - \| \tilde \Sigma_M \theta\|_{\tilde \Sigma_M}^2 |$ is $O_p(M^{-3.5}).$
\end{lemma}
\begin{proof}
Recall that we can rearrange \Cref{eqn:logistic_U} to see that $U_M$ satisfies
\(
\|(I+\tilde \Sigma_M\inv)\inv\hat \theta_M\|_{\tilde \Sigma_M}^2 &= U_M + 2\|\tilde\Sigma_M^2(I_N + \tilde\Sigma_M)\inv\|_F^2 +  {} \\
&\sqrt{\|\tilde \Sigma_M^4(I_N + \tilde \Sigma_M)^2\|_F^2 + 4 \opnorm{\tilde \Sigma_M^2(I_N+\tilde \Sigma_M)\inv}^2 U_M}
\)
where we have simplified $\tilde\Sigma_M\msqrt(I_N + \tilde\Sigma_M\inv)\inv\tilde\Sigma_M\msqrt$ to 
$\tilde \Sigma_M^2(I_N +\tilde \Sigma_M)\inv.$

We next further simplify the condition above by replacing two quantities with simplifying approximations plus lower order terms.
First note that we may write 
\(
\|(I+\tilde \Sigma_M\inv)\inv\hat \theta_M\|_{\tilde \Sigma_M}^2 &= \|\tilde \Sigma_M \hat \theta_M - \tilde \Sigma_M^2(I_N + \tilde \Sigma_M)\inv\hat \theta_M\|_{\tilde \Sigma_M}^2 \\
                                                                 &= \|\tilde \Sigma_M \hat \theta_M\|_{\tilde \Sigma_M}^2 + \|\tilde \Sigma_M^2(I_N + \tilde \Sigma_M)\inv\hat \theta_M\|_{\tilde \Sigma_M}^2 - 
2 \hat \theta_M^\top \tilde \Sigma_M^4(I_N + \tilde \Sigma_M)\inv\hat\theta_M \\
&= \|\tilde \Sigma_M (\theta + \epsilon_M)\|_{\tilde \Sigma_M}^2 + O_p(M^{-4})\\
&= \|\tilde \Sigma_M \theta\|_{\tilde \Sigma_M}^2 + \|\tilde \Sigma_M \epsilon_M\|_{\tilde \Sigma_M}^2 + 2\epsilon_M^\top\tilde\Sigma_M^3 \theta + O_p(M^{-4})\\
&= \|\tilde \Sigma_M \theta\|_{\tilde \Sigma_M}^2 + O_p(M^{-3.5}).
\)
Second, we write
\(
\sqrt{\|\tilde \Sigma_M^4 (I_N + \tilde \Sigma_M)^2\|_F^2 + 4 \opnorm{\tilde \Sigma_M^2(I_N+\tilde \Sigma_M)\inv}^2 U_M}
                         &=\sqrt{O_p(M^{-8})+ 4 \opnorm{\tilde \Sigma_M^2(I_N+\tilde \Sigma_M)\inv}^2 U_M} \\
                         &=2\opnorm{\tilde \Sigma_M^2(I_N+\tilde \Sigma_M)\inv}\sqrt{U_M} + O_p(M^{-4}).
\)
As such, we may see that $U_M$ satisfies
\begin{align}\label{eqn:logistic_norm_U_diff}
\begin{split}
\|\tilde \Sigma_M \theta_M\|_{\tilde \Sigma_M}^2 - U_M &= 2\|\tilde\Sigma_M^2(I_N + \tilde\Sigma_M)\inv\|_F^2 + 
2\opnorm{\tilde \Sigma_M^2(I_N+\tilde \Sigma_M)\inv}\sqrt{U_M} +O_p(M^{-3.5})\\
&= 2\opnorm{\tilde \Sigma_M^2(I_N+\tilde \Sigma_M)\inv}\sqrt{U_M} +O_p(M^{-3.5})
\end{split}
\end{align}
where we have dropped $2\|\tilde\Sigma_M^2(I_N + \tilde\Sigma_M)\inv\|_F^2$ since it is $O_p(M^{-4}).$

We next observe that $U_M$ must be $O_p(M^{-3}).$
Otherwise, the event that
$\|\tilde \Sigma_M \theta_M\|^2_{\tilde \Sigma_M} - U_M <0$ must occur infinitely often (since $\|\tilde \Sigma_M\theta\|_{\tilde \Sigma_M}^2$ is $O_p(M^{-3})$);
in turn, this condition would imply that $\opnorm{\tilde \Sigma_M^2(I_N + \tilde \Sigma_M)\inv }\sqrt{U_M}<0$ occurs infinitely often, 
which provides a contradiction.

Finally, in tangent with \Cref{eqn:logistic_norm_U_diff}, that $U_M$ is $O_p(M^{-3})$ allows us to see that $\left |U_M -  \|\Sigma\theta\|^2_{\tilde \Sigma_M}\right|$ is $O_p(M^{-3.5})$,
as desired.
\end{proof}

\begin{lemma}\label{lemma:logistic_W_b_dist}
Let $\alpha \in (0, 1)$ and $\theta \in \R^N.$
Consider the sequence of wins, $W_M(\theta)$, and bounds, $b_M(\alpha),$ computed for logistic regression.
Then
$$
M^{1.5}(W_M(\theta)-b_M(\alpha)) \overset{d}{\rightarrow} 2\sqrt{\theta^\top H(\theta)^{-3} \theta}(\delta - z_{\frac{1-\alpha}{2}}), 
$$
where $\delta \sim \mathcal{N}(0, 1).$
\end{lemma}
\begin{proof}
We prove the lemma by first writing $W_M$ and $b_M$ using simplifying approximations and lower order terms.
The result is obtained by manipulating a scaling of the difference between the two expressions and considering the limit in $M.$

Note first that we may write
\(
W_M(\theta) :&= 2\epsilon^\top (\theta^*_M - \hat \theta_M)- \|\theta^*_M - \hat \theta_M\|^2 \\
&= 2\epsilon^\top (\tilde \theta^*_M - \hat \theta_M)- \|\tilde \theta^*_M - \hat \theta_M\|^2 + O_p(M^{-2})\\
&= 2\epsilon^\top (I_N + \tilde \Sigma_M\inv)\inv \hat \theta_M - \|\tilde \theta^*_M - \hat \theta_M\|^2 + O_p(M^{-2})\\
&= 2\epsilon^\top \tilde \Sigma_M \hat \theta_M - \|\tilde \theta^*_M - \hat \theta_M\|^2 + O_p(M^{-2})\\
&= 2\epsilon^\top \tilde \Sigma_M \theta - \|\tilde \theta^*_M - \hat \theta_M\|^2 + O_p(M^{-2}).
\)

Next we write
\(
b_M(\alpha) &= 2\trace\left[(I_N + \tilde \Sigma_M\inv)\inv\tilde \Sigma_M\right]+ 2 z_{\frac{1-\alpha}{2}}\sqrt{
U_M + 2 \| \tilde \Sigma_M^2(I_N + \tilde \Sigma_M)\inv\|_F^2} - \|\tilde \theta_M^* - \hat \theta_M\|^2\\
&= 2 z_{\frac{1-\alpha}{2}}\sqrt{
\|\tilde \Sigma_M\theta\|_{\tilde \Sigma_M}^2 + O_p(M^{-3.5})} - \|\tilde \theta_M^* - \hat \theta_M\|^2 + O_p(M^{-2}) \\
&= 2 z_{\frac{1-\alpha}{2}}\|\tilde \Sigma_M\theta\|_{\tilde \Sigma_M}- \|\tilde \theta_M^* - \hat \theta_M\|^2 + O_p(M^{-2}).
\)
where the second line uses \Cref{lemma:U_convergence_rate}.

By considering a scaled difference between these two terms we find,
\(
M^{1.5}(W_M(\theta) - b_M(\alpha)) &=  2M^{1.5}\epsilon^\top \tilde \Sigma_M \theta 
- 2 M^{1.5}z_{\frac{1-\alpha}{2}}\|\tilde \Sigma_M\theta\|_{\tilde \Sigma_M}
+ O_p(M\nsqrt) \\
&\overset{d}{\rightarrow}
2M^{1.5} \|\tilde \Sigma_M\theta\|_{\tilde \Sigma_M}(\delta - z_{\frac{1-\alpha}{2}})
\)
for $\delta \sim \mathcal{N}(0, 1),$
by recognizing that $\epsilon_M$ is asymptotically normal with mean zero and covariance $\Sigma_M,$
and therefore that $2\epsilon^\top \tilde \Sigma_M \theta$ is asymptotically normal with variance $\|\tilde \Sigma_M\theta\|_{\tilde \Sigma_M}^2.$

Finally, the result obtains by noting that \Cref{lemm:uniform_law} implies that 
$$
M^{1.5} \|\tilde \Sigma_M\theta\|_{\tilde \Sigma_M} = \sqrt{\theta^\top (H_M(\hat \theta) / M)^{-3} \theta} 
\overset{a.s.}{\rightarrow} \sqrt{\theta^\top H(\theta)^{-3} \theta}.
$$

\end{proof}

\subsection{Empirical validation of logistic regression bound in simulation}\label{sec:logreg_simulation_details}
\begin{figure}[H]
    \centering
     \begin{subfigure}[b]{0.325\textwidth}
        \centering
        \includegraphics[width=\textwidth]{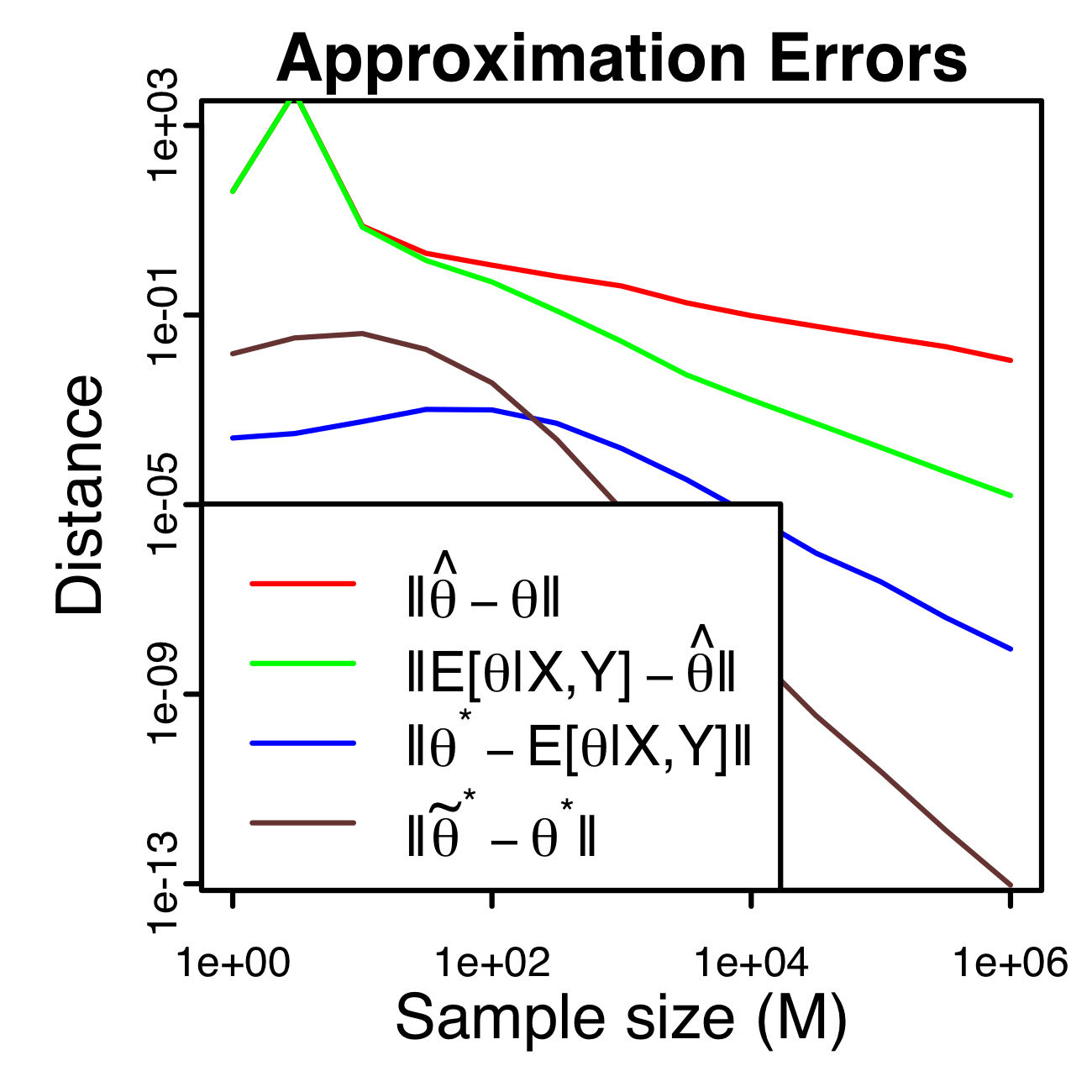}
        \caption{}\label{fig:log_reg_convergence}
     \end{subfigure}
     \begin{subfigure}[b]{0.325\textwidth}
        \centering
        \includegraphics[width=\textwidth]{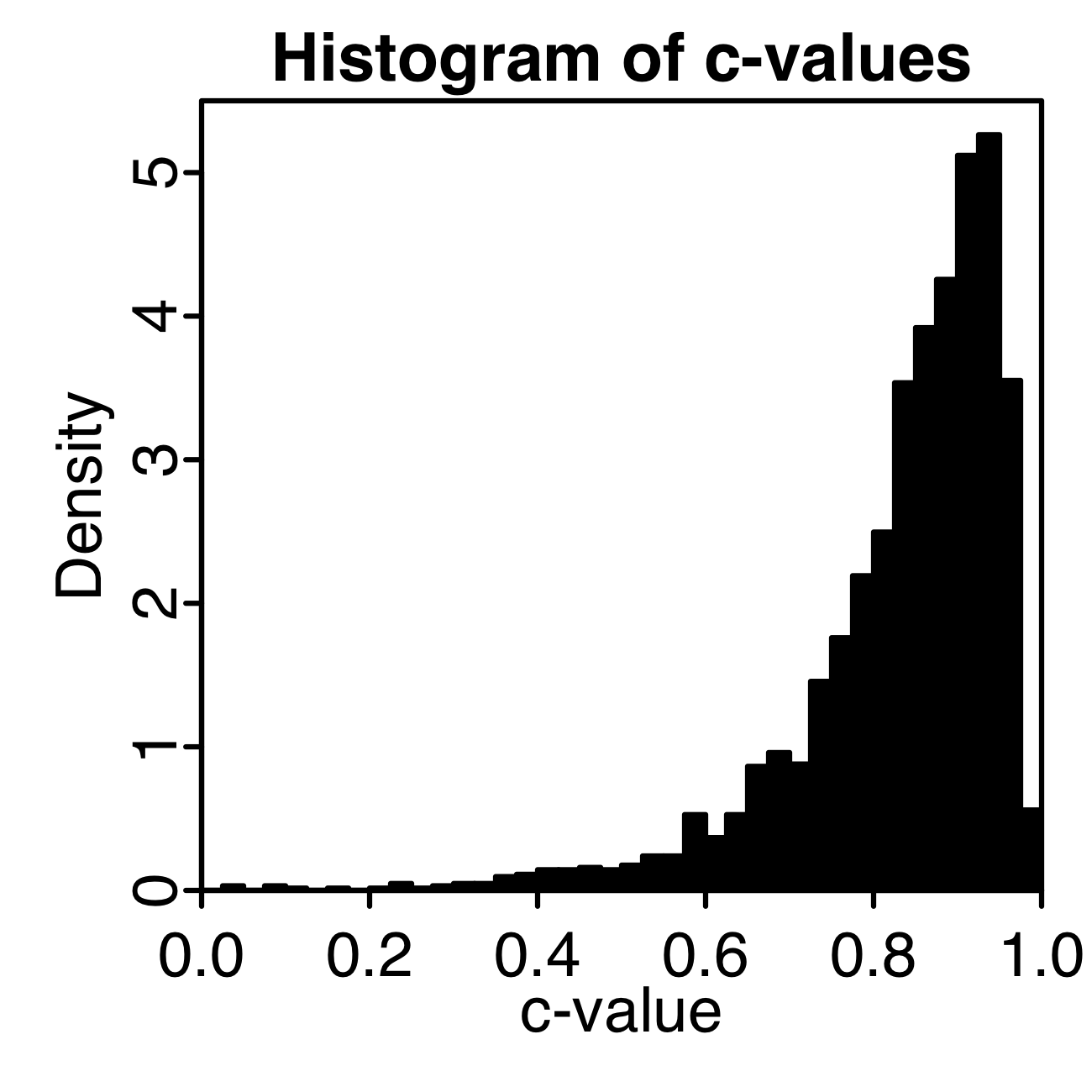}
        \caption{}\label{fig:log_reg_c_value_hist}
    \end{subfigure}
     \begin{subfigure}[b]{0.325\textwidth}
        \centering
        \includegraphics[width=\textwidth]{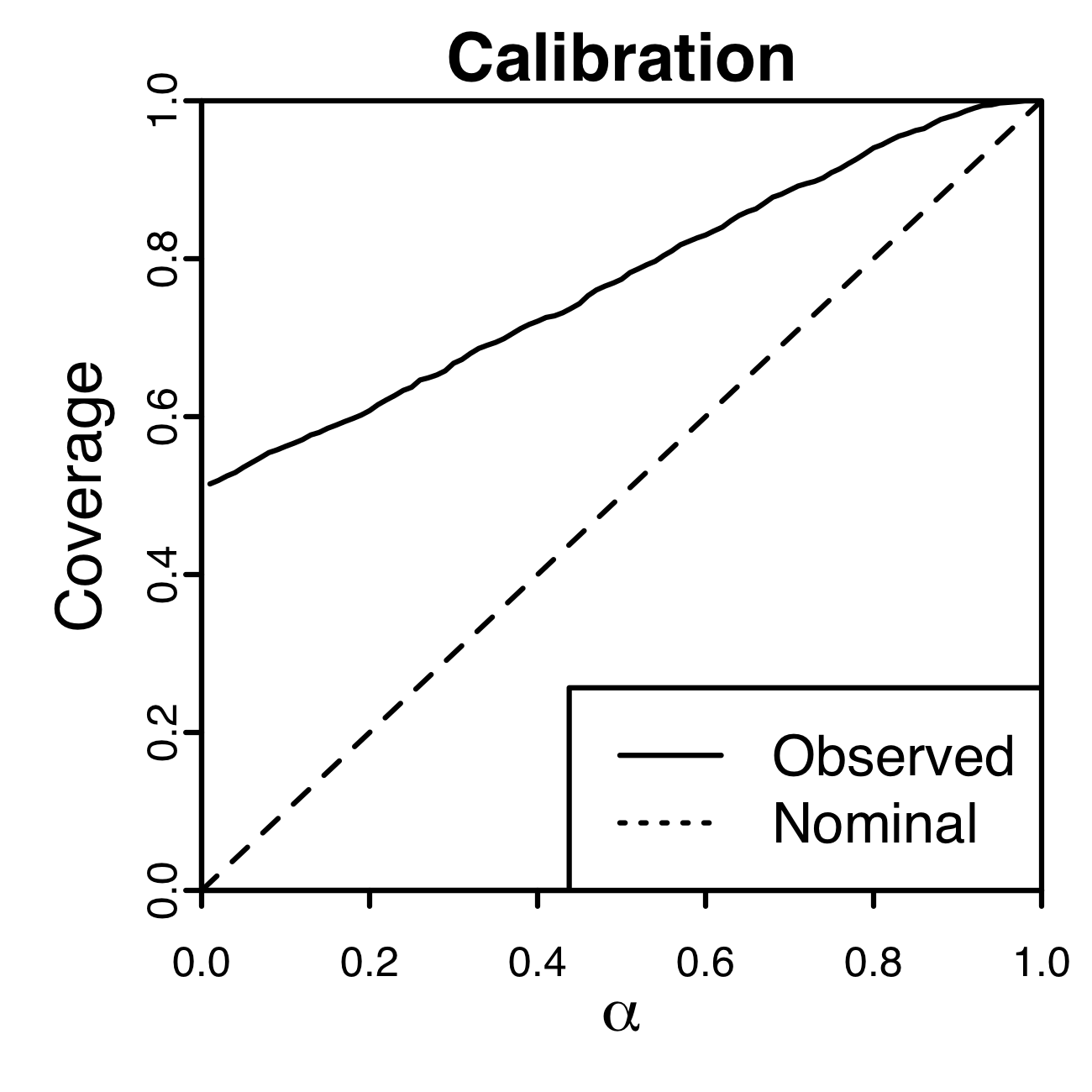}
        \caption{}\label{fig:log_reg_coverage}
    \end{subfigure}
\caption{c-values for logistic regression in two sets of simulations.
With $N=2,$ (a) empirical rates of convergence of distances amongst various estimates and the true parameter.
With $N=25$ and $M=1000$ (b) c-values are able to detect improvements, sometimes with high confidence, and
(c) the approximate bound has greater than nominal coverage.
See \Cref{sec:logreg_simulation_details} for details.
}
\end{figure}

We here demonstrate the fast convergence of our approximation to the MAP in logistic regression on simulated data.
We also include supplementary results illustrating the favorable performance of c-values in this setting, which is made possible by this fast convergence.
\Cref{fig:log_reg_convergence} shows the distance between various estimates and the true parameter for a range of sample sizes in simulation.
Due to the log-log scale, the slopes of the series in this plot reflect the polynomial rates of convergence.
Notably we see the fast $O_p(M^{-2})$ rate of convergence of our approximation to the MAP estimate, $\tilde \theta^*_M,$ to the exact MAP estimate, $\theta^*_M.$

\Cref{fig:log_reg_c_value_hist} demonstrates that our approach is able to detect improvements (i.e.\ we can obtain high c-values).
Furthermore, our proposed bound has similar coverage properties as in the Gaussian case (\Cref{fig:log_reg_coverage}).
In the experiments for \Cref{fig:log_reg_c_value_hist,fig:log_reg_coverage}, we simulated the parameter as $\theta\sim \mathcal{N}(0, \frac{1}{2}I_N)$ 
and, in each replicate, simulated the covariates for each data point, indexed by $m,$ as $x_m \overset{i.i.d.}{\sim}\mathcal{N}(0, N^{-2}I_N).$

Two of the series in \Cref{fig:log_reg_convergence} are distances between the posterior mean of $\theta$ and other estimates,
$\E[\theta| X, Y]  = \int p(\theta|X, Y) \theta d\theta.$
Because this model is non-conjugate, the estimate does not have an analytic form.
As such we approximated these quantities with Gauss-Hermite quadrature.
For each sample size $M,$ we performed $25$ replicate simulations.

In the experiments that went into \Cref{fig:log_reg_c_value_hist,fig:log_reg_coverage},
we used $N=25$ and $M=1000.$ 
See \texttt{logistic\_regression\_approximations.ipynb} and 
\linebreak 
\texttt{logistic\_regression\_c\_values\_and\_operating\_characteristics.ipynb} for details.
\section{Additional details on applications}\label{sec:applications_supp}
In this section, we provide additional details associated with the applications in \Cref{sec:applications}.
    \subsection{Estimation from educational testing data}\label{sec:small_area_estimation_supp}

\paragraph{Conservatism of c-values with the empirical Bayes step.}
The application in \Cref{sec:small_area_education} diverges from the scenarios covered by our theory in \Cref{sec:simpler_cases,sec:affine} in its use of the empirical Bayes step to estimate $\beta, \tau,$ and $\sigma.$
As a result, our theory does provide that $c(y)$ satisfies the guarantee of \Cref{thm:c_values}.
However, given the favorable asymptotic and empirical properties of the empirical Bayes procedure established in \Cref{sec:empirical_bayes}, we conjectured that the looseness in the lower bound $b(y,\alpha)$ would be sufficiently large to compensate for any error introduced by these departures from the assumptions of our theory.
\begin{figure}
    \centering
    \includegraphics[width=0.4\textwidth]{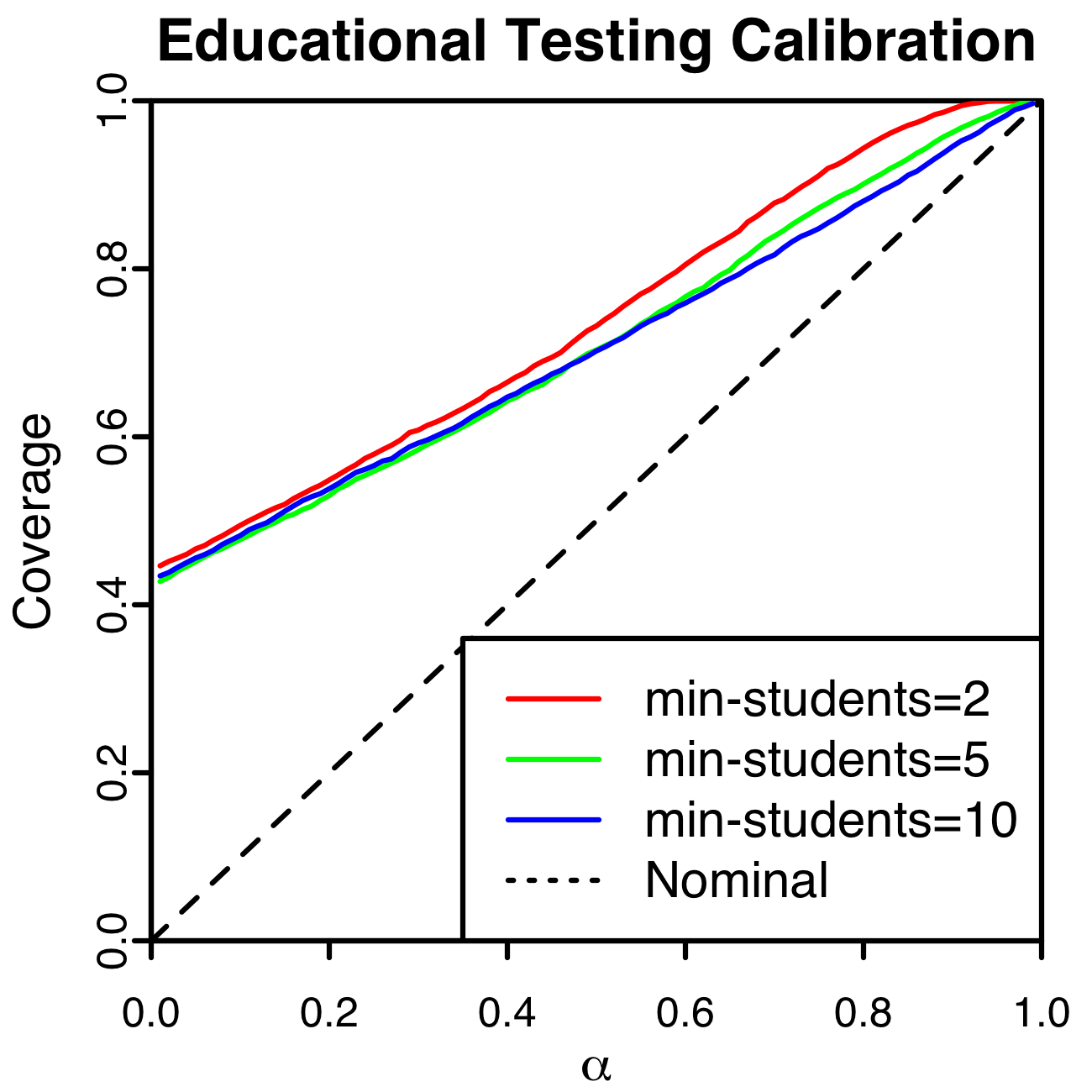}
\caption{
Calibration of the lower bounds $b(y, \alpha)$ in small area inference with an empirical Bayes step ($5000$ replicates).
The coverage on the y-axis is a Monte Carlo estimate of $\pr_\theta\left[W(\theta, y) \ge b(y, \alpha)\right].$
Each series corresponds to a set of simulations within which we excluded a different subset of schools based on a minimum number of students tested.
}\label{fig:els_calibration}
\end{figure}
To investigate this, we performed a simulation study in which we used this empirical Bayes step and confirmed that the c-values retained at least nominal coverage (\Cref{fig:els_calibration}).
To ensure that the simulated data had similar characteristics to the real data,
we simulated $5000$ datasets by drawing hypothetical school level means according the assumed generative model with the parameters ($\beta, \tau$ and $\sigma$) fit on the real dataset.
In each simulation, we re-estimated the fixed effects and variances (again using \texttt{lme4}), and computed the associated MLE, Bayes estimates, and bounds across a range of confidence levels.
We then computed the empirical coverage of these bounds and found them to be conservative across all tested levels.

\paragraph{Additional preprocessing and calibration details.}
\citet{hoff2019smaller} considered only schools at which $2$ or more students took the reading test.
We excluded an additional $8$ schools with fewer than $5$ students tested because we expected that the high variance in these observations could introduce too much slack into our bound as result of the poor conditioning of $\Sigma^{\frac{1}{2}} (A-C)\Sigma^{\frac{1}{2}}$ (recall the operator norm bound in \Cref{eqn:affine_norm_upper_bound}, derived in \Cref{eqn:affine_norm_upper_bound_derivation}).
Consistent with this hypothesis we computed a c-value of $0.88$ when we included these additional schools,
and when we further restricted to the $657$ schools with at least $10$ students tested we computed a c-value $0.999992.$
To further validate this hypothesis of increased conservatism we simulated additional datasets with these different thresholds on school size and evaluated the calibration of computed bounds (\Cref{fig:els_calibration}).
We observed the coverage for the simulations with smallest threshold was noticeably higher at large $\alpha$, in agreement with this hypothesis.
    \subsection{Estimation of violent crime rates in Philadelphia}\label{sec:philadelphia_supp}

\paragraph{Dependence on the order in which estimates are compared.}
In \Cref{sec:philadelphia} we chose to report one among three estimates as described in \Cref{remark:comparing_3_estimates}.
We note however that this paradigm is sensitive to the order in which the different estimates are considered.
For this set of three models, if we had first compared $\theta^\circ(y)$ as the alternative to $\hat \theta(y)$ as the default we would have rejected $\hat \theta(y)$  (with $c=0.99942$), and then again sided against updating our estimate a second time with a low c-value ($c=0.0$) for comparing $\theta^*(y)$ as the alternative against $\theta^{\circ}(y)$ as the default.
The potential cost of ending up with a worse estimate as a result of considering these estimates in sequence may be understood as a cost of looking at the data an additional time.

\paragraph{Selection of prior parameters from historical data.}
The parameters $\sigma_\delta^2, \sigma_z^2, \sigma_y^2$ were selected based on historical data.
Specifically, we estimated $\sigma_y^2$ and $\sigma_z^2$ as the averages of the sample variances of the violent and non-violent report rates, respectively,
computed within each census block in the preceding years.
For the first model described in \Cref{sec:philadelphia}, we then estimated $\sigma_\delta^2$ using these same historical data to reflect the prior belief that half of the variability across the unknown rates is common across the two response types.

For the second model considered,
we selected the signal variance and length scale of this covariance function by 
drawing hypothetical datasets of crime levels from the prior predictive distributions 
and selecting those which produced the most reasonable looking patterns.
In particular, we chose the length scale to be one sixth of the maximum distance between the centroids of census blocks,
and the signal variance to reflect the prior belief that one third of the variability in the unknown rates was explained by the spatial component.
In addition, we choose a smaller value for $\sigma_\delta^2$ in this second model, so that the total implied variance would be the same.
See supplementary code in \texttt{Philly\_reported\_crime\_estimation.ipynb} for additional details.

\paragraph{Derivation of $\theta^*$ (posterior mean in the first model).}
As mentioned in the main text, since the prior and likelihoods for this model are independent across each census block we can compute the posterior mean for each block independently.

Let $\pi(\cdot)$ denote the joint density of all variables.
Then, since $z_n \indep y_n \big| \theta_n$, we have that
\(
\pi(\theta_n | y_n, z_n) &\propto \pi(\theta_n|z_n)\pi(y_n | \theta_n, z_n) \\
&= \pi(\theta_n|z_n)\pi(y_n | \theta_n).
\)

Next observe that by construction,
 $z_n -\theta_n = \epsilon^z_n +\delta_n^z - \delta_n^y \sim \mathcal{N}(0,2\sigma_\delta^2 + \sigma_z^2 )$
and so $\theta_n| z_n \sim \mathcal{N}(z_n, 2\sigma_\delta^2 + \sigma_z^2 )$.
Since again by construction we have that $y_n |\theta_n \sim \mathcal{N}(\theta_n, \sigma_y^2),$
Gaussian conjugacy provides that
$$
\theta_n | y_n, z_n \sim \mathcal{N}(\E[\theta_n| y_n, z_n], \Var[\theta_n | y_n, z_n]),
$$
where 
\(
\Var[\theta_n | y_n, z_n]&=\frac{1}{\sigma_y^{-2} + (2\sigma_\delta^2 + \sigma_z^2)\inv}\\
&=\frac{\sigma_y^2(2\sigma_\delta^2 + \sigma_z^2)}{\sigma_y^2 + 2\sigma_\delta^2 + \sigma_z^2}
\) and 
\(
\E[\theta_n| y_n, z_n] &= \Var[\theta_n | y_n, z_n] (\Var[\theta_n | z_n]\inv \E[\theta_n | z_n] + \Var[y_n | \theta_n]\inv y_n) \\
&=\frac{\sigma_y^2(2\sigma_\delta^2 + \sigma_z^2)}{\sigma_y^2 + 2\sigma_\delta^2 + \sigma_z^2}
\left[(2\sigma_\delta^2 + \sigma_z^2)\inv z_n + \sigma^{-2}y_n\right] \\
&=\frac{2\sigma_\delta^2 + \sigma_z^2}{2\sigma_\delta^2 + \sigma_y^2 + \sigma_z^2}y_n + 
\frac{\sigma_y^2}{ 2\sigma_\delta^2 +\sigma_y^2 + \sigma_z^2}z_n
\)
as desired.

Analogously, for the second model considered in \Cref{sec:philadelphia} we find the posterior mean as
$$\theta^\circ(y) = \left[ I_N + \sigma_y^2(2K + 2\sigma_\delta^2 I_N + \sigma_z^2 I_N)\inv\right]\inv y + 
\left[ I_N + \sigma_y^{-2}(2K + 2\sigma_\delta^2 I_N + \sigma_z^2 I_N)\right]\inv z.$$

\paragraph{Additional dataset details.}
The data considered in this application are counts of police responses categorized as associated with violent crimes and violent crimes in October 2018.
These were obtained from \href{https://www.opendataphilly.org/}{opendataphilly.org}.
The observed data we model are the inverse hyperbolic sine transform of the number of recorded police responses per square mile. 
For all practical purposes, these values can be interpreted as log densities (see, e.g., \citet{Burbidge1988}).
    \subsection{Gaussian process kernel selection for estimation of ocean currents}\label{sec:submesoscale_supp}
We here provide additional details of the Gaussian process covariance functions used in \Cref{sec:submesoscale}.
The first covariance function described, which incorporated covariation at two scale is defined, for both the longitudinal and latitudinal components ($i$ in $\{1, 2\}$) and for each pair of buoys $n$ and $n^\prime$, as
\(
k(\theta^{(i)}_n, \theta^{(i)}_{n^\prime}) = &\sigma_1^2 \exp\left\{-\frac{1}{2}
\left[\frac{(\text{lat}_n - \text{lat}_{n^\prime})^2}{r_{1, \text{lat}}^2}  + 
    \frac{(\text{lon}_n - \text{lon}_{n^\prime})^2}{r_{1, \text{lon}}^{2}} +
    \frac{(t_n - t_{n^\prime})^2}{r_{1,t}^{2}} 
    \right]\right\} \\
+ &\sigma_2^2 \exp\left\{-\frac{1}{2}
\left[\frac{(\text{lat}_n - \text{lat}_{n^\prime})^2}{r_{2, \text{lat}}^2}  + 
    \frac{(\text{lon}_n - \text{lon}_{n^\prime})^2}{r_{2, \text{lon}}^{2}} +
    \frac{(t_n - t_{n^\prime})^2}{r_{2,t}^{2}} 
    \right]\right\},
\)
where $\sigma_1^2, r_{1, \text{lat}},  r_{1, \text{lon}}$ and $r_{1, t}$ parameterize the mesoscale variation in currents whereas
$\sigma_2^2, r_{2, \text{lat}},  r_{2, \text{lon}}$ and $r_{2, t}$ parameterize the submesoscale variation.
As in \citet{lodise2020investigating}, the latitudinal and longitudinal components of $F$ are modeled as a priori independent.
We choose these parameters by maximal marginal likelihood \citep[Chapter 5]{rasmussen2006gaussian} on an independent subset of the GLAD dataset.
Estimates of the underlying currents are obtained as the posterior mean of $F$ under this model, which we take as the alternative, $\theta^*(y)$.

The second covariance function captures covariation among observations only at the mesoscale.
In this case, the Gaussian process prior has covariance function
$$
k(\theta^{(i)}_n, \theta^{(i)}_{n^\prime}) = \sigma_1^2 \exp\left\{-\frac{1}{2}
\left[\frac{(\text{lat}_n - \text{lat}_{n^\prime})^2}{r_{1, \text{lat}}^2}  + 
    \frac{(\text{lon}_n - \text{lon}_{n^\prime})^2}{r_{1, \text{lon}}^{2}} +
    \frac{(t_n - t_{n^\prime})^2}{r_{1,t}^{2}} 
    \right]\right\} + \sigma_2^2 \mathbbm{1}[n=n^\prime], 
$$
which maintains the same marginal variance but excludes submesoscale covariances.
We take the posterior mean under this model as the default estimate $\hat \theta(y)$.
See \linebreak \texttt{submesoscale\_GP\_c\_value.ipynb} for further implementation details.
\end{document}